\documentclass[12pt]{article}
\usepackage[T1]{fontenc}
\usepackage[utf8]{inputenc}
\usepackage[letterpaper]{geometry}
\usepackage{lmodern}
\usepackage{jheppub}

\usepackage{parskip}
\usepackage{cleveref}
\crefformat{section}{#2§#1#3}
\crefformat{subsection}{#2§#1#3}
\crefname{appendix}{App.}{App.}
\crefname{figure}{Fig.}{Fig.}

\usepackage{tikz}
\usepackage{tikz-cd}
\usetikzlibrary{calc, backgrounds, shapes.symbols}

\usepackage{extarrows}
\usepackage{mathtools}
\usepackage{amsthm}
\usepackage{braket}
\usepackage{stackrel}
\usepackage{slashed}
\usepackage{stmaryrd}
\usepackage{bm}
\usepackage{mdframed}

\newcommand{\ph}{\mathrm{ph}}
\newcommand{\cA}{\mathcal{A}}
\newcommand{\cQ}{\mathcal{Q}}
\newcommand{\bZ}{\mathbb{Z}}
\newcommand{\bR}{\mathbb{R}}

\newcommand{\bF}{\mathbb{F}}
\newcommand{\cD}{\mathcal{D}}

\newcommand{\cZ}{\mathcal{Z}}
\newcommand{\cT}{\mathcal{T}}
\newcommand{\Arf}{\mathrm{Arf}}
\newcommand{\B}{\mathit{B}}

\newcommand{\RP}{\mathbb{RP}}
\newcommand{\CP}{\mathbb{CP}}
\newcommand{\HP}{\mathbb{HP}}

\newcommand{\PD}{\mathrm{PD}}
\newcommand{\pt}{\mathrm{pt}}
\newcommand{\MT}{\mathit{MT}}
\newcommand{\M}{\mathit{M}}
\newcommand{\SO}{\mathrm{SO}}
\newcommand{\Oo}{\mathrm{O}}
\newcommand{\KO}{\mathit{KO}}
\newcommand{\ko}{\mathit{ko}}
\newcommand{\Spin}{\mathrm{Spin}}
\newcommand{\Pin}{\mathrm{Pin}}
\newcommand{\pin}{\mathrm{pin}}
\newcommand{\NS}{\mathrm{NS}}
\newcommand{\R}{\mathrm{R}}
\newcommand{\CS}{\mathrm{CS}}

\DeclareMathOperator{\Sq}{Sq}
\DeclareMathOperator{\im}{im}
\DeclareMathOperator{\coker}{coker}
\DeclareMathOperator{\Hom}{Hom}
\DeclareMathOperator{\Tr}{Tr}
\DeclareMathOperator{\extpow}{{\wedge}}

\colorlet{a}{green!50!black}
\colorlet{b}{a}
\colorlet{c}{a}
\colorlet{d}{a}
\colorlet{e}{blue}
\colorlet{f}{cyan!80!black}

\newcommand{\mathtextover}[3][c]{\mathmakebox[\widthof{\(#3\)}][#1]{#2}}

\newcommand{\example}{\noindent \textbf{Example:}~}
\newenvironment{shadebox}{\begin{mdframed}[backgroundcolor=black!4,hidealllines=true,skipabove=
\topskip,skipbelow=\topskip]\noindent\slshape\ignorespaces}{\end{mdframed}}

\title{Bosonisation Cohomology: \\ Spin Structure Summation in Every Dimension}

\author[a,b]{Philip Boyle Smith}
\author[c]{and Joe Davighi}

\affiliation[a]{SISSA, via Bonomea 265, 34136 Trieste, Italy}
\affiliation[b]{INFN, Sezione di Trieste, via Valerio 2, 34127 Trieste, Italy}
\affiliation[c]{CERN, Esplanade des Particules 1, 1217 Meyrin, Switzerland}

\emailAdd{philip.boyle.smith@sissa.it}
\emailAdd{joseph.davighi@cern.ch}

\abstract{
Gauging fermion parity and summing over spin structures are subtly distinct operations.
We introduce `bosonisation cohomology' groups $H_B^{d+2}(X)$ to capture this difference, for theories in spacetime dimension $d$ equipped with maps to some $X$. Non-trivial classes in $H_B^{d+2}(X)$ contain theories for which $(-1)^F$ is anomaly-free, but spin structure summation is anomalous.
We formulate a sequence of cobordism groups whose failure to be exact is measured by $H_B^{d+2}(X)$, and from here we compute it for $X=\pt$.
The result is non-trivial only in dimensions $d\in 4\bZ+2$, being due to the presence of gravitational anomalies.
The first few are $H_B^4=\bZ_2$, probed by a theory of 8 Majorana--Weyl fermions in $d=2$, then $H_B^8=\bZ_8$, $H_B^{12}=\bZ_{16}\times \bZ_2$. We rigorously derive a general formula extending this to every spacetime dimension.
Along the way, we compile many general facts about (fermionic and bosonic) anomaly polynomials, and about spin and pin$^-$ (co)bordism generators, that we hope might serve as a useful reference for physicists working with these objects.
We briefly discuss some physics applications, including how the $H_B^{12}$ class is trivialised in supergravity. Despite the name, and notation, we make no claim that $H_B^\bullet(X)$ actually defines a cohomology theory (in the Eilenberg--Steenrod sense).
}

\begin{document}

\maketitle
\newpage

\section{Introduction} \label{sec:int}

Bosonisation is a transformation from a fermionic theory to a bosonic theory that was originally defined in two dimensions \cite{Coleman:1974bu, Mandelstam:1975hb}. Since then, it has been reformulated, refined and generalised in various ways,
to provide a web of dualities that help us understand the space of physically inequivalent quantum field theories -- an important challenge in theoretical physics.

In the beginning, bosonisation referred to an exact equivalence between two particular quantum field theories: a $(-1)^F$-gauged Dirac fermion and a free compact boson. An equivalent viewpoint is that of a transformation: starting from a Dirac fermion, one gauges the fermion parity symmetry $(-1)^F$ which converts it into a compact boson.

The above idea has since been extended to arbitrary spacetime dimension $d$, and to arbitrary input theories. The step to three dimensions was made with the beautiful story of \emph{flux attachment}, where one instead gauges a $U(1)$ fermion number symmetry, and a Chern--Simons term is responsible for transmuting the statistics of the particles from fermions to bosons \cite{Karch:2016sxi}. An analogous mechanism using $\bZ_2$ gauge fields also exists \cite{Chen:2017fvr}, which gives a slightly different version of 3d bosonisation that looks a lot more like its 2d counterpart. This latter mechanism admits a uniform generalisation to all dimensions \cite{Gaiotto:2015zta}, both in the continuum \cite{Chen:2019wlx} and on the lattice \cite{Tantivasadakarn:2020lhq, Chen:2021ppt}, with the lattice version generalising the Jordan--Wigner transformation in two dimensions. See also \cite{Cappelli:2025ecm} for a recent review.

These analyses typically assume that the theory to be bosonised is free of anomalies. In this paper, we are interested in the additional richness brought to the story when this is not the case. Even in the simplest case of two dimensions, it is known that something interesting and rather subtle can occur: a fermionic theory can carry a perturbative gravitational anomaly that still allows for gauging $(-1)^F$, but this anomaly cannot be carried by any bosonic theory. The $(-1)^F$-gauged theory therefore has no choice but to be fermionic, typically a new theory \cite{bb}. In $d=3$ by contrast, things are simpler, for lack of a suitable gravitational anomaly \cite{lorentzfrac}. But what happens in other dimensions? Perturbative gravitational anomalies exist in all dimensions $d \in 4\bZ + 2$, and one would expect them to have a similar effect. Global (discrete) gravitational anomalies also exist in various dimensions, but are more subtle to analyse. Can they also play a role? Our goal will be to answer such questions.

One additional motivation for our investigation is the connection between bosonisation and higher symmetry structures. For example, the Kramers--Wannier defect is the prototypical example of a non-invertible symmetry; under bosonisation it is dual to stacking an invertible fermionic phase, and under a lattice version of bosonisation, it is also understood as being emanant from a lattice translation \cite{Seiberg_2024}. This connection has been generalised to $d$ dimensions to produce various Kramers--Wannier-like defects \cite{Su:2025wbe}. Such defects naturally form part of the symmetry structure of a $(d+1)$-dimensional TQFT known as the SymTFT, which for non-anomalous bosonisation is a twisted $\bZ_2^{[d-2]}$ gauge theory. Our interest is in how this symmetry structure shifts when we add gravitational anomalies. Although we defer a fuller investigation to further work, we will write down the corresponding SymTFT, and comment on some of its properties that follow from our results. Other applications we will consider include the cancellation of fermionic anomalies by bosonic fields, particularly in 10d supergravity, where it turns out these bosonic fields must include secret fermionic dependence, as has recently been explained from a different point of view in \cite{Hsieh:2020jpj, Hosseini:2025oka}.

\subsection{A sequence of anomalies} \label{sec:int:seqanom}

In this paper the key questions we are interested in asking are: given a fermionic theory $\cT_F$ in $d$ spacetime dimensions, when can we gauge fermion parity $(-1)^F$? When can we sum over spin structures to bosonise? And when do these conditions differ?

The answers to all of the above questions are governed by anomalies. In recent years it has come to be understood that all anomalies in quantum field theory, including those for $(-1)^F$ and spin structure summation, are classified by a mathematical gadget called \emph{cobordism} (at least for unitary theories, which we assume). Roughly speaking (see \cref{sec:bordism} for slightly more precision), \emph{bordism} is an equivalence relation between manifolds equipped with certain structures -- like a spin structure or a background gauge field. Two such manifolds are equivalent under bordism if they can be smoothly connected by an interpolating manifold in one dimension higher. Anomalies, both perturbative and non-perturbative ones, are detected by invariants of manifolds that are constant on such bordism classes, in dimension $d+2$ (for perturbative anomalies) and $d+1$ (for non-perturbative ones). For example, a 2d perturbative anomaly for a $U(1)$ symmetry is captured by a 4d instanton number, which is quantised and does not change under a smooth interpolation of field configurations.
Mathematically, this means the anomaly is an element in an abelian group dual to bordism, namely a \emph{co}bordism group that we denote by $\mho_H^{d+2}(X)$ for theories with spacetime symmetry type $H$ (typically $\Spin$ or $\SO$) and equipped with maps to some $X$ (typically the classifying space $\B G$ for some group $G$).
This has been shown to follow, rigorously, from quite general properties of the anomaly theory~\cite{Kapustin:2014dxa,Witten_2016,Freed:2016rqq,Grady:2023sav}.

To get started, we first need to answer: when is $(-1)^F$ anomaly-free, and thus gaugable? Let us, for now, restrict ourselves to unitary theories $\cT_F$ in $d$ dimensions in which fermions are defined using a spin structure, and suppose we do not couple to any background gauge fields. If we think of $(-1)^F$ as an independent $\bZ_2$ symmetry, then its anomaly is an element in the cobordism group
\begin{equation} \label{eq:int:anomfparity}
    \text{Anomaly}[(-1)^F, \cT_F] \; \in \; \widetilde{\mho}^{d+2}_{\Spin}(\B\bZ_2)
\end{equation}
The symmetry is then gaugable if the anomaly is zero. For instance, in $d=2$, we have $\widetilde\mho^{4}_{\Spin}(B\bZ_2) \cong \bZ_8$; a valid choice of generator for this cobordism group is the anomaly for a single Majorana--Weyl fermion, and the mod-8 nature means that collections of such fermions can in fact be gapped (without breaking the $\bZ_2$) in groups of 8, matching the well-known symmetric mass generation story uncovered by Fidkowski and Kitaev~\cite{Fidkowski:2009dba, Ryu:2012he}.

Of course, fermion parity is not, in fact, an independent $\bZ_2$ symmetry. Because $(-1)^F \subseteq \Spin(d)$, its anomaly must be fully determined by the gravitational anomaly associated with $\Spin(d)$. Again assuming our fermions are defined with a spin structure (and turning off any other background gauge fields), the most general fermionic anomaly is captured by the spin cobordism group
\begin{equation} \label{eq:int:anomtf}
    \text{Gravitational-Anomaly}[\cT_F] \; \in \; \mho_\Spin^{d+2}(\pt)
\end{equation}
For instance, for $d=2$ we have $\mho_\Spin^{4}(\pt) \cong \bZ$, generated by $\sigma/16$ where $\sigma$ is the signature, which happens to equal the anomaly polynomial of a single Majorana--Weyl. We moreover expect there should be a map
\begin{equation} \label{eq:int:q}
    q \; : \; \mho_\Spin^{d+2}(\pt) \; \longrightarrow \; \widetilde\mho^{d+2}_{\Spin}(\B\bZ_2)
\end{equation}
that sends a gravitational anomaly for a particular fermionic system to the induced anomaly in $(-1)^F$. Explicitly, we can model $q$ by a map of partition functions $\mathcal{Z}$ for the associated $(d+1)$-dimensional anomaly theories, which are functions of the choice of spin structure $\rho$: we have $q : \cZ[\rho] \mapsto \frac{\cZ[\rho+A]}{\cZ[\rho]}$ where $A$ is a background gauge field for the $\bZ_2$. (We suppress an additional metric dependence of $\cZ$.)

The fermion parity symmetry of $\cT_F$ can thus be gauged when the map \eqref{eq:int:q} sends the gravitational anomaly \eqref{eq:int:anomtf} to the zero element in the cobordism group \eqref{eq:int:anomfparity} that classifies $\bZ_2$ anomalies. Thus, these theories form the kernel of the map $q$.

On the other hand, we also wish to know whether $\cT_F$ can be bosonised. In general, bosonisation can be defined as gauging $(-1)^F$ followed by stacking with a topological counterterm to remove the spin structure dependence.\footnote{Or, at the lattice level, applying a `disentangling unitary' \cite{Su:2025wbe}.} \cite{Gaiotto:2015zta} So the question is: when can this procedure produce a purely bosonic theory $\cT_B$? The key point is that, if so, then this process can be reversed (by first stacking the inverse counterterm, then gauging the quantum symmetry) to recover $\cT_F$ from $\cT_B$, a process known as \emph{fermionisation}. Since neither of these steps modifies the gravitational anomaly, this means the gravitational anomaly of $\cT_F$ must be induced from that of $\cT_B$.

A bosonic theory is defined not with a spin structure, but simply (we assume) with an orientation. Bosonic anomaly theories are therefore detected by the cobordism group
\begin{equation} \label{eq:int:anomtb}
    \text{Gravitational-Anomaly}[\cT_B] \; \in \; \mho_\SO^{d+2}(\pt)
\end{equation}
In order to be consistent on all oriented manifolds, an element in this cobordism group must make sense on a much larger set of manifolds than the corresponding element in spin cobordism. Because oriented manifolds are defined with strictly less structure than spin manifolds, an element of \eqref{eq:int:anomtb} will also make sense on a spin manifold (even though it may, of course, evaluate to zero on all spin manifolds). This means there is always a map
\begin{equation} \label{eq:int:p}
    p \; : \; \mho_\SO^{d+2}(\pt) \; \longrightarrow \; \mho_\Spin^{d+2}(\pt)
\end{equation}
defined by evaluating each oriented cobordism element only on spin manifolds. It can be interpreted as the map of gravitational anomalies induced by fermionisation.\footnote{The bosonised theory $\cT_B$ also has a quantum $\bZ_2^{[d-2]}$ symmetry with a fixed anomaly $(-1)^{\int A_{d-1} w_2}$ \cite{Gaiotto:2015zta, Cappelli:2025ecm}. Since it is fixed, and not part of the pure gravitational anomaly, for us it will play no role. It would be interesting to revisit this claim in view of symmetry fractionalisation \cite{lorentzfrac}.}
To get a feel for this, let us return to our illustrative example of $d=2$. The generator $\sigma/16$ we described above for spin cobordism only made sense because Rokhlin's theorem tells us that $\sigma(M_\Spin)\in 16 \bZ$ for all spin manifolds. However, a merely oriented 4-manifold can have $\sigma\in \bZ$ in general; for example, $\sigma(\CP^2)=1$. This means the corresponding generator of oriented cobordism is 16 times the generator of spin cobordism, \emph{i.e.}\ we deduce that $p:\mho_\SO^4(\pt) \to \mho_\Spin^4(\pt)$ sends $1 \mapsto 16$. So the map $p$ is certainly not surjective. Nor, in general, is it injective: there can, in particular, be additional torsion elements in oriented cobordism, thanks to the non-vanishing of the Stiefel--Whitney number $w_2$ which we can use to build bordism invariants: any such element would map to zero under $p$ because $w_2(M_\Spin)$ is trivial.

We argued above that a fermionic theory $\cT_F$ can be bosonised if and only if its gravitational anomaly \eqref{eq:int:anomtf} is induced from a bosonic anomaly \eqref{eq:int:anomtb} under the map \eqref{eq:int:p}. Thus, bosonisable theories form the image of the map $p$.

The astute reader will notice that we can chain together the maps $p$ and $q$ to form the following sequence:
\begin{equation} \label{eq:int:seq}
    \mho^{d+2}_\SO(\pt) \xrightarrow[\hspace{5ex}]{p} \mho^{d+2}_\Spin(\pt) \xrightarrow[\hspace{5ex}]{q} \; \widetilde{\mho}^{d+2}_\Spin(\B\bZ_2)
\end{equation}
To reiterate, the first map $p$ sends bosonic anomalies to their image in spin cobordism, while the second map $q$ sends a fermionic gravitational anomaly to the induced anomaly in fermion parity $(-1)^F$. This sequence, which we shall mathematically justify in \cref{sec:math_seq}, is our primary tool for investigating gravitationally-anomalous bosonisation in any dimension. It is clear that
\begin{equation}
    q \circ p = 0
\end{equation}
because $p$ hits only the bosonic theories, for which $(-1)^F$ is trivial and so necessarily anomaly-free. Thus, $\im(p) \subseteq \ker(q)$. We also know that this sequence need not be exact, as we can already see from our example in $d=2$ for which the sequence reads
\begin{equation}
    \bZ \xrightarrow[\hspace{5ex}]{\times 16} \bZ \xrightarrow[\hspace{5ex}]{\text{mod~}8} \bZ_8
\end{equation}
so $\im(p) = 16\bZ \neq \ker(q) = 8\bZ$.
We can distil the failure of exactness into a set of abelian groups, defined by
\begin{equation} \label{eq:int:hb}
    H_B^{d+2}(\pt) \coloneqq \ker(q)/\im(p)
\end{equation}
that we shall refer to as `bosonisation cohomology'. Its non-trivial classes contain all theories for which $(-1)^F$ can be gauged, but that are nonetheless intrinsically fermionic \emph{i.e.}\ they are not the `fermionisation' of any bosonic theory. For example, $H_B^4(\pt) = 8\bZ / 16\bZ = \bZ_2$. In \cref{sec:defining} we also show how the construction generalises to the presence of a non-zero internal symmetry $G$, or indeed for maps to any $X$, which lets us define $H_B^\bullet(X)$ for a topological space $X$.

\subsection{Summary of results} \label{sec:intro_summary}

Our main goal is to calculate the bosonisation cohomology groups \eqref{eq:int:hb} for all $d$. To this end, we will need to determine the structure of the cobordism sequence \eqref{eq:int:seq}. This is carried out in \cref{sec:computing}.

To give a brief preview of the results, in \cref{fig:int:map} we have depicted the cobordism sequence in low degrees $n \leq 13$. (We will always use $d$ for the spacetime dimension of the QFT, and $n = d + 2$ for the cobordism degree of its anomaly.) The sequence is composed of several building blocks, which can roughly be divided into two types: those involving only discrete anomalies, and those involving continuous (perturbative) anomalies. Below, we list all the building blocks that can occur, and their contributions to $H_B^n(\pt)$:

\begin{figure}
    \centering
    \begin{tikzcd}[
        row sep=0,
        nodes in empty cells,
        every matrix/.append style = {name=m},
        execute at end picture={
            \begin{scope}[on background layer]
                \fill[black!3] (m-1-2.west|-m-2-2.north) rectangle (m-1-4.east|-m-2-2.south);
                \fill[black!3] (m-1-2.west|-m-3-1.north) rectangle (m-1-4.east|-m-3-1.south);
                \fill[black!3] (m-1-2.west|-m-4-3.north) rectangle (m-1-4.east|-m-4-3.south);
                \fill[black!3] (m-1-2.west|-m-5-3.north) rectangle (m-1-4.east|-m-5-3.south);
                \fill[black!3] (m-1-2.west|-m-6-4.north) rectangle (m-1-4.east|-m-6-4.south);
                \fill[black!3] (m-1-2.west|-m-7-1.north) rectangle (m-1-4.east|-m-7-1.south);
                \fill[black!3] (m-1-2.west|-m-8-2.north) rectangle (m-1-4.east|-m-8-2.south);
                \fill[black!3] (m-1-2.west|-m-9-1.north) rectangle (m-1-4.east|-m-9-1.south);
                \fill[black!3] (m-1-2.west|-m-10-2.north) rectangle (m-1-4.east|-m-11-2.south);
                \fill[black!3] (m-1-2.west|-m-12-1.north) rectangle (m-1-4.east|-m-12-1.south);
                \fill[black!3] (m-1-2.west|-m-13-2.north) rectangle (m-1-4.east|-m-13-2.south);
                \fill[black!3] (m-1-2.west|-m-14-3.north) rectangle (m-1-4.east|-m-15-3.south);
                \fill[black!3] (m-1-2.west|-m-16-2.north) rectangle (m-1-4.east|-m-19-2.south);
                \fill[black!3] (m-1-2.west|-m-20-4.north) rectangle (m-1-4.east|-m-20-4.south);
                \draw
                    ($(m-1-1.south)!0.5!(m-2-1.north)$) coordinate (l)
                    (m.west|-l) -- (m.east|-l);
                \draw[dashed]
                    ($(m-9-1.south)!0.5!(m-10-2.north)$) coordinate (l)
                    (m.west|-l) -- (m.east|-l);
                \node at ($(m-10-1.center)!0.5!(m-11-1.center)$) {$8$};
                \node at ($(m-14-1.center)!0.5!(m-15-1.center)$) {$11$};
                \node at ($(m-16-1.center)!0.5!(m-19-1.center)$) {$12$};
            \end{scope}
        }
    ]
    n & \mho^n_\SO(\pt) \arrow[r, "p"] & \mho^n_\Spin(\pt) \arrow[r, "q"] & \widetilde{\mho}^n_\Spin(\B\bZ_2) \\[1em]
    0 & \textcolor{e}{\bZ} \arrow[r, color=e] & \textcolor{e}{\bZ} & \\[0.2em]
    1 \\[0.2em]
    2 & & \textcolor{c}{\bZ_2} \arrow[r, color=c] & \textcolor{c}{\bZ_2} \\[0.2em]
    3 & & \textcolor{c}{\bZ_2} \arrow[r, color=c] & \textcolor{c}{\bZ_2} \\[0.2em]
    4 & \textcolor{e}{\bZ} \arrow[r, color=e, "16"] & \textcolor{e}{\bZ} \arrow[r, color=e] & \textcolor{e}{\bZ_8} \\[0.2em]
    5 \\[0.2em]
    6 & \textcolor{a}{\bZ_2} \\[0.2em]
    7 \\[1em]
    & \textcolor{e}{\bZ} \arrow[r, color=e, "128"] & \textcolor{e}{\bZ} \arrow[r, color=e] & \textcolor{e}{\bZ_{16}} \\
    & \textcolor{e}{\bZ} \arrow[r, color=e] & \textcolor{e}{\bZ} \\[0.2em]
    9 \\[0.2em]
    10 & \textcolor{a}{\bZ_2^2} & \textcolor{c}{\bZ_2^2} \arrow[r, color=c] & \textcolor{c}{\bZ_2^2} \\[0.2em]
    & & \textcolor{c}{\bZ_2^2} \arrow[r, color=c] & \textcolor{c}{\bZ_2^2} \\
    & \textcolor{b}{\bZ_2} \arrow[r, color=b] & \textcolor{b}{\bZ_2} \\[0.2em]
    & \textcolor{e}{\bZ} \arrow[r, color=e, "2048"] & \textcolor{e}{\bZ} \arrow[r, color=e] & \textcolor{e}{\bZ_{128}} \\
    & \textcolor{e}{\bZ} \arrow[r, color=e, "16"] & \textcolor{e}{\bZ} \arrow[r, color=e] & \textcolor{e}{\bZ_8} \\
    & \textcolor{f}{\bZ} \arrow[r, color=f, "2"] & \textcolor{f}{\bZ} \arrow[r, color=f] & \textcolor{f}{\bZ_2} \\
    & \textcolor{a}{\bZ_2} \\[0.2em]
    13 & & & \textcolor{d}{\bZ_2}
    \end{tikzcd}
    \caption{The cobordism sequence \eqref{eq:int:seq} in low degrees $n \leq 13$.}
    \label{fig:int:map}
\end{figure}

\paragraph{Discrete anomalies} \

\noindent Among discrete anomalies, the possible building blocks are as follows:

\begin{itemize}

\item Isolated bosonic anomalies:
\[
    \begin{tikzcd}[
        every matrix/.append style = {name=m},
        execute at end picture={
            \begin{scope}[on background layer]
                \fill[black!3] (m-1-2.west|-m-1-2.north) rectangle (m-1-4.east|-m-1-2.south);
            \end{scope}
        }
    ]
    n & \mathtextover{\textcolor{a}{\bZ_2}}{\mho^n_\SO(\pt)} & \phantom{\mho^n_\Spin(\pt)} & \phantom{\widetilde{\mho}^n_\Spin(\B\bZ_2)}
    \end{tikzcd}
\]
These are anomalies of bosonic systems that disappear on a spin manifold. They are destroyed by fermionisation. Conversely, under bosonisation, they represent an ambiguity; there are different choices of bosonisation procedures differing by such anomalies, and it is not always possible to canonically choose one preferred bosonisation procedure over the others. These anomalies play no role in the computation of $H_B^n(\pt)$. The dimensions $n$ where they occur are counted by the partition function \eqref{eq:countingfn1}
\[
    t^6 + 2 t^{10} + t^{12} + 4t^{14} + 2t^{15} + \dots
\]
The first term represents the $(-1)^{\int w_2 w_3}$ anomaly of all-fermion electrodynamics. \cite{Wang:2018qoy} In general, all later anomalies are also given by Stiefel--Whitney numbers.

\item Fermionic anomalies that are images of bosonic anomalies:
\[
    \begin{tikzcd}[
        every matrix/.append style = {name=m},
        execute at end picture={
            \begin{scope}[on background layer]
                \fill[black!3] (m-1-2.west|-m-1-2.north) rectangle (m-1-4.east|-m-1-2.south);
            \end{scope}
        }
    ]
    n & \mathtextover{\textcolor{b}{\bZ_2}}{\mho^n_\SO(\pt)} \arrow[r, color=b] & \mathtextover{\textcolor{b}{\bZ_2}}{\mho^n_\Spin(\pt)} & \phantom{\widetilde{\mho}^n_\Spin(\B\bZ_2)}
    \end{tikzcd}
\]
A fermionic theory with this anomaly is both $(-1)^F$-gaugable and spin-structure-summable. It therefore makes no contribution to $H_B^n(\pt)$. The partition function that counts these anomalies is \eqref{eq:countingfn2}
\[
    t^{11} + 2t^{19} + t^{21} + t^{23} + 4t^{27} + \dots
\]
The first term represents the anomaly $(-1)^{\int w_4 w_6}$. In general, all later anomalies are also given by Stiefel--Whitney numbers.

\item Fermionic anomalies that induce $(-1)^F$ anomalies:
\[
    \begin{tikzcd}[
        every matrix/.append style = {name=m},
        execute at end picture={
            \begin{scope}[on background layer]
                \fill[black!3] (m-1-2.west|-m-1-2.north) rectangle (m-1-4.east|-m-1-2.south);
            \end{scope}
        }
    ]
    n & \phantom{\mho^n_\SO(\pt)} & \mathtextover{\textcolor{c}{\bZ_2}}{\mho^n_\Spin(\pt)} \arrow[r, color=c] & \mathtextover{\textcolor{c}{\bZ_2}}{\widetilde{\mho}^n_\Spin(\B\bZ_2)}
    \end{tikzcd}
\]
A fermionic theory with this anomaly is neither spin-structure summable nor $(-1)^F$-gaugable. It therefore makes no contribution to $H_B^n(\pt)$. The partition function counting these anomalies is \eqref{eq:countingfn3}
\[
    t^2 + t^3 + 2t^{10} + 2t^{11} + 5t^{18} + 5t^{19} + 11t^{26} + \dots
\]
The $t^3$ term represents the well-known fermion parity anomaly of an unpaired quantum-mechanical Majorana fermion. It is captured by an anomaly theory of $(-1)^{\Arf[\rho]} = (-1)^{\text{mod-2-index}(\slashed{D}_\rho)}$. In general, all other anomalies are also captured by mod-2 indexes of Dirac operators.

\item Isolated $\bZ_2$ anomalies:
\[
    \begin{tikzcd}[
        every matrix/.append style = {name=m},
        execute at end picture={
            \begin{scope}[on background layer]
                \fill[black!3] (m-1-2.west|-m-1-2.north) rectangle (m-1-4.east|-m-1-2.south);
            \end{scope}
        }
    ]
    n & \phantom{\mho^n_\SO(\pt)} & \phantom{\mho^n_\Spin(\pt)} & \mathtextover{\textcolor{d}{\bZ_2}}{\widetilde{\mho}^n_\Spin(\B\bZ_2)}
    \end{tikzcd}
\]
These are anomalies of a $\bZ_2$ symmetry of a fermionic system that can never be realised as the anomaly of $(-1)^F$. They play no role in the computation of $H_B^n(\pt)$. The partition function counting these anomalies is \eqref{eq:countingfn4}
\[
    t^{13} + t^{14} + t^{17} + t^{18} + 3t^{21} + \dots
\]
The first term represents the anomaly $(-1)^{\int A^2 w_4 w_6}$.

\end{itemize}
Note that there is one more possible building block that could appear, but does not:
\[
    \begin{tikzcd}[
        every matrix/.append style = {name=m},
        execute at end picture={
            \begin{scope}[on background layer]
                \fill[red!3] (m-1-2.west|-m-1-2.north) rectangle (m-1-4.east|-m-1-2.south);
            \end{scope}
            \begin{scope}[opacity=0.5,transparency group]
                \node at (m-1-3.center) [correct forbidden sign,line width=1,draw=red,inner xsep=2ex] {};
            \end{scope}
        }
    ]
    \textcolor{red}{n} & \phantom{\mho^n_\SO(\pt)} & \mathtextover{\textcolor{red}{\bZ_2}}{\mho^n_\Spin(\pt)} & \phantom{\widetilde{\mho}^n_\Spin(\B\bZ_2)}
    \end{tikzcd}
\]
Such a theory would be $(-1)^F$-gaugable, but not spin-structure summable, thus contributing a $\bZ_2$ to $H_B^n(\pt)$. But it does not exist. We conclude that $H_B^n(\pt)$ receives no contributions from discrete anomalies.

\paragraph{Continuous anomalies} \

\noindent
The interesting physics, if any, must therefore be entirely due to continuous anomalies. To see if there is any, let us list the building blocks in this case. There are two:

\begin{itemize}

\item There is an infinite structure
\[
    \begin{tikzcd}[
        row sep=0,
        every matrix/.append style = {name=m},
        execute at end picture={
            \begin{scope}[on background layer]
                \fill[black!3] (m-4-2.west|-m-1-2.north) rectangle (m-4-4.east|-m-1-2.south);
                \fill[black!3] (m-4-2.west|-m-2-2.north) rectangle (m-4-4.east|-m-2-4.south);
                \fill[black!3] (m-4-2.west|-m-3-2.north) rectangle (m-4-4.east|-m-3-4.south);
                \shade[top color=black!3, bottom color=white] (m-4-2.west|-m-4-1.north) rectangle (m-4-4.east|-m-4-1.south);
            \end{scope}
        }
    ]
    \mathtextover[l]{n}{n + 4} & \textcolor{e}{\bZ} \arrow[r, color=e] & \textcolor{e}{\bZ} & \\[0.2em]
    n + 4 & \textcolor{e}{\bZ} \arrow[r, color=e, "16"] & \textcolor{e}{\bZ} \arrow[r, color=e] & \textcolor{e}{\bZ_8} \\[0.2em]
    n + 8 & \textcolor{e}{\bZ} \arrow[r, color=e, "128"] & \textcolor{e}{\bZ} \arrow[r, color=e] & \textcolor{e}{\bZ_{16}} \\[0.2em]
    \vdots & \hphantom{\mho^n_\SO(\pt)} & \hphantom{\mho^n_\Spin(\pt)} & \hphantom{\widetilde{\mho}^n_\Spin(\B\bZ_2)}
    \end{tikzcd}
\]
The dimensions $n$ where it starts are counted by the partition function \eqref{eq:countingfn5}
\[
    1 + t^8 + 2t^{16} + 4t^{24} + \dots
\]
The first term represents a free Dirac fermion. In general, later terms also represent various other types of free fermions which include twisting by the tangent bundle---for example Rarita--Schwinger-like fields.

\item There is another infinite structure
\[
    \begin{tikzcd}[
        row sep=0,
        every matrix/.append style = {name=m},
        execute at end picture={
            \begin{scope}[on background layer]
                \fill[black!3] (m-4-2.west|-m-1-2.north) rectangle (m-4-4.east|-m-1-2.south);
                \fill[black!3] (m-4-2.west|-m-2-2.north) rectangle (m-4-4.east|-m-2-4.south);
                \fill[black!3] (m-4-2.west|-m-3-2.north) rectangle (m-4-4.east|-m-3-4.south);
                \shade[top color=black!3, bottom color=white] (m-4-2.west|-m-4-1.north) rectangle (m-4-4.east|-m-4-1.south);
            \end{scope}
        }
    ]
    \mathtextover[l]{n}{n + 4} & \textcolor{f}{\bZ} \arrow[r, color=f, "2"] & \textcolor{f}{\bZ} \arrow[r, color=f] & \textcolor{f}{\bZ_2} \\[0.2em]
    n + 4 & \textcolor{f}{\bZ} \arrow[r, color=f, "8"] & \textcolor{f}{\bZ} \arrow[r, color=f] & \textcolor{f}{\bZ_4} \\[0.2em]
    n + 8 & \textcolor{f}{\bZ} \arrow[r, color=f, "256"] & \textcolor{f}{\bZ} \arrow[r, color=f] & \textcolor{f}{\bZ_{32}} \\[0.2em]
    \vdots & \hphantom{\mho^n_\SO(\pt)} & \hphantom{\mho^n_\Spin(\pt)} & \hphantom{\widetilde{\mho}^n_\Spin(\B\bZ_2)}
    \end{tikzcd}
\]
The dimensions $n$ where it starts are counted by the partition function \eqref{eq:countingfn6}
\[
    t^{12} + 2t^{20} + 4t^{28} + \dots
\]
Again, all such anomalies represent various types of free fermions.

\end{itemize}

We can now read off the bosonisation cohomology groups \eqref{eq:int:hb}. Each infinite structure contributes an identical sequence of cyclic groups
\[
\bZ_2 \;,\quad \bZ_8 \;,\quad \bZ_{16} \;,\quad \bZ_{128} \;,\quad \bZ_{256} \;,\quad \bZ_{1024} \;,\quad \cdots
\]
in degrees $n + 4, n + 8, n + 12, \dots$. The orders of these groups are \href{https://oeis.org/A046161}{A046161} in OEIS, or $2^{2k - \text{bitcount}(k)}$, where $\text{bitcount}(k)$ counts the number of 1s appearing in the binary expansion of the integer $k$.

Thus, the bosonisation cohomology groups $H_B^\bullet(\pt)$ are supported only in degree $4\bZ$, and the first few nonzero groups are
\begin{equation}
\begin{alignedat}{7}
    H_B^4(\pt) &= \bZ_2 \\
    H_B^8(\pt) &= \bZ_8 \\
    H_B^{12}(\pt) &= \bZ_{16} \;&\times&\; \bZ_2 \\
    H_B^{16}(\pt) &= \bZ_{128} \;&\times&\; \bZ_8 \;&\times&\; \bZ_2 \\
    H_B^{20}(\pt) &= \bZ_{256} \;&\times&\; \bZ_{16} \;&\times&\; \bZ_8 \;&\times&\; \bZ_2^2 \\
    H_B^{24}(\pt) &= \bZ_{1024} \;&\times&\; \bZ_{128} \;&\times&\; \bZ_{16} \;&\times&\; \bZ_8^2 \;&\times&\; \bZ_2^2
\end{alignedat}
\end{equation}
The first column is due to a free Dirac fermion, while subsequent columns are due to increasingly-complicated Rarita--Schwinger-like fields. As we go on, we shall develop a concrete picture of exactly what all these anomalies are in terms of twisted Dirac operators, and show how to do explicit physics calculations with them.

\subsection{The plan of the paper}

The rest of the paper is structured as follows. In \cref{sec:defining} we derive the central sequence of anomaly theories~\eqref{eq:int:seq}, and from there we define the bosonisation cohomology groups $H_B^\bullet(\pt)$. We explain in \cref{sec:smashX} how this can be generalised in the presence of other structures, showing there is a well-defined construction of $H_B^\bullet(X)$ for some $X$. In \cref{sec:low-dim} we compute the bosonisation cohomology groups in low-degrees (up to $d+2 \leq 7$) by drawing on known results, to illustrate the general idea.

We then move onto the main body of our work in \cref{sec:computing}, devoted to a general computation of $H_B^\bullet(\pt)$ which we are able to carry out rigorously in all dimensions, by piecing together a variety of powerful facts about the oriented, spin, and pin bordism groups and relations between them. In more detail: in \cref{sec:4k} we show that $H_B^n(\pt)$ is zero unless $n \in 4\bZ$; in \cref{sec:phi} we explain how to build complete bases of fermionic and bosonic anomaly polynomials that generate the free part of $\mho_{\Spin,\SO}^{4k}(\pt)$, from which we compute the map $p$ defined in~\eqref{eq:int:seq} and its image; in \cref{sec:pin} we review pin bordism, a necessary ingredient to compute the kernel of the map $q$ in \cref{sec:q}; we put things together to compute $H_B(\pt)=\ker(q)/\im(p)$ in \cref{sec:result}. In \cref{sec:examples} we break down this result in dimensions $d=2$, $6$, and $10$; for $d=10$, we discuss how our result sheds light on the non-trivial way that spin structure anomalies cancel in supergravity. We conclude in \cref{sec:discussion} with comments on SymTFT and other open questions.

\paragraph{Ancillary code} \

\noindent
For completeness, we include three computer programs as ancillary files in the \texttt{arXiv} submission of this paper, designed to automate various calculational tasks:
\begin{enumerate}
    \item \texttt{counting-fns.nb}, a \texttt{Mathematica} notebook that generates the Hilbert--Poincaré series encoding various bordism groups (oriented, spin, and $\pin^-$) up to a chosen high degree; 
    \item \texttt{stiefel-whitneys.ipynb}, a \texttt{Sage} notebook that implements the Wu relations to compute non-trivial Stiefel--Whitney numbers (related to interacting SPT phases) for different types of tangential structure up to high degree;    
    \item \texttt{anom-polys.ipynb}, a \texttt{Sage} notebook designed to compute fermionic and bosonic anomaly polynomials to higher-order.
\end{enumerate}
These tasks roughly correspond to the material in appendices~\ref{app:gen},~\ref{app:stiefelwhitneys}, and~\ref{app:proofs} respectively.

\section{Defining Bosonisation Cohomology} \label{sec:defining}

Spin structure summation and gauging $(-1)^F$ are inequivalent when there is a gravitational anomaly. We capture this inequivalence mathematically by constructing a sequence of spectra that gives rise to the anomaly sequence~\eqref{eq:int:seq} that we anticipated in the introduction.
Before doing so, we begin by recalling the motivation for our modern understanding of anomalies in quantum field theory.

\subsection{Bordism, cobordism, and anomalies} \label{sec:bordism}

In the introduction we sketched how anomalies are classified by so-called cobordism groups, without really explaining what cobordism groups are. In this section we give a definition of bordism and cobordism, and motivate in more detail why it is these objects that classify anomalies.

The key physics hypothesis that unlocked the algebraic classification of anomaly theories is that of \emph{anomaly inflow}~\cite{Callan:1984sa}, which asserts that anomalies can themselves be identified with quantum field theories in one dimension higher (and satisfying certain special properties). To see how this works in its most pedestrian setting, let's consider the case of a perturbative chiral fermion anomaly in $d=4$ dimensions à la Adler, Bell, and Jackiw (ABJ)~\cite{Adler:1969gk,Bell:1969ts}. Under a chiral $U(1)_A$ background gauge transformation $A\mapsto A +d\alpha$, the fermion path integral transforms by a phase $Z[A]\mapsto Z[A]\exp \left(i n \int_\Sigma \alpha \frac{dA\wedge dA}{8\pi^2}\right)$, for some anomaly coefficient $n\in \bZ$. Inflow says that this anomalous transformation can be captured by the variation of a classical Chern--Simons action defined on a bulk 5-manifold $X$ whose boundary is the physical spacetime, $Z_\CS[A]=\exp\left( i n \int_X \frac{1}{8\pi^2} A \wedge dA\wedge dA \right)$. In condensed matter systems this bulk may be part of the physical system; famously, the $d=2$ version of this abelian perturbative anomaly inflow describes the integer quantum Hall effect.

For a perturbative anomaly such as this, the gauge-invariant object that determines the anomaly is called an \emph{anomaly polynomial}, a creature that will play a central role in this paper. Here, it is $\Phi_6 =\frac{n}{8\pi^2}dA\wedge dA\wedge dA$, the curvature of the Chern--Simons form. Being a closed differential $(d+2)$-form, its integral on any manifold that is a boundary, to which we assume all requisite structures (such as an orientation, the spin structure and the gauge field) can be extended, will vanish. This property makes the anomaly polynomial a simple example of a \emph{bordism invariant} in degree $d+2$.

We pause to recall a few mathematical notions needed to understand this last statement. Firstly, \emph{bordism} is an equivalence relation on $k$-dimensional manifolds equipped with given structures, whereby $M_0 \sim M_1$ if there exists some $(k+1)$-dimensional smooth manifold $X$ such that $\partial X = M_0 \sqcup (-M_1)$ and to which all the structures smoothly extend, where $-M_1$ denotes the orientation reversal of $M_1$ (assuming our manifolds come with an orientation). This equivalence relation partitions the set of manifolds with given structures into classes, which can moreover be given an abelian group structure under disjoint union of manifolds, thus defining \emph{bordism groups}. For the case of manifolds with spin structure and $G$-gauge bundles, the bordism groups are denoted $\Omega_k^\Spin(\B G)$, where $BG$ is the classifying space of $G$.\footnote{Recall that $BG$ is a topological space such that, given some manifold $M$ and some group $G$, principal $G$-bundles over $M$ (without connection) are classified by homotopy classes of maps from $M$ into $BG$.}
The zero class in $\Omega_k^\Spin(BG)$ contains all manifolds that are boundaries. From this description, one can appreciate that bordism is a kind of homology theory that works directly with manifolds (rather than cycles), and which incorporates all the structures we need to define quantum field theories. Any quantity that vanishes on all boundaries is a \emph{bordism invariant}, meaning it will moreover be constant on each non-zero bordism class: the anomaly polynomial therefore defines an element in the group
\begin{equation} \label{eq:local}
    \Hom(\Omega_{d+2}^\Spin(\B G), \bZ) \qquad \{\text{local anomalies} \}
\end{equation}
via its integration, given also the quantisation condition on the anomaly coefficient.

Witten taught us that not all anomalies can, like the ABJ example, be detected by Feynman diagrams~\cite{Witten:1982fp}. For example, a 4d Weyl fermion transforming in the doublet representation of $SU(2)$ carries such a non-perturbative (or `global') anomaly: under certain $SU(2)$ background gauge transformations~\cite{Witten:1982fp,Wang:2018qoy}, the partition function flips sign. It was long suspected~\cite{Witten:1985xe,Dai:1994kq,Witten_2016} and is now proven~\cite{Witten:2019bou} that any spin-$\frac{1}{2}$ chiral fermion anomaly -- perturbative or non-perturbative -- can be captured via inflow.\footnote{There is, to our knowledge, no comparable proof pertaining to higher-spin fermionic fields. Part of the `problem' is that interacting higher-spin particles typically require additional states for consistency. While we consider the anomaly theories of higher-spin fermions throughout this paper, we do not expect this issue to pose a problem for our rigorous results, because the anomalies in question persist in the limit of a free theory and the anomaly theories themselves remain well-defined quantum field theories.}
The anomaly theory in question is a generalisation of Chern--Simons called the \emph{eta invariant}~\cite{Atiyah:1975jf,Atiyah:1976jg,Atiyah:1976qjr} for the Dirac operator $\cD_X$ corresponding to our fermion spectrum, appropriately extended (with particular boundary conditions) to the $(d+1)$-dimensional bulk $X$:
\begin{equation} \label{eq:etadef}
    \eta(\cD_X) \coloneqq \frac{1}{2} \sum_{\lambda \in \text{Spec}(\cD_X)} \frac{\text{sign}(\lambda)}{|\lambda|^s} \bigg|_{\substack{\text{analytically} \\ \text{continued to } s = 0}}
    \quad \text{with} \quad
    \text{sign}(0) \coloneqq +1
\end{equation}
A remarkable theorem of Atiyah, Patodi and Singer (APS)~\cite{Atiyah:1975jf,Atiyah:1976jg,Atiyah:1976qjr} tells us that this eta invariant is closely related to the corresponding anomaly polynomial $\Phi_{d+2}$ for that Dirac operator,
\begin{equation}
    \mathrm{index}(\cD_Y) = \int_Y \Phi_{d+2} - \eta(\cD_X), \qquad \text{for~} \partial Y = X
\end{equation}
The anomalous variation of the fermion partition function can be computed by evaluating the exponentiated eta invariant on a mapping torus that interpolates between two field configurations. In a perturbative context, a mapping torus $X$ can always be bounded by some $Y$; thence, the APS theorem tells us the anomaly is determined by the anomaly polynomial.

The APS index theorem tells us something more: that, when the perturbative anomaly vanishes, the exponentiated eta invariant is trivial on all boundaries. Thus, when $\Phi_{d+2}=0$, any residual (hence non-perturbative) anomaly is a bordism invariant, but this time in degree $d+1$. A non-perturbative anomaly is an element in the group
\begin{equation} \label{eq:global}
    \Hom \left(\mathrm{Tor}\, \Omega_{d+1}^\Spin(\B G), U(1) \right) \qquad \{\text{global anomalies} \}
\end{equation}
The restriction to the torsion subgroup is because any integral class in $\Omega_{d+1}$ can be cancelled by a local counter-term~\cite{Witten:2019bou}.
Putting~\eqref{eq:local} and~\eqref{eq:global} together, we learn that chiral fermion anomalies belong to a group $\mho^{d+2}_\Spin(\B G)$ that sits in the middle of the following short exact sequence
\begin{equation} \label{eq:UCT}
    \Hom \left(\mathrm{Tor}\, \Omega_{d+1}^\Spin(\B G), U(1) \right)
    \hookrightarrow \mho^{d+2}_\Spin(\B G)
    \twoheadrightarrow \Hom(\Omega_{d+2}^\Spin(\B G), \bZ)
\end{equation}
Exactness in the middle means that the global anomalies (the image of the left map) are precisely what remains when the perturbative anomalies cancel.

A sequence like~\eqref{eq:UCT} is familiar to any algebraic topologist: it is a universal coefficient sequence, and the objects in the middle define a generalised cohomology theory dual to bordism. We call this a \emph{cobordism} group. Technically speaking, what we call $\mho_H^\bullet$ is the Anderson dual to the bordism groups. We shall give a definition of $\mho^\bullet_H$ in terms of spectra in \cref{sec:math_seq}.

The fact that anomalies in $d$-dimensions are determined, via inflow, by certain quantum field theories in $d+1$ dimensions that live in $\mho_H^{d+2}$ goes beyond the chiral fermion anomalies to which our discussion has so far been restricted.
Remarkably, the group $\mho_H^{d+2}$ classifies \emph{all} anomaly theories, suitably defined, that pertain to $d$-dimensional quantum field theories with symmetry type $H$ (where the symmetry type includes spacetime and internal symmetries, \emph{e.g.}\ $H=\Spin\times G$ for the case~\eqref{eq:UCT}). This includes all the free fermion anomalies, but also other intrinsically interacting anomaly theories that one can write down -- some examples of which were touched upon in \cref{sec:intro_summary}.
The key property of an anomaly theory is \emph{invertibility}~\cite{Freed:2004yc,Freed:2012bs,Freed:2014iua}. Freed and Hopkins conjectured~\cite{Freed:2016rqq} (proven in its more general form by Grady~\cite{Grady:2023sav}) that reflection positive (\emph{i.e.}\ unitary), invertible quantum field theories in $(d+1)$-dimensions are classified by cobordism groups.

\subsection{Deriving the anomaly sequence} \label{sec:math_seq}

With this background, one can appreciate better the meaning of the cobordism groups discussed in the introduction, that were relevant for describing $(-1)^F$ anomalies, fermionic anomalies in general, and bosonic anomalies.
We are almost ready to derive the sequence of anomaly theories ~\eqref{eq:int:seq} built from these objects, that we reprise here for ease of reference:
\begin{equation*}
    \mho_\SO^{d+2}(\pt) \xrightarrow[\hspace{5ex}]{p} \mho_\Spin^{d+2}(\pt) \xrightarrow[\hspace{5ex}]{q} \widetilde{\mho}_\Spin^{d+2}(\B\bZ_2)
\end{equation*}
Recall the group on the left captures all bosonic anomalies, the group in the middle all fermionic anomalies (with the map $p$ sending each bosonic anomaly to its image in spin cobordism), and the group on the right captures anomalies in a $\bZ_2$ symmetry (with the map $q$ sending a fermionic anomaly to the induced anomaly in $(-1)^F$).

To show that there is indeed such a sequence of cobordism groups (without relying on the physics arguments put forth in the introduction), we need a little more mathematical machinery concerning the idea of \emph{spectra}, which we shall describe at a level appropriate for our naïveté.
For a complete account of the material introduced here, we recommend~\cite{freed2019lectures}.
For readers happy to take the sequence~\eqref{eq:int:seq} on trust rather than brush up on their spectra, we recommend skipping ahead to \cref{sec:low-dim}.

\subsubsection*{A rough guide to spectra}

A spectrum is something like a topological space, but one that is decomposed into distinct components spread across infinitely many dimensions. Spectra provide an ideal setting for describing things like cohomology groups (which also are defined in every dimension $d\in \bZ_{\geq 0}$) in a unified way, and they enable mathematicians to discuss \emph{stable} behaviour that emerges only in some limit where $d$ becomes large.

\begin{figure}
    \centering
    \scalebox{1.5}{
    \begin{tikzpicture}
        \newcommand{\X}{1}
        \newcommand{\Y}{1.9}
        \newcommand{\T}{0.25}
        \filldraw[black,fill=cyan!30]
            (0,\Y/2) circle [x radius=\X/2, y radius=0.3*\X/2]
            (0,-\Y/2) circle [x radius=\X/2, y radius=0.3*\X/2];
        \filldraw[black,fill=cyan!40]
            (0,0) circle [x radius=\X, y radius=0.3*\X];
        \draw[cyan!50!black]
            (0,-\Y) -- (\X,0) -- (0,\Y) -- (-\X,0) -- cycle;
        \node at (-0.1*\X,0) {$X$};
        \fill[cyan!50!black]
            (0,-\Y) circle [x radius=0.05, y radius=0.05]
            (0,\Y) circle [x radius=0.05, y radius=0.05];
        \draw[red]
            (0,-\Y) .. controls (\X+\T*\X,\T*\Y) and (-\X,0) .. (0,\Y);
    \end{tikzpicture}
    }
    \caption{An illustration of the suspension $\Sigma X$ for some space $X$ indicated by the blue disc. One can think of `hanging' $X$ between two endpoints. For \emph{reduced} suspension, all the basepoints, indicated by the red line, are also identified.}
    \label{fig:suspension}
\end{figure}
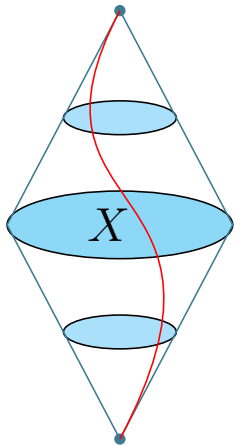

More precisely, a spectrum is a set of topological spaces $\{E_n\}$, for each $n\in \bZ_{\geq 0}$, together with homeomorphisms relating them $\sigma_n : \Sigma E_n \to E_{n+1}$, where $\Sigma$ denotes the suspension operation -- illustrated in Fig.~\ref{fig:suspension} -- defined as $\Sigma X = (X \times I)/{\sim}$, where $(x_1,0)\sim (x_2,0)$, $(x_1,1) \sim (x_2,1)$, $\forall x_1, x_2 \in X$.
The simplest example is the sphere spectrum $S^0$, for which $E_n = S^n$ and the $\sigma_n$ are the canonical homeomorphisms. (It should be clear from Fig.~\ref{fig:suspension} that the suspension of $S^n$ is homeomorphic to $S^{n+1}$, with the endpoints of the interval $I$ ending up as the North and South pole.) The sphere spectrum is not only the simplest example of a spectrum; for mathematicians, it is the monoidal unit in the category of spectra, so it is sometimes denoted $S^0=1$.
We can also generate examples of spectra \emph{ad libitum} just by iterating the suspension operation. To wit, given some topological space $E$, define the \emph{suspension spectrum} of $E$ (often also denoted $E$, a notation that is clear enough once one has arrived in the category of spectra) whose component spaces are $\Sigma^n E$, with $\sigma_n$ being identity maps.

As we anticipated, spectra provide a mathematical tool for unifying the description of cohomology theories.
Key to this is a powerful theorem of Brown~\cite{brown1965abstract}, which says that any generalised cohomology theory $E^d(X)$ can always be \emph{represented} by a target spectrum $E$, meaning that
\begin{equation} \label{eq:coh}
    E^d(X) \cong [X_+, E_d]
\end{equation}
where $[A,B]$ is the space of homotopy classes of based maps from $A$ to $B$, and $X_+$ denotes $X$ with a disjoint basepoint.
Let us illustrate how this works with familiar examples of cohomology.
First, consider the \emph{Eilenberg--MacLane spectrum} $H\bZ_2$. This has as its building blocks $E_n \cong K(\bZ_2,n)$ the Eilenberg--MacLane spaces, that is $H\bZ_2 = \{ \bZ_2, B\bZ_2 \cong \RP^\infty, K(\bZ_2, 2), \dots \}$, and it represents ordinary cohomology with mod 2 coefficients, in that $H^n(X; \bZ_2) \cong [X, K(\bZ_2, n)]$.
In a similar way, the Eilenberg--MacLane spectrum with spaces $H\bZ = \{\bZ, \B\bZ \cong U(1), \B U(1) \cong \CP^\infty, K(\bZ, 3), \dots\}$ represents ordinary integral cohomology.
Shortly, we will see how a similar definition can be given for the cobordism groups that classify anomalies.

\subsubsection*{Spectra for symmetry types of QFTs}

Turning our attention to physics, spectra are a useful gadget when it comes to describing and/or classifying local quantum field theories, including those that capture anomalies via inflow.
Perhaps the most basic reason for this is that locality demands that field theories give us well-defined maps out of manifolds in different dimensions -- we explicitly saw an example of this in our description of anomaly inflow above, for which manifolds in dimensions $d$, $d+1$, and $d+2$ all feature.
For this `extension' of field theories~\cite{Freed:2012bs} to make sense, the various structures we need to formulate our field theory -- like spin structures or gauge connections -- must also be extended on manifolds of different dimensions.
Spectra, and maps between them, can carry all the information pertaining to the symmetry type of an extended field theory at once~\cite{Freed:2016rqq}.

We now describe the particular spectrum that we need to capture the symmetry type and sketch a cohomological formulation, as per~\eqref{eq:coh}, of the cobordism groups that classify deformation classes of reflection positive invertible field theories \emph{ergo} anomalies~\cite{Freed:2016rqq}.
First, we want to put our description of tangential structures (like orientation, spin structure) on a \emph{stable} footing. Consider a quantum field theory defined only with an orientation, meaning the structure group of the tangent bundle is the special orthogonal group $\SO(m)$.
Recall that its classifying space $B\SO(m)$ classifies all real, oriented vector bundles of rank $m$.
Now, building up from the natural inclusions $\SO(m) \hookrightarrow \SO(m+1)\,:\, R \mapsto \begin{psmallmatrix} R & \\ & 1 \end{psmallmatrix}$, it makes sense to consider the limiting space (strictly a `colimit') $\SO=\lim_{m\to \infty} \SO(m)$.\footnote{The notion of a \emph{colimit} may deserve a little more explanation for our physicist readers. As a set, it is the union $\SO = \bigcup_{m=1}^\infty \SO(m)$, given the inclusions defined in the main text. It is equipped with the direct limit topology, whereby a subset $\mathcal{U} \subset \SO$ is open iff $\mathcal{U} \cap \SO(m)$ is open in $\SO(m)$ for all $m$. Thus defined, $\SO$ is a topological space, and moreover a topological group.} Moreover, the limit of classifying spaces $\B\SO \coloneqq \lim_{m \to \infty} \B\SO(m)$ can also be defined by taking limits of Grassmannians.
This $\B\SO$ is an infinite-dimensional topological space that classifies all real, oriented vector bundles of \emph{arbitrary} rank, up to adding/removing trivial bundles. A particular such bundle $\pi:V\to X$ would be specified (up to stable equivalence) by a classifying map $X\to \B\SO$.
In a similar way, if we wish to do quantum field theories with spinors, as in this paper, then we need a spin structure (or a variant thereof). And so, taking double covers, one defines $\Spin=\lim_{m\to\infty} \Spin(m)$ and $\B\Spin=\lim_{m\to \infty} \B\Spin(m)$ in a similar way to before.
From here, we are a hop, skip, and a jump away from defining the spectrum we need to classify anomalies: given a tangential structure such as an orientation, $\xi:\B\SO\to \B\Oo$, first form its inverse (as a virtual vector bundle) $-\xi$;
then take the Thom space; from there take the suspension spectrum as defined above to form the Madsen--Tillmann spectrum $\MT\SO$, and likewise to get $\MT\Spin$.\footnote{If we had not done step 1, we would have arrived at the somewhat more familiar \emph{Thom spectrum} $\M\SO$, $\M\Spin$. In fact, $\M H=\MT H$ for $H=\SO,\Spin$ while $\M\Pin^\pm = \MT\Pin^\mp$. \label{foot:bordism}}

It is straightforward to extend this to theories equipped also with maps to some $X$, such as $X=\B G$ if we are interested in theories with internal symmetry $G$. We take, say, a colimit of spaces $H_m = \Spin(m) \times G \to H = \Spin \times G$, for the case of theories with spin structure. When we come to taking Thom (or $\MT$) spectra, the direct product of groups becomes a smash product,\footnote{When manipulating spectra in this section, we make frequent use of common operations on topological spaces, like `wedge sums' and `smash products'. The wedge sum $X \vee Y$ of two pointed spaces is defined as the quotient of $X \sqcup Y$ by $x_0 \sim y_0$; the smash product $X \wedge Y$ is then $X\times Y / X \vee Y$.} resulting in $\MT\Spin \wedge (\B G)_+$.

\subsubsection*{Representing cobordism}

The bordism groups introduced above can be obtained as the stable homotopy groups of these Madsen--Tillmann spectra, \emph{viz.}\ $\Omega_n^\SO(\pt) \cong \pi_n(\MT\SO)$ and $\Omega_n^\Spin(\pt) = \pi_n(\MT\Spin)$.

The cobordism groups we have been referring to, that classify anomalies, are then defined as classes of maps from such a Madsen--Tillmann spectrum into a universal target spectrum, for example
\begin{equation} \label{eq:mho_def}
    \mho_\Spin^n(\pt) \cong [\MT\Spin, \Sigma^n I_\bZ]
\end{equation}
is the spin cobordism group. Including also our internal symmetry $G$, we have
\begin{equation} \label{eq:mhoG_def}
    \mho_\Spin^n(\B G) \cong [\MT\Spin \wedge (\B G)_+, \Sigma^n I_\bZ]
\end{equation}
In all cases, the target spectrum is (the suspension shift of) the so-called \emph{Anderson dual} $I_\bZ$ of the sphere spectrum we defined above. The Anderson dual of a spectrum $E$~\cite{anderson1970universal} is another spectrum denoted $I_\bZ E$. We do not need to know its technical definition, only that it satisfies a key universal property such that the group appearing on the RHS of~\eqref{eq:mho_def} sits inside the precise short exact sequence~\eqref{eq:UCT} that we saw in our physics-led classification of general anomalies.

The eagle-eyed reader will notice that we have not yet fulfilled our promise to write cobordism groups as a cohomology theory à la Brown. For completeness, we can recast~\eqref{eq:mhoG_def}
in the form~\eqref{eq:coh}. This can be done with the tool of \emph{function spectra}, which is, colloquially, the `hom' between pairs of spectra.
This has a nice \emph{adjunction} property
$[E_1 \wedge E_2, E_3] = [E_1, F(E_2, E_3)]$ where $E_i$ are any three spectra,
that lets us move the $\MT\Spin$ dependence in~\eqref{eq:mhoG_def} to the right.
We deduce that the representing spectrum is
$F(\MT\Spin, I_\bZ)$
in the case of spin cobordism. That is,
\begin{equation}
    \mho_\Spin^n(X) \cong [X_+,\, F(\MT\Spin, I_\bZ)]
\end{equation}

\subsubsection*{The sequence of spectra for bosonisation }

With all this spectral machinery in place, it becomes very simple to prove that our conjectured sequence~\eqref{eq:int:seq} of anomaly theories follows from a sequence of groups.
The three cobordism groups that appear in our all-important non-exact sequence~\eqref{eq:int:seq} can all be defined in these terms. We have
\begin{align}
    \mho^{d+2}_\SO(\pt) &= [\MT\SO, \Sigma^{d+2} I_\bZ] \\
    \mho^{d+2}_\Spin(\pt) &= [\MT\Spin, \Sigma^{d+2} I_\bZ] \\
    \widetilde{\mho}^{d+2}_\Spin(\B\bZ_2) &= [\MT\Spin \wedge \B\bZ_2, \Sigma^{d+2} I_\bZ]
\end{align}
Here we show that the sequence of cobordism groups \eqref{eq:int:seq} is induced by a sequence of its underlying spectra
\begin{equation} \label{eq:seqspec}
    \MT \Spin \wedge \B\bZ_2 \xrightarrow[\hspace{5ex}]{q} \MT\Spin \xrightarrow[\hspace{5ex}]{p} \MT\SO
\end{equation}
with $pq$ homotopic to the constant map, and construct this sequence explicitly.

Everything essentially follows from the diagram of groups
\begin{center} \label{eq:group_seq}
\begin{tikzcd}
    \arrow[loop, out=-165, in=165, distance=5em, "s"] \Spin(n) \times \bZ_2 \arrow[r, yshift=1ex, "\pi"] &
    \Spin(n) \arrow[r, "p"] \arrow[l, yshift=-1ex, "i"] &
    \SO(n)
\end{tikzcd}
\end{center}
where $s(g, z) = (gz, z)$ is the `twist' (with $z = \pm 1$), $\pi(g, z) = g$ is the projection, $i(g) = (g, 1)$ is the inclusion, and $p(g) = \{\pm g\}$ is the double cover. It doesn't commute, but satisfies relations $s^2 = 1$, $s i = i$, $p \pi s = p \pi$, $\pi i = 1$.

Following the steps described above but applied to the whole sequence~\eqref{eq:group_seq}, we (i) apply the $B$ functor, (ii) take the direct limit over $n$, then (iii) apply the $\MT$ functor to produce Madsen--Tillmann spectra. For our purposes here, it is important to know two facts about this construction:
\begin{enumerate}

\item It is a functor in an appropriate sense.

\item It takes $H \times G \rightleftharpoons H$ to $\MT H \wedge (\B G)_+ \rightleftharpoons \MT H$, with all maps the obvious ones.

\end{enumerate}
(This first statement very concisely sweeps an enormous amount of mathematical subtlety under the rug.) Applying the functor to our diagram of groups yields
\begin{center}
\begin{tikzcd}
    \arrow[loop, out=-165, in=165, distance=5em, xshift=-1em, "s"] \MT\Spin \wedge (\B\bZ_2)_+ \arrow[r, yshift=1ex, "\pi"] &
    \MT\Spin \arrow[r, "p"] \arrow[l, yshift=-1ex, "i"] &
    \MT\SO
\end{tikzcd}
\end{center}
which is almost what we want, but for the $+$.
To remove the $+$, we need to recall that there is an equivalence of spectra
\begin{equation}
    \MT\Spin \wedge (\B\bZ_2)_+ \cong (\MT\Spin \wedge \B\bZ_2) \vee \MT\Spin
\end{equation}
where recall $\vee$ denotes the wedge sum. This equivalence is how one defines the splitting of a generalised cohomology theory into reduced cohomology times the cohomology of a point.
We splice the above equivalence into our diagram of spectra to get
\begin{center}
\begin{tikzcd}
    \arrow[loop, out=-165, in=165, distance=5em, xshift=-3em, "s"] (\MT\Spin \wedge \B\bZ_2) \vee \MT\Spin \arrow[r, yshift=1ex, "\pi"] &
    \MT\Spin \arrow[r, "p"] \arrow[l, yshift=-1ex, "i"] &
    \MT\SO
\end{tikzcd}
\end{center}
Now the important point is that the maps $\pi$ and $i$ coincide, up to homotopy, with the obvious maps defined using the wedge sum structure. This follows by fact (2).

The sequence of spectra we want is then hiding in here: it is
\begin{center}
\begin{tikzcd}
    \MT\Spin \wedge \B\bZ_2 \arrow[r, "\pi s"] &
    \MT\Spin \arrow[r, "p"] &
    \MT\SO
\end{tikzcd}
\end{center}
To see that it forms a complex, we use the relation $p \pi s \simeq p \pi$, and the fact that $\pi$ maps $\MT\Spin \wedge \B\bZ_2$ to a point. This establishes what we wanted to show.

\subsubsection*{The Smith isomorphism}

The first term of our sequence of spectra \eqref{eq:seqspec} can be simplified using
\begin{equation} \label{eq:smith}
    \MT\Spin \wedge \B\bZ_2 \cong \Sigma \MT\Pin^-
\end{equation}
This is known as a \emph{Smith isomorphism}. This version was established by Anderson, Brown and Peterson in~\cite{anderson1969pin}
(see also \emph{e.g.}~\cite{gilkey1989geometry}). 
It will stand us in good stead for \cref{sec:q} to elaborate more on what this isomorphism does, 
for which we largely follow the geometric description given in~\cite[Sec. 3.1]{Tachikawa_2019}. 

\begin{figure}
    \centering
    \begin{tikzpicture}
        \newcommand{\SEP}{3.0}
        \newcommand{\X}{0.6}
        \newcommand{\RAD}{1.1}
        \newcommand{\ANGLE}{55}

        \begin{scope}[xshift=-\SEP cm-\X*0.5 cm]
            \node at (0,\SEP) {$(M_n, \rho, A)$};
            \begin{scope}[yshift=\X cm, rotate=\ANGLE]
                \filldraw[cyan!40!black,fill=cyan!5] (-\RAD,-\X) rectangle (\RAD,\X);
                \filldraw[black,fill=cyan!30] (-\RAD,0) coordinate(A) circle [radius=\X];
                \filldraw[black!70,fill=cyan!15] (\RAD,0) coordinate(B) circle [radius=\X];
            \end{scope}
            \begin{scope}[opacity=0.1,transparency group]
                \foreach \t in {A,B} {
                    \foreach \i in {-30,0,30} {
                        \draw[{Stealth}-{Stealth}, black] ($(\t)+(180-\i:\X)$) -- ($(\t)+(\i:\X)$);
                    }
                }
            \end{scope}
            \draw[red,thick] (\ANGLE:\RAD) -- (\ANGLE:-\RAD);
            \fill[red!50!black]
                (\ANGLE:\RAD) circle [radius=0.05]
                (\ANGLE:-\RAD) circle [radius=0.05];
            \node at ($(A)+(135:\X)$) [anchor=south east,inner sep=0] {\scriptsize \textcolor{cyan!40!black}{$S^1$}};
        \end{scope}

        \begin{scope}[xshift=\SEP cm]
            \node at (0,\SEP) {$(M_{n-1}, \rho)$};
            \draw[red,thick] (\ANGLE:\RAD) -- (\ANGLE:-\RAD);
            \fill[red!50!black]
                (\ANGLE:\RAD) circle [radius=0.05]
                (\ANGLE:-\RAD) circle [radius=0.05];
            \node at (0,0) [anchor=north west] {\scriptsize \textcolor{red!50!black}{$M_{n-1}$}};
        \end{scope}

        \node at (0,\SEP) {$\rightleftharpoons$};
        \draw[->, thick] (-\SEP*0.4,0.3) -- (\SEP*0.4,0.3) node[midway, above] {$\text{PD}(A)$};
        \draw[->, thick] (\SEP*0.4,-0.3) -- (-\SEP*0.4,-0.3) node[midway, below] {$M_{n-1} \times_\xi S^1$};
    \end{tikzpicture}
    \caption{Illustration of how to go left and right through the Smith isomorphism.}
    \label{fig:smith}
\end{figure}
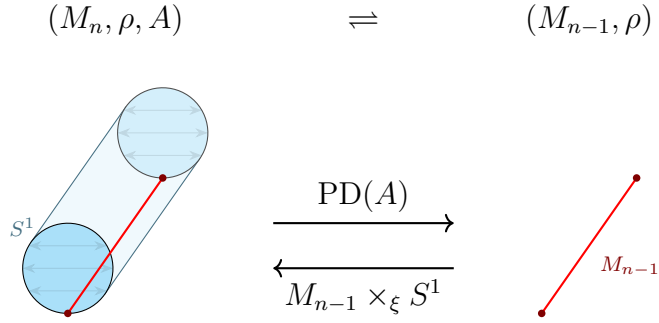

The Smith isomorphism is usually discussed after taking homotopy groups, where it becomes an isomorphism of bordism groups
\begin{equation}
    \widetilde{\Omega}_n^\Spin(\B\bZ_2) \cong \Omega_{n-1}^{\Pin^-}(\pt)
\end{equation}
In Fig.~\ref{fig:smith} we depict a cartoon showing how to go in both directions, at the level of manifolds (with structures).
To go right, form the Poincaré dual $\PD(A)$ in $M_n$. This is a codimension-1 submanifold, and inherits a $\pin^-$ structure. To go left, form the circle bundle $S^1 \times_\xi M_{n-1}$, where $\xi$ is the orientation bundle of $M_{n-1}$, and the $\bZ_2$ structure group of $\xi$ acts on $S^1$ as a reflection. This is a manifold of one higher dimension, and inherits a spin structure. We take the $\bZ_2$ gauge field $A$ to have nontrivial holonomy around the $S^1$ fibre. It can then be checked that both directions are well-defined at the level of bordism, and are inverses.

We draw attention to the relation of tangent bundles $TM_n|_{\PD(A)} = TM_{n-1} \oplus \xi$. This formula is useful when translating various Dirac indexes across the Smith isomorphism, which we shall return to in \cref{sec:q}.

From now on, we shall be happy to freely use the Smith isomorphism~\eqref{eq:smith} whenever convenient.

\subsection{Generalisation for internal symmetry} \label{sec:smashX}

So far our discussion has been limited to purely gravitational theories. That means we have not demanded our theories have any internal symmetry, and if they had any, we ignored it.

In general we can ask exactly the questions we have asked but preserving some internal symmetry. Let us consider $d$-dimensional fermionic theories equipped with an internal, invertible symmetry classified by maps to some space $X$ (for example $X = \B G$ for a 0-form symmetry group $G$). Then the sequence of spectra governing spin structure summation and gauging $(-1)^F$ while preserving the symmetry is simply
\begin{equation} \label{eq:seqspecX}
    \MT \Spin \wedge \B\bZ_2 \wedge X_+ \xrightarrow[\hspace{5ex}]{q \wedge \text{id}} \MT\Spin \wedge X_+ \xrightarrow[\hspace{5ex}]{p \wedge \text{id}} \MT\SO \wedge X_+
\end{equation}
or \eqref{eq:seqspec} smashed with $X_+$. To justify this, let us examine the induced sequence of cobordism groups:
\begin{equation} \label{eq:seqcobordX}
    \mho_\SO^{d+2}(X) \xrightarrow[\hspace{5ex}]{p} \mho_\Spin^{d+2}(X) \xrightarrow[\hspace{5ex}]{q} \frac{\mho_\Spin^{d+2}(\B\bZ_2 \times X)}{\mho_\Spin^{d+2}(X)}
\end{equation}
It is clear that $\im(p)$ consists of the theories that are bosonisable while preserving symmetry $X$. To interpret the last term, note that a theory with combined $\bZ_2$ and $X$ symmetry can have (1) pure gravitational anomalies, (2) pure $\bZ_2$ anomalies, (3) pure $X$ anomalies, and (4) mixed $\bZ_2$-$X$ anomalies. We require (2) to vanish for $(-1)^F$ to be gaugable, and (4) to vanish for gauging $(-1)^F$ to preserve symmetry $X$. Thus the last term of \eqref{eq:seqcobordX}, in which (1) and (3) are divided out, is exactly what must vanish in order for $(-1)^F$ to be gaugable while preserving $X$. This construction allows us to define $H^{d+2}_B(X)$ for any space $X$.

Perhaps ironically, $H_B^\bullet(X)$ does \emph{not} define a generalised cohomology theory, in the Eilenberg--Steenrod sense. Although it satisfies all of the Eilenberg--Steenrod axioms bar one, there is no reason to believe it obeys the exactness axiom. Thus the `bosonisation cohomology' groups that we define are only a cohomology theory at the Kindergarten-level understanding of the word, that is, as a measure of inexactness associated to a complex.

\subsection{Examples in low dimensions} \label{sec:low-dim}

Let's determine what the cobordism sequence \eqref{eq:int:seq} looks like in various low dimensions. Using the information in \cref{app:gen}, we have
\begin{center}
    \begin{tikzcd}[row sep=tiny, nodes in empty cells,execute at end picture={
        \draw
            ($(\tikzcdmatrixname-1-1.south)!0.5!(\tikzcdmatrixname-2-1.north)$) coordinate (lx)
            (\tikzcdmatrixname.west|-lx) -- (\tikzcdmatrixname.east|-lx);
        \draw
            ($(\tikzcdmatrixname-1-1.east)!0.5!(\tikzcdmatrixname-1-2.west)$) coordinate (ly)
            (ly|-\tikzcdmatrixname.north) -- (ly|-\tikzcdmatrixname.south);
        }]
        n = d+2 & \mho^n_\SO(\pt) \arrow[r, "p"] & \mho^n_\Spin(\pt) \arrow[r, "q"] & \widetilde{\mho}^n_\Spin(\B\bZ_2) \\
        0 & \bZ \arrow[r] & \bZ & 0 \\
        1 & 0 & 0 & 0 \\
        2 & 0 & \bZ_2 \arrow[r] & \bZ_2 \\
        3 & 0 & \bZ_2 \arrow[r] & \bZ_2 \\
        4 & \bZ \arrow[r] & \bZ \arrow[r] & \bZ_8 \\
        5 & 0 & 0 & 0 \\
        6 & \bZ_2 & 0 & 0 \\
        7 & 0 & 0 & 0
    \end{tikzcd}
\end{center}
We would like to determine the maps $p$ and $q$. In the above, we have drawn arrows where the maps are not trivially zero, and need to be determined. We do this on a case-by-case basis:

\paragraph{$\bm{d = -2}$} This is a very degenerate case, but we include it for completeness. The cobordism groups simplify to $\Omega_H^0(\pt) \cong \Hom(\Omega_0^H(\pt), \bZ) = \bZ$ for both $H=\SO$ and $H=\Spin$, generated by a point with an orientation; the map between them is $p=1$.

\paragraph{$\bm{d = 0}$} The generators of $\mho_\Spin^2(\pt) \cong \bZ_2$ and $\widetilde{\mho}_\Spin^2(\B\bZ_2) \cong \bZ_2$ are objects $\cZ_1$ and $\cZ_2$ that can be evaluated on 1-manifolds with various structures. Since the only interesting 1-manifold is a circle, it is sufficient to specify them on $S^1$. We have
\begin{equation}
    \cZ_1[\rho] = \begin{cases}
        +1 & \rho = \rho_\NS \\
        -1 & \rho = \rho_\R
    \end{cases}
    \qquad
    \cZ_2[\rho, A] = (-1)^{\int_{S^1} A}
\end{equation}
with $\rho_\NS$ the Neveu--Schwarz (or antiperiodic, or bounding) spin structure on $S^1$. It is clear that $q(\cZ_1)[\rho, A] \coloneqq \frac{\cZ_1[\rho + A]}{\cZ_1[\rho]} = \cZ_2[\rho, A]$, because adding a nontrivial gauge field to a spin structure changes the spin structure. This shows that $q$ maps the generator to the generator, so $q = 1$.

\paragraph{$\bm{d = 1}$} This case is structurally identical to the last case, except our cobordism generators are now defined on 2-manifolds. They are
\begin{equation}
    \cZ_1[\rho] = (-1)^{\Arf[\rho]}
    \qquad
    \cZ_2[\rho, A] = (-1)^{\Arf[\rho + A] + \Arf[\rho]}
\end{equation}
Here $\Arf[\rho]$ is the \emph{Arf invariant}, defined as the number of zero modes of either $\slashed{D}_{\rho, L}$ or $\slashed{D}_{\rho, R}$, both being the same, modulo 2. The reduction mod 2 ensures this quantity is a topological invariant independent of the metric. Computing $\frac{\cZ_1[\rho + A]}{\cZ_1[\rho]} = \cZ_2[\rho, A]$, we again learn that $q = 1$.

\paragraph{$\bm{d = 2}$} This is the first interesting case. As reviewed in \cref{sec:int}, the generator $\sigma$ of bosonic anomaly polynomials is $16$ times the generator $\frac{1}{2}\hat{A}$ of fermionic anomaly polynomials, so $p = 16$. Meanwhile a 2d Majorana--Weyl fermion has a $(-1)^F$ anomaly of 1 mod 8, so $q = 1$. For more details, see \cite{bb}.

\noindent
Putting these results together, we can fill in the table:
\begin{center}
    \begin{tikzcd}[ampersand replacement=\&, row sep=tiny, nodes in empty cells,execute at end picture={
        \draw
            ($(\tikzcdmatrixname-1-1.south)!0.5!(\tikzcdmatrixname-2-1.north)$) coordinate (lx)
            (\tikzcdmatrixname.west|-lx) -- (\tikzcdmatrixname.east|-lx);
        \draw
            ($(\tikzcdmatrixname-1-1.east)!0.5!(\tikzcdmatrixname-1-2.west)$) coordinate (ly)
            (ly|-\tikzcdmatrixname.north) -- (ly|-\tikzcdmatrixname.south);
        \draw
            ($(\tikzcdmatrixname-1-4.east)!0.5!(\tikzcdmatrixname-1-5.west)$) coordinate (ly)
            (ly|-\tikzcdmatrixname.north) -- (ly|-\tikzcdmatrixname.south);
        }]
        n = d+2 \& \mho^n_\SO(\pt) \arrow[r, "p"] \& \mho^n_\Spin(\pt) \arrow[r, "q"] \& \widetilde{\mho}^n_\Spin(\B\bZ_2) \& H^n_B(\pt) \\
        0 \& \bZ \arrow[r, "1"] \& \bZ \& 0 \& 0 \\
        1 \& 0 \& 0 \& 0 \& 0 \\
        2 \& 0 \& \bZ_2 \arrow[r, "1"] \& \bZ_2 \& 0 \\
        3 \& 0 \& \bZ_2 \arrow[r, "1"] \& \bZ_2 \& 0 \\
        4 \& \bZ \arrow[r, "16"] \& \bZ \arrow[r, "1"] \& \bZ_8 \& \bZ_2 \\
        5 \& 0 \& 0 \& 0 \& 0 \\
        6 \& \bZ_2 \& 0 \& 0 \& 0 \\
        7 \& 0 \& 0 \& 0 \& 0
    \end{tikzcd}
\end{center}
It is then straightforward to compute $\ker(q) / \im(p) = H_B^n(\pt)$, as shown in the final column.

\section{Computing Bosonisation Cohomology} \label{sec:computing}

In the previous sections we gathered together some low-degree results. We observe, for instance, that bosonisation cohomology appears to be concentrated in degrees $n \in 4\bZ$. We now use more powerful tools to prove this holds, and eventually to compute the bosonisation cohomology groups in every dimension.

\subsection{Concentration of $H_B^n(\pt)$ in degrees $n \in 4\bZ$} \label{sec:4k}

The cobordism groups $\mho^\bullet$ appearing in the all-important anomaly sequence~\eqref{eq:int:seq} are themselves built from bordism groups $\Omega_\bullet$ via the universal coefficient sequence~\eqref{eq:UCT}. By taking the stable homotopy of the sequence of spectra \eqref{eq:seqspec}, we get
the following sequence of bordism groups
\begin{equation} \label{eq:seq_bord}
    \widetilde{\Omega}_n^\Spin(\B\bZ_2) \xrightarrow[\hspace{5ex}]{q_n} \Omega_n^\Spin(\pt) \xrightarrow[\hspace{5ex}]{p_n} \Omega_n^\SO(\pt)
\end{equation}
The question we'd like to ask here is: how much does this sequence tell us about the sequence of \emph{cobordism} groups \eqref{eq:int:seq} that we're actually interested in? Note that the arrows go `the other way' for the bordism sequence \emph{vs.}\ the cobordism sequence.

To answer this question, we will split the above sequence into its torsion and free parts. Recall from our discussion in \cref{sec:bordism} that the torsion part is related to global anomalies, while the free part (in one degree higher) determines the local anomaly. The splitting of~\eqref{eq:seq_bord} looks like:
\begin{equation} \label{eq:bord_grid}
\begin{tikzcd}[row sep=huge, column sep=large]
    T_n \arrow[r, "q_n^\text{(tor)}"] &
    \bZ_2^{F_n} \arrow[r, "p_n^\text{(tor)}"] &
    \bZ_2^{B_n} \\
    &
    \bZ^{A_n} \arrow[r, "p_n^\text{(free)}"] \arrow[ru, "p_n^\text{(diag)}" {xshift=2ex}] &
    \bZ^{A_n}
\end{tikzcd}
\end{equation}
The top row is the torsion part and the bottom row is the free part.
Note that we have invoked the known structure of all the groups involved, as gathered in \cref{app:gen} for the reader's convenience. In particular, we draw attention to the fact that $T_n = \widetilde{\Omega}_n^\Spin(\B\bZ_2)$ is pure torsion, while the torsion in spin bordism and oriented bordism is always a power of $\bZ_2$. The integers $A_n$, $B_n$ and $F_n$ are calculable, and given in \cref{app:gen} as the coefficients of certain power series.

The maps appearing in this diagram are all well-defined except for the diagonal maps $p_n^\text{(diag)}$ that mix torsion and non-torsion. These tricky maps depend on the choice of splitting.

Passing to the cobordism version of this splitting diagram via the UCT~\eqref{eq:UCT}, we get information about the morphisms between cobordism groups, hence between anomaly theories. Specifically, \eqref{eq:int:seq} splits as
\begin{equation} \label{eq:cobord_grid}
\begin{tikzcd}[row sep=huge, column sep=large]
    \bZ_2^{B_{n-1}} \arrow[r, "(p_{n-1}^\text{(tor)})^*"] &
    \bZ_2^{F_{n-1}} \arrow[r, "(q_{n-1}^\text{(tor)})^*"] &
    T_{n-1} \\
    \bZ^{A_n} \arrow[r, "(p_n^\text{(free)})^*"] \arrow[ru, "?"] &
    \bZ^{A_n} \arrow[ru, "?"]
\end{tikzcd}
\end{equation}
The horizontal maps are determined by the bordism sequence \eqref{eq:seq_bord}, but the diagonal maps are not. If we need them, we shall have to calculate them some other way. However, before investing any effort into doing so, we should first try to get as much mileage out of \eqref{eq:cobord_grid} as possible. We invoke the following facts, inferred from the seminal work of Anderson, Brown, and Peterson~\cite{anderson1966spin,spin}:
\begin{itemize}

\item $A_n = 0$ for $n \neq 0$ mod 4.

This corresponds, physically, to the statement that perturbative gravitational anomalies occur only in dimensions $d$ with $d+2\in 4\bZ$.
Mathematically, it can be read off from the structure of the oriented and spin bordism rings that we summarise in \cref{app:gen}; in particular, see the Hilbert--Poincaré series~\eqref{eq:app_Om_SO} and~\eqref{eq:app_Om_Spin}.
Thus for $n \neq 0$ mod 4, the bottom row is simply missing. The cobordism sequence \eqref{eq:cobord_grid} reduces to
\begin{equation}
\begin{tikzcd}[column sep=large]
    \bZ_2^{B_{n-1}} \arrow[r, "(p_{n-1})^*"] &
    \bZ_2^{F_{n-1}} \arrow[r, "(q_{n-1})^*"] &
    \bZ_2^{G_{n-2}}
\end{tikzcd}
\end{equation}
which \emph{is} fully determined by the bordism sequence. Furthermore, $T_{n-1}$ is a power of $\bZ_2$ in this dimension.

\item $p_n$ is injective for $n \neq 1, 2$ mod 8.

This is a strong statement, as it implies $q_n = 0$. What's more, $p_n^\text{(tor)}$ is a linear map between $\bF_2$-vector spaces, so it implies $(p_n^\text{(tor)})^*$ is surjective.

Thus for $n \neq 2, 3$ mod 8, the cobordism sequence reduces to
\begin{equation} \label{eq:cobordcase2}
\begin{tikzcd}
    \bZ_2^{B_{n-1}} \arrow[r, "\text{surjective}"] &[3ex]
    \bZ_2^{F_{n-1}} &
    T_{n-1} \\
    \bZ^{A_n} \arrow[r, "(p_n^\text{(free)})^*"] &
    \bZ^{A_n} \arrow[ru, "?"]
\end{tikzcd}
\end{equation}
where we have made use of the freedom of choice in the splitting to set one of the diagonal maps to zero.

\end{itemize}
Every $n$ falls into at least one of the above situations. Below, we analyse the different possibilities in turn, depending on the value of $n$ mod 8.

\subsubsection*{$\bm{n \text{ mod 8} \in \{1, 5, 6, 7\}}$}

Here both situations above apply, and the cobordism sequence simply reads
\begin{equation}
\begin{tikzcd}
    \bZ_2^{B_{n-1}} \arrow[r, "\text{surjective}"] &[3ex]
    \bZ_2^{F_{n-1}} &
    \bZ_2^{G_{n-2}}
\end{tikzcd}
\end{equation}
So in these dimensions, fermion parity is always gaugable, and bosonisation is always possible. Hence $H_B^n(\pt) = 0$.

\subsubsection*{$\bm{n \text{ mod 8} \in \{2, 3\}}$}

This case is similar to before, except the first map is no longer surjective. We shall show the cobordism sequence reads
\begin{equation} \label{eq:seqcase2}
\begin{tikzcd}[row sep=0]
    \bZ_2^{B_{n-1}} \arrow[r, "\text{surjective}"] &[3ex]
    \bZ_2^{F_{n-1} - A_{8m}} \\
    & \bZ_2^{A_{8m}} \arrow[r, "\text{injective}"] &[3ex] \bZ_2^{G_{n-2}}
\end{tikzcd}
\end{equation}
where we have written $n = 8m + i + 1$ with $i \in \{1, 2\}$. So again $H^n_B(\pt) = 0$.

To do this, we invoke one more fact from \cite{spin} pertaining to \eqref{eq:seq_bord}:
\begin{equation}
    \ker(p_{n-1}) = \bZ_2^{A_{8m}} = \braket{ u_r (S^1_\R)^i : r = 1 \dots A_{8m} }
\end{equation}
Here the $u_r$ are any manifolds that generate $\Omega_{8m}^\Spin(\pt)/\text{torsion} \cong \bZ^{A_{8m}}$, and $S^1_R$ is the circle with the Ramond (periodic) spin structure. Thus we not only know $\ker(p_{n-1})$, but have an explicit basis for it. Taking the $\bF_2$-vector space dual, we learn
\begin{equation} \label{eq:impcond}
    \cZ \in \im((p_{n-1})^*) \quad \iff \quad \cZ[u_r (S^1_R)^i] = 1 \text{ for all } r
\end{equation}
But $(p_{n-1})^*$ is the first map in the cobordism sequence. So we have proved the first half of \eqref{eq:seqcase2}.

It remains to show that \eqref{eq:impcond} is implied by $\cZ$ having no $(-1)^F$ anomaly. To do this recall that $\cZ$ has vanishing $(-1)^F$ anomaly if and only if $\cZ[M, \rho]$ is independent of spin structure. If so, then $\cZ[u_r (S^1_\R)^i] = \cZ[u_r (S^1_\NS)^i]$ since these manifolds only differ by their spin structure. But the latter is 1, since $S^1_\NS$ is null-bordant and $\cZ$ is a bordism invariant. This proves the claim, and the rest of \eqref{eq:seqcase2}.

\subsubsection*{$\bm{n \text{ mod 8} \in \{0, 4\}}$}

Here the situation of \eqref{eq:cobordcase2} applies. Discarding an irrelevant piece that cannot contribute to $H_B^n(\pt)$, the cobordism sequence reads
\begin{equation}
\begin{tikzcd}
    &[3ex] & T_{n-1} \\
    \bZ^{A_n} \arrow[r, "(p_n^\text{(free)})^*"] &
    \bZ^{A_n} \arrow[ru, "?"]
\end{tikzcd}
\end{equation}
Sadly, this is where \eqref{eq:seq_bord} finally runs out of steam. The second map is an undetermined ``anomaly interplay'' map \cite{anominter}, \emph{i.e.}\ one that relates a local anomaly to a global one for a different symmetry type. This map affects the answer. Thus we will have to determine it using different techniques.

We have learned that $H_B^n(\pt) \neq 0$ is a phenomenon exclusive to dimensions $d=n-2$ in which there are perturbative gravitational anomalies, and requires $n = 0$ mod 4. Thus, the remaining challenge is to calculate $H_B^{4k}(\pt)$ for all integer $k$. Restating the challenge, and using the information we have learnt, $H_B^{4k}(\pt)$ is given by the homology of the following sequence
\begin{equation} \label{eq:seq_4k}
\begin{tikzcd}
    \underbrace{\Hom(\Omega_{4k}^\SO(\pt), \bZ)}_{\mathclap{\substack{\text{bosonic anomaly polynomials} \\ \text{in dimension $4k$}}}} \arrow[r, "p"] \; & \;
    \underbrace{\Hom(\Omega_{4k}^\Spin(\pt), \bZ)}_{\mathclap{\substack{\text{fermionic anomaly polynomials} \\ \text{in dimension $4k$}}}} \; \arrow[r, "q"] &
    \; \underbrace{\Hom(\Omega_{4k-2}^{\Pin^-}(\pt), U(1))}_{\mathclap{\substack{\text{$\pin^-$ bordism invariants} \\ \text{in dimension $4k - 2$}}}}
\end{tikzcd}
\end{equation}
in the middle. In the following subsections we will explain how the maps $p$ and $q$ are defined and computed.

\subsection{Bosonic \emph{vs.}\ fermionic anomaly polynomials} \label{sec:phi}

We now focus our attention on the first map,
\begin{equation} \label{eq:seq_4kp}
\begin{tikzcd}
    \underbrace{\Hom(\Omega_{4k}^\SO(\pt), \bZ)}_{\mathclap{\substack{\text{bosonic anomaly polynomials} \\ \text{in dimension $4k$}}}} \arrow[r, "p"] \; & \;
    \underbrace{\Hom(\Omega_{4k}^\Spin(\pt), \bZ)}_{\mathclap{\substack{\text{fermionic anomaly polynomials} \\ \text{in dimension $4k$}}}}
\end{tikzcd}
\end{equation}
that is, the map from bosonic anomaly polynomials into the fermionic ones.
Recall from the introduction that a bosonic anomaly polynomial must be consistently defined on a larger set of manifolds (namely, the oriented ones) than a fermionic anomaly polynomial, which need only be well-defined on all oriented manifolds which are moreover spin. Recall that in $n=4$, we deduced this map $p$ sends $1 \mapsto 16$, in that case a straightforward consequence of Rokhlin's signature theorem.

In order to calculate the image of the map $p$, which is all we need to know about it to compute $H_B^{4k}(\pt)$, we need to know how full bases of bosonic and fermionic anomaly polynomials are calculated in a general dimension. These problems were tackled by Stong~\cite{hs1,hs2}; we also refer the reader to the appendices of~\cite{freed2021} for a more recent review, as well as~\cite{expliciths} for further explicit formulae.

\subsubsection*{Fermionic anomaly polynomials}

The key idea is that the group $\Hom(\Omega_{4k}^\Spin(\pt), \bZ)$ is generated by anomaly polynomials for some set of twisted Dirac operators, which are themselves associated to an underlying set of polynomials. To wit, for each polynomial $\theta \in \bZ[\pi^1, \pi^2, \dots]$ in abstract variables $\pi^1, \pi^2, \dots$, define a Dirac operator
\begin{equation}
    \slashed{D}_{\theta} \coloneqq \slashed{D}_{\theta(\pi^r(TM))}
\end{equation}
where the operations $\pi^r : \KO^0(M) \rightarrow \KO^0(M)$, $r \geq 1$ are called $\KO$-Pontryagin classes.\footnote{See \cite{freed2021} for a recent review.  The notation $\pi^r$ that we adopt for the $KO$-Pontryagin classes was originally introduced in \cite{KOchar}. Their Pontryagin characters are denoted $e_r$ in \cite{hs1}. One can also express the $\pi^r$ in terms of the operations $\gamma_t$ defined in \cite{expliciths}, via $\gamma_t(\widetilde{V}) = \pi_{t(1-t)}(V)$.} We apply these $\pi^r$ operations to the tangent bundle $TM$, feed them to the polynomial $\theta(\pi^1, \pi^2, \dots)$, and use the result to twist the Dirac operator. By way of explicit formulae, the first few of the $\pi^r$ operations are, for some vector bundle $E$,
\begin{align}
    \pi^1(E) &= \widetilde{E}, \\
    \pi^2(E) &= \extpow^2(\widetilde{E}) + 2\widetilde{E}, \nonumber \\
    \pi^3(E) &= \extpow^3 (\widetilde{E}) + 4\extpow^2(\widetilde{E})+5\widetilde{E}, \nonumber\\
    \pi^4(E) &= \extpow^4(\widetilde{E})+6\extpow^3(\widetilde{E})+14\extpow^2(\widetilde{E})+14\widetilde{E} \nonumber
\end{align}
where $\widetilde{E} = E - \dim(E)$ denotes the projection of $E$ to reduced $\KO$ theory; for instance in four dimensions, this means that $\slashed{D}_{\theta = \pi^1} = \slashed{D}_{\pi^1(TM)} = \slashed{D}_{TM - 4} = \slashed{D}_{TM} - 4\slashed{D}$, a formal difference of Dirac operators. In general, the decomposition of $\pi^r(E)$ into exterior product bundles is given by the following formula:
\begin{equation} \label{eq:pir_general}
    \pi^r(E) = \frac{1}{r} \sum_{k=1}^r k \tbinom{2r}{r-k} \extpow^k(\widetilde{E})
\end{equation}
The inverse formula can already be found in~\cite[Proposition 5.1]{anderson1966spin}.

We are now ready to state how the free part of spin bordism in degrees $4k$ is generated by anomaly polynomials of such twisted Dirac operators. We have
\begin{shadebox}
\textbf{Theorem:} (Fermionic Hattori--Stong) The set of all fermionic anomaly polynomials in dimension $4k$ is \cite{hs1, hs2}
\begin{equation} \label{eq:HS-1}
    \Hom(\Omega_{4k}^\Spin(\pt), \bZ)
    =
    \left\{ \frac{\Phi_{4k}(i\slashed{D}_\theta)}{\gcd(2,k+1)} : \deg(\theta) \leq k \right\}
\end{equation}
where in $\deg(\theta)$, we regard $\pi^r$ as having formal degree $r$.
\end{shadebox}
The Atiyah-Singer index theorem guarantees all polynomials on the right hand side are correctly normalised to give integers on a spin manifold. This is because $\text{index}(i\slashed{D}_\theta)$ is an integer, and moreover an even integer when $k$ is odd, due to the existence of Majorana-Weyl fermions. The Hattori--Stong theorem then provides a converse to this statement. The reason for the degree bound on $\theta$ is that all other anomaly polynomials are zero, as we discuss next.

The anomaly polynomials that appear in~\eqref{eq:HS-1}, for twisted Dirac operators, take the form
\begin{equation}
    \Phi(i\slashed{D}_\theta) = \hat{A}(TM) \, \theta(\ph(\pi^r(TM)))
\end{equation}
with $\Phi_{4k}(i \slashed{D}_\theta)$ obtained by taking the differential form in degree $4k$ from the infinite series $\Phi(i\slashed{D}_\theta)$.
Here $\hat{A}(TM)$ is called the Dirac `$A$-hat genus', defined by
\begin{equation} \label{eq:Ahat_gen}
    \hat{A}(TM) = \prod_{i=1}^\infty \frac{x_i/2}{\sinh(x_i/2)}    
\end{equation}
where $x_i$ are the Chern roots of the vector bundle.\footnote{The fact that the product runs over infinitely-many Chern roots reflects the stable setting of our preferred formalism; for bundles over a particular base space $M$ of dimension $n$, there are $2n$ distinct Chern roots.} 
This can be expanded out as a Taylor series and thence expressed in terms of the Pontryagin classes $p_r\in H^{4r}(M; \bR)$, which are straightforwardly related to these Chern roots via $p_r = e_r ( \{x_i^2\})$ where $e_r$ denotes the $r^\text{th}$ elementary symmetric polynomial. The series begins
\begin{equation} \label{eq:Ahat}
    \hat{A}(TM) = 1 - \frac{p_1}{24} + \frac{7 p_1^2 - 4 p_2}{5760} + \frac{-31 p_1^3 + 44 p_2 p_1 - 16 p_3}{967680} + \dots
\end{equation}
in low degrees.

The object $\ph(E)$, for some vector bundle $E$, is the \emph{Pontryagin character}  (not to be confused with the Pontryagin \emph{classes}). 
As for $\hat{A}(TM)$, these objects can be expressed via a generating function, see \emph{e.g.}~\cite[App. B]{freed2021}, \emph{viz.}\
\begin{equation} \label{eq:ph-gen}
     \ph(\pi_t(TM)) = \prod_{i=1}^\infty \left(1 +t\left(e^{x_i}+e^{-x_i}-2 \right) \right) = \prod_{i=1}^\infty \left( 1+2t(\cosh x_i - 1)\right)
\end{equation}
where $\pi_t(E) \coloneqq \sum_{r=0}^\infty \pi^r(E) t^r$, and $x_i$ are the same Chern roots as before. Note that, unlike~\eqref{eq:Ahat_gen}, this function is a polynomial in variable $t$, whose $r^{\text{th}}$ coefficient tells us the $\pi^r(E)$ operation on vector bundles.
To get an idea, the first few are given by
\begin{align} \label{eq:ph-1}
    \ph(\pi^1(TM)) &= p_1 + \frac{1}{3!} \left(\frac{p_1^2}{2} - p_2\right) + \frac{1}{5!} \left(\frac{p_1^3}{3} - p_1 p_2 + p_3\right) + \dots \\
    \ph(\pi^2(TM)) &= p_2 + \frac{1}{4} \left(\frac{p_1 p_2}{3} - p_3\right) + \dots \nonumber \\
    \ph(\pi^3(TM)) &= p_3 + \dots \nonumber
\end{align}
For general $r$, we can extract the $t^r$ coefficient 
from~\eqref{eq:ph-gen}, 
\begin{align}
     \ph(\pi^r(TM)) = 2^r e_r(\{\cosh x_i-1\}) = p_r + \mathcal{O}(x^{2r+2})
\end{align}
matching the explicit equations~\eqref{eq:ph-1}, and illustrating how one can obtain general results about the Pontryagin characters. 
In App.~\ref{app:proofs}
we derive further properties about both $\hat{A}(TM)$ and $\ph(\pi^r(TM))$ that are needed to compute the general structure of the bosonisation cohomology groups $H_B^{4k}(\pt)$.

Let's now see how to use the Hattori--Stong theorem, in the form quoted in~\eqref{eq:HS-1}, by way of the simplest non-trivial example.

\example Consider $k=1$, corresponding to degree-4 anomaly polynomials (the first interesting case). Then the only possible monomials with $\deg(\theta) \leq 1$ are $1$ and $\pi^1$. So $\tfrac{1}{2}\Phi_4(i\slashed{D})$ and $\tfrac{1}{2}\Phi_4(i\slashed{D}_{\pi^1})$ generate all fermionic anomaly polynomials. Using later results, these are $-\frac{p_1}{48}$ and $\frac{p_1}{2}$. So the single generator $\frac{p_1}{48}$ suffices.

This example illustrates that in general, the Hattori--Stong theorem gives an overcomplete basis. There are $p(0) + \dots + p(k)$ possible monomials $\theta$ with $\deg(\theta) \leq k$, but only $p(k)$ linearly independent fermionic anomaly polynomials. (Here $p(k)$ are the partition numbers.) Hence to get a basis, we must take a basis refinement. While this yields an effective algorithm to compute a basis, it does not give an explicit formula for it, and in general no such formula exists.

\subsubsection*{From fermionic to bosonic anomaly polynomials}

We have given an entirely explicit account of general fermionic anomaly polynomials, and how these can be taken to generate the spin cobordism group appearing in our all-important bosonisation sequence~\eqref{eq:seq_4k}. Recall that the map $p$ in this sequence sends a general bosonic anomaly polynomial to its image in the (larger) set of fermionic anomaly polynomials.

From Stong~\cite{hs1}, there is a simple recipe to determine all bosonic anomaly polynomials from all fermionic ones:
\begin{shadebox}
\textbf{Theorem:} (Bosonic Hattori--Stong) The set of all bosonic anomaly polynomials in dimension $4k$ is \cite{hs1}
\begin{equation} \label{eq:HS-2}
    \Hom(\Omega_{4k}^\SO(\pt), \bZ)
    =
    \{ 2^{d(\Phi)} \Phi : \Phi \in \Hom(\Omega_{4k}^\Spin(\pt), \bZ) \}
\end{equation}
where $2^{d(\Phi)}$ is the smallest power of 2 that clears all powers of 2 in the denominator of $\Phi$, regarded as a rational polynomial in the Pontryagin classes.
\end{shadebox}
Before we unpack this in general dimensions, let us illustrate how this recipe is implemented through an example.

\begin{example}
If $k=1$, the generator of fermionic anomaly polynomials is $\Phi = \frac{p_1}{48}$. It has $2^{d(\Phi)} = 2^4$, so $2^4 \Phi = \frac{p_1}{3}$ is the generator of bosonic anomaly polynomials. This chimes with our discussion in the introduction, whereby the relative factor of 16 was explained via the Rokhlin signature theorem.
\end{example}

\subsection{The map $p$ and $2$-localisation} \label{sec:p2loc}

Having described how the spaces of bosonic and fermionic anomaly polynomials are calculated, we now describe how to pick bases thereon such that the map $p$ between them takes a particularly simple (diagonal) form. This motivates introducing an algebraic process known as $2$-localisation, which throws away a large amount of information irrelevant to the computation of $H_B^{4k}(\pt)$; the resulting simplification is essential and will allow us to eventually compute the explicit forms of $p$ and $q$.

Consider the special subset of anomaly polynomials built from the Pontryagin classes with integer coefficients. Explicitly, we can write this as $\text{span}\{ p^J : n(J) = k \}$, where we have taken the opportunity to shoehorn in some new notation: $J=(j_1,j_2,\dots)$ is a sequence of integers, usually called a \emph{partition}, $n(J) = j_1 + j_2 + \dots$ is the size of the partition, and $p^J = p_{j_1} p_{j_2} \dots$ is the associated Pontryagin monomial. The constraint $n(J) = k$ ensures that $p^J$ has differential form degree $4k$. Since every $p^J$ is a fermionic anomaly polynomial, we have the inclusion of lattices
\begin{equation}
    \text{span}\{ p^J : n(J) = k \} \subseteq \Hom(\Omega_{4k}^\Spin(\pt), \bZ)
\end{equation}
where both sides are isomorphic to $\bZ^{p(k)}$.

An inclusion of lattices is ripe for applying the Smith normal form. This furnishes us with a basis $\{P_1, \dots, P_{p(k)}\}$ for the left hand side such that $\{\frac{1}{\lambda_1} P_1, \dots, \frac{1}{\lambda_{p(k)}} P_{p(k)}\}$ is a basis for the right hand side, with the $\lambda_i$ integers. Factorising these integers as $\lambda_i = 2^{d_i} r_i$ where the $r_i$ are odd numbers, we deduce that
\begin{equation} \label{eq:bfbases}
\begin{alignedat}{3}
    \Hom(\Omega_{4k}^\Spin(\pt), \bZ) &= \text{span}\bigg\{& \frac{1}{2^{d_i} r_i} &P_i : i = 1, \dots, p(k) \bigg\} \\
    \Hom(\Omega_{4k}^\SO(\pt), \bZ) &= \text{span}\bigg\{& \frac{1}{r_i} &P_i : i = 1, \dots, p(k) \bigg\}
\end{alignedat}
\end{equation}
where the bottom line follows from the oriented Hattori-Stong theorem \eqref{eq:HS-2}. Thus, with respect to these bases, the inclusion map $p$ takes the simple diagonal form
\begin{equation} \label{eq:pdiag}
    p = \begin{pmatrix}
        2^{d_1} & 0 & \cdots & 0 \\
        0 & 2^{d_2} & \cdots & 0 \\
        \vdots & \vdots & \ddots & \vdots \\
        0 & 0 & \cdots & 2^{d_{p(k)}}
    \end{pmatrix}
\end{equation}
Such a map is said to be \emph{invertible away from $2$}. (In fact, our exposition is backwards: invertibility away from $2$ is proven first, and \eqref{eq:HS-2} is a corollary).

The above result implies the bosonisation cohomology groups $H_B^{4k}(\pt)$ are always 2-torsion, meaning $H_B^{4k}(\pt)$ is always isomorphic to a product of cyclic groups each of order a power of 2. To prove this, consider localising the sequence~\eqref{eq:seq_4k} away from 2, i.e.\ tensoring it with the dyadic rationals $\bZ[\tfrac{1}{2}]$. Because $\pin^-$ bordism groups are always 2-torsion, tensoring with $\bZ[\tfrac{1}{2}]$ kills the third group. Meanwhile, the map $p$ is invertible away from 2, so it becomes an isomorphism. Thus, localising away from 2 transforms the sequence into
\begin{equation}
\begin{tikzcd}
    \bZ[\tfrac{1}{2}]^{p(k)} \arrow[r, "1"] &
    \bZ[\tfrac{1}{2}]^{p(k)} \arrow[r, "0"] &
    0
\end{tikzcd}
\end{equation}
This is a very boring sequence. It has trivial homology. Since taking homology commutes with localisation, this proves the claim. All the information in the sequence~\eqref{eq:seq_4k} is therefore contained in the 2-localisation, i.e.\ the result of tensoring with $\bZ_{(2)}$, the ring of all rationals with odd denominator.

\subsubsection*{2-local bases of anomaly polynomials}

In view of the previous result, it suffices to work with 2-local bases of fermionic and bosonic anomaly polynomials. By relaxing the notion of basis, we can get more explicit answers, including the integers $d_i$ appearing on the diagonal of \eqref{eq:pdiag}.

Given a lattice, we define a ``2-local basis'' as a linearly independent set which spans a sublattice of odd index. (This is equivalent to a set which forms a basis after taking the 2-localisation.) For example, $\{(3,0), (0,5)\}$ forms a 2-local basis of $\bZ^2$. An explicit 2-local basis for the space of fermionic anomaly polynomials is then
\begin{shadebox}
\textbf{Theorem:} (2-local fermionic Hattori--Stong) A 2-local basis for the set of all fermionic anomaly polynomials in dimension $4k$ is \cite{spin}
\begin{equation} \label{eq:HS-3}
    \Hom(\Omega_{4k}^\Spin(\pt), \bZ) \otimes \bZ_{(2)}
    =
    \mathrm{span}\left\{ \frac{\Phi_{4k}(i\slashed{D}_{\pi^J})}{\gcd(2,k+1)} : n(J) \leq k \text{ and $1 \notin J$}\right\} \otimes \bZ_{(2)}
\end{equation}
\end{shadebox}
Note that the number of partitions $J$ is exactly $p(k)$, matching the dimension of the space of anomaly polynomials.

To derive the analogous result for bosonic anomaly polynomials, we appeal to a fact proved in \cref{app:proofs}. Let $J$ be a partition occurring in~\eqref{eq:HS-3}. Then
\begin{equation} \label{eq:phitop}
    2^{f(k - n(J))} \Phi_{4k}(i\slashed{D}_{\pi^J}) = p^{J'} \text{ mod } 2 \bZ_{(2)}[p_1, p_2, \dots]
\end{equation}
where $f(n) = 4n - \text{bitcount}(n)$, or sequence \href{https://oeis.org/A120738}{A120738} in OEIS, and $J'$ denotes the partition obtained from $J$ by topping it up with enough $1$s to make $n(J') = k$. This result states that the smallest power of 2 that clears all powers of 2 in the denominator of $\Phi_{4k}(i\slashed{D}_{\pi^J})$ is $2^{f(k - n(J))}$, hence by \eqref{eq:HS-2} the left hand side is a bosonic anomaly polynomial. As $J$ runs over all allowed partitions, the right hand side runs over a 2-local basis of bosonic anomaly polynomials.\footnote{For the proof, \eqref{eq:bfbases} implies that a 2-local basis of bosonic anomaly polynomials is $\{p^J : n(J) = k\}$, also a well-known result. This 2-local basis differs from the right hand side of \eqref{eq:phitop} by a matrix with determinant 1 mod $2\bZ_{(2)}$, hence it forms 2-local basis too.} Thus we have proved
\begin{shadebox}
\textbf{Theorem:} (2-local bosonic Hattori--Stong) A 2-local basis for the set of all bosonic anomaly polynomials in dimension $4k$ is
\begin{equation} \label{eq:HS-4}
    \Hom(\Omega_{4k}^\SO(\pt), \bZ) \otimes \bZ_{(2)}
    =
    \mathrm{span}\Big\{ 2^{f(k - n(J))} \Phi_{4k}(i\slashed{D}_{\pi^J}): n(J) \leq k \text{ and $1 \notin J$}\Big\} \otimes \bZ_{(2)}
\end{equation}
where $f(n) = 4n - \mathrm{bitcount}(n)$.
\end{shadebox}
Working with respect to the 2-local bases \eqref{eq:HS-3} and \eqref{eq:HS-4}, the inclusion map $p$ becomes diagonal. It takes the form
\begin{equation}
    p = \text{diag}\big(\{ 2^{f(k-n(J)) + (k \text{ mod } 2)} : n(J) \leq k \text{ and } 1 \notin J \}\big)
\end{equation}
From this we can read off the powers of 2 occurring on the diagonal of \eqref{eq:pdiag}.

\example When $k = 1$, the only contributing partition is $J = \{\}$. We get $p = 2^{f(1) + 1}$, $f(1) = 4 - \text{bitcount}(1) = 3$, so $p = 16$.

This concludes our analysis of the map $p$ in every relevant dimension.

\subsection{Pin${}^-$ bordism} \label{sec:pin}

Having dealt with the first map $p$ in our sequence~\eqref{eq:seq_4k} of cobordism groups, we now turn to the second map $q$:
\begin{equation} \label{eq:seq_4kq}
\begin{tikzcd}
    \underbrace{\Hom(\Omega_{4k}^\Spin(\pt), \bZ)}_{\mathclap{\substack{\text{fermionic anomaly polynomials} \\ \text{in dimension $4k$}}}} \; \arrow[r, "q"] &
    \; \underbrace{\Hom(\Omega_{4k-2}^{\Pin^-}(\pt), U(1))}_{\mathclap{\substack{\text{$\pin^-$ bordism invariants} \\ \text{in dimension $4k - 2$}}}}
\end{tikzcd}
\end{equation}
Before proceeding, let us recall the basic physics idea behind this map. The group on the right-hand-side is isomorphic, via Smith, to the group $\widetilde{\mho}_{\Spin}^{4k}(\B\bZ_2)$ that classifies anomalies for QFTs with spin structure and $\bZ_2$ symmetry, in dimension $d=4k-2$. For the particular case of fermion parity, the potential $\bZ_2$ anomaly is fully determined by the gravitational anomaly associated with $\Spin$, and this is captured by the map $q$.

We have already given a complete description of the group on the left-hand-side, which is the group of fermionic anomaly polynomials. Thanks to the Smith isomorphism (see \cref{sec:math_seq}), we can elucidate the structure of the group on the right by appealing to known results for the classification of $\pin^-$ bordism, which we only need in degrees 2 mod 4. This is our next task.

Roughly speaking, $\pin^-$ bordism generators in the relevant degrees can be obtained by taking the direct product of a spin bordism generator with an $\RP^{4l-2}$ manifold equipped with $\pin^-$ structure~\cite{anderson1969pin, kirbytaylor}. In much the same way as for anomaly polynomials, these generators are labelled by integer partitions $J = (j_1, j_2, \dots)$ with $1 \notin J$. The generators come in three different flavours:
\begin{enumerate}

\item For each partition with $1 \notin J$ and $n(J)$ even, there are generators
\begin{equation} \label{tab:pin1}
\begin{tabular}{c|c|c}
manifold & dimension & order \\
\hline
$\RP^2\times M_J$ & $4n(J)+2$ & $8$ \\
$\RP^6\times M_J$ & $4n(J)+6$ & $16$ \\
$\RP^{10}\times M_J$ & $4n(J)+10$ & $8\cdot16$ \\
$\RP^{14}\times M_J$ & $4n(J)+14$ & $16^2$ \\
$\vdots$ & $\vdots$ & $\vdots$ \\
\end{tabular}
\end{equation}
The $M_J$ are certain manifolds in the classification of spin bordism \cite{spin}.

\item For each partition with $1 \notin J$ and $n(J)$ odd, there are generators
\begin{equation} \label{tab:pin2}
\begin{tabular}{c|c|c}
manifold & dimension & order \\
\hline
? & $4n(J)-2$ & $2$ \\
$\tfrac{1}{2}(\RP^2\times M_J)$ & $4n(J)+2$ & $4$ \\
$\tfrac{1}{2}(\RP^6\times M_J)$ & $4n(J)+6$ & $2\cdot16$ \\
$\tfrac{1}{2}(\RP^{10}\times M_J)$ & $4n(J)+10$ & $4\cdot16$ \\
$\tfrac{1}{2}(\RP^{14}\times M_J)$ & $4n(J)+14$ & $2\cdot16^2$ \\
$\vdots$ & $\vdots$ & $\vdots$ \\
\end{tabular}
\end{equation}
The notation $\tfrac{1}{2}M$ denotes any manifold $N$ such that $2N = M$ in $\pin^-$ bordism, and also asserts that such a manifold exists.

Here we encounter an issue: the manifold labelled ? appearing in the first row of the above table is known to exist abstractly, but an explicit representative is not known \cite{kirbytaylor}. This mystery manifold will not affect our calculation in \cref{sec:result}. In fact, we will unmask it along the way as being $N_J$, another manifold from the classification of spin bordism \cite{spin}.

\item There are also various further manifolds of order 2 appearing in $\pin^-$ bordism,
\begin{equation} \label{tab:pin3}
\begin{tabular}{c|c|c}
manifold & dimension & order\tabularnewline
\hline
? & sporadic $n$ & $2$
\end{tabular}
\end{equation}
Again, although such manifolds exist abstractly, no explicit representatives are known. The number of manifolds in each dimension can be counted (see \cref{app:gen}); the lowest-dimensional example in this category has $n = 22$.

\end{enumerate}
Of course, $\pin^-$ bordism also has generators in other degrees not equal to 2 mod 4; these are not listed here, since they play no part in our story.

\example To compute $\pin^-$ bordism in dimension $10$, one simply collects together all the generators of the right dimension, from the three sets described above. From the even partition $J = \{\}$ we get $\RP^{10}$ of order $128$, from the even partition $J=\{2\}$ we get $\RP^2 \times M_{\{2\}}$ of order 8, and from the odd partition $J=\{3\}$ we get a ? generator of order 2 (which is secretly $N_{\{3\}}$). So $\Omega^{\Pin^-}_{10}(\pt) \cong \bZ_{128} \times \bZ_8 \times \bZ_2$.

\subsection{The map $q$ and eta invariants} \label{sec:q}

We now turn to the map $q$, which recall sends a perturbative fermionic anomaly to the induced anomaly for gauging $(-1)^F \in \Spin(d)$. To compute how $q$ acts, we work at the level of partition functions of anomaly theories, as discussed in \cref{sec:int}.

\noindent \textbf{Recipe for $q$:}
\begin{itemize}

\item Start with a fermionic anomaly polynomial $\Phi_{4k} \in \Hom(\Omega_{4k}^\Spin(\pt), \bZ)$.

\item Pick any fermionic anomaly theory $\cZ_{4k-1}[g, \rho]$ with perturbative anomaly $\Phi_{4k}$. The choice is not unique, but does not matter by results of \cref{sec:4k}.

\item Consider the metric-independent ratio
\[
    \frac{\cZ_{4k-1}[g, \rho + A]}{\cZ_{4k-1}[g, \rho]} \in \Hom(\widetilde{\Omega}_{4k-1}^{\Spin}(\B\bZ_2), U(1))
\]

\item Apply the Smith isomorphism, landing in $\Hom(\Omega_{4k-2}^{\Pin^-}(\pt), U(1))$.

\end{itemize}

We will need to evaluate $\big( q(\Phi_{4k}) \big)(M_{4k-2}) \in U(1)$ for various anomaly polynomials and pin manifolds. This can be done by expressing $q(\Phi_{4k})$ as an eta invariant, and then using various tools for eta invariants to evaluate it.

For our purposes, the eta invariant of a Hermitian operator $\cD$ is defined using the convention \eqref{eq:etadef}, which ensures that it jumps by $\pm1$ as an eigenvalue of $\cD$ passes through zero. Thus, usually, only $e^{2 \pi i \eta(\cD)}$ is continuous as an eigenvalue goes through zero. But if it happens that $\cD$ has a doubly degenerate spectrum, then actually $e^{2 \pi i \eta(\cD) / 2}$ is continuous.

Let's start with the simplest case of $\theta = 1$, and work up to the general case in steps. So consider the fermionic anomaly polynomial $\frac{\Phi_{4k}(i\slashed{D})}{\gcd(2,k+1)}$. We describe how to apply $q$ to it. Running the recipe,
\begin{itemize}

\item The corresponding fermionic anomaly theory is the exponentiated eta invariant of a massive fermion,
\begin{equation}
    \cZ_{4k-1}[g, \rho] = \exp\!\left( 2\pi i \frac{\eta(i\slashed{D})}{\gcd(2,k+1)} \right)
\end{equation}

\item We consider the difference
\begin{equation}
    \exp\!\left( 2\pi i \frac{\eta(i\slashed{D}_{\rho+A}) - \eta(i\slashed{D}_{\rho})}{\gcd(2,k+1)} \right) \in \Hom(\widetilde{\Omega}_{4k-1}^{\text{Spin}}(\bZ_2), U(1))
\end{equation}
where the subscripts here indicate spin structures (not $\theta$).

\item Under the Smith isomorphism, this eta invariant becomes \cite[Lemma 3.3a]{bahri1987eta}
\begin{equation}
    \exp\!\left( 2\pi i \frac{\eta(\bar{\gamma}\slashed{D})}{\gcd(2,k+1)} \right) \in \Hom(\Omega_{4k-2}^{\Pin^-}(\pt), U(1))
\end{equation}
where $\bar{\gamma}$ is the chirality matrix, needed to make the Dirac operator well-defined on an even-dimensional pin manifold.

\end{itemize}
Thus the final answer is
\begin{equation} \label{eq:qplaindirac}
    q\!\left(\frac{\Phi_{4k}(i\slashed{D})}{\gcd(2,k+1)}\right)
    =
    \exp\!\left( 2\pi i \frac{\eta(\bar{\gamma}\slashed{D})}{\gcd(2,k+1)} \right)
\end{equation}
The right-hand side is a quantity that can be evaluated on any $4k-2$ dimensional $\pin^-$ manifold, and is a bordism invariant.

Note that in all cases, the factors of $\gcd(2, k + 1)$ are either 1 or 2, and take the value 2 precisely when the Dirac operator in question has a doubly degenerate spectrum. So all exponentiated eta invariants are well-defined.

We must now confront the task of evaluating eta invariants on various $\pin^-$ manifolds in the different classes set out above. To make our life easier, there are two tricks we employ, as follows.

\subsubsection*{Trick 1: orbifolding}

Let us consider evaluating \eqref{eq:qplaindirac} on the simplest $4k-2$ dimensional $\pin^-$ manifold, $\RP^{4k-2}$. This can be done using the ``orbifolding trick'' (see \cite{Witten_2016, Tachikawa_2019} for exposition). 
To wit, consider the orbifold $T^{4k-1}/\bZ_2$ where the torus $T^{4k-1}$ has the periodic spin structure and the $\bZ_2$ action that we quotient by sends $x \mapsto -x$. This group action has $2^{4k-1}$ fixed points, and cutting out balls around them gives $2^{4k-1}$ $\RP^{4k-2}$ boundaries. The APS index theorem says
\begin{equation}
    2^{4k-1} \eta(\bar{\gamma} \slashed{D}, \RP^{4k-2}) = \text{index}\left(i \slashed{D}, \tfrac{T^{4k-1} \backslash \text{balls}}{\bZ_2}\right)
\end{equation}
and the latter is given by the number of $\bZ_2 = +1$ zero modes of $\slashed{D}$ on $T^{4k-1}$ minus the number of $\bZ_2 = -1$ zero modes. Since the zero modes are constant Dirac spinors, they all have $\bZ_2 = +1$ and there are $2^{2k-1}$ of them. So the eta invariant is
\begin{equation} \label{eq:eta_RP}
    \eta(\bar{\gamma} \slashed{D}, \RP^{4k-2}) = \frac{2^{2k-1}}{2^{4k-1}} = \frac{1}{2^{2k}}
\end{equation}
Combined with the previous result, this gives
\begin{equation} \label{eq:q_RP}
    q\!\left(\frac{\Phi_{4k}(i\slashed{D})}{\gcd(2,k+1)}\right)(\RP^{4k-2})
    =
    \exp\!\left( \frac{2 \pi i}{2^{2k+(k \text{ mod } 2)}} \right)
\end{equation}

\example If $k = 1$, this gives $e^{2 \pi i / 8}$.

\subsubsection*{Trick 2: factorisation}

To generalise this calculation to the manifolds $M_J \times \RP^{4l-2}$ we need another trick: namely, a factorisation formula for evaluating eta invariants on product spaces, due to Bahri and Gilkey \cite{bahri1987eta}. 
For the ordinary (untwisted) Dirac operator, this factorisation formula takes the guise
\begin{equation}
    \eta({\bar{\gamma} \slashed{D}}, M_J \times \RP^{4l-2})
    =
    \mathrm{index}(i \slashed{D}, M_J)
    \,
    \eta({\bar{\gamma} \slashed{D}}, \RP^{4l-2})
\end{equation}
Here we take a moment to point out the slightly subtle way the spectrum doubling works out in this formula. When the Dirac operator on the left has a doubled spectrum (\emph{i.e.}\ when $n(J) + l$ is odd), then exactly one of the Dirac operators on the right also has a doubled spectrum (when $n(J)$ is odd \emph{or} $l$ is odd). This ensures that the exponentiated eta invariant on the left can always be expressed in terms of that on the right, without a sign ambiguity.

Applied to our case, and using \eqref{eq:eta_RP} to evaluate the eta invariant on the $\RP^{4l-2}$ part, 
we find the appropriate generalisation of the evaluation of~\eqref{eq:q_RP} is
\begin{equation}
    q\!\left(\frac{\Phi_{4k}(i\slashed{D})}{\gcd(2,k+1)}\right)(M_J \times \RP^{4l-2})
    =
    \exp\!\left( \frac{2 \pi i}{2^{2l+(k \text{ mod } 2)}} \int_{M_J} \Phi_{4(k-l)}(i \slashed{D}) \right)
\end{equation}

\subsubsection*{Generalisation for twisted Dirac operators}

On the other hand, we need a further generalisation of \eqref{eq:qplaindirac} to the case with a twisting bundle, \emph{i.e.}\ $\theta = \theta(\pi^1, \pi^2, \dots) \neq 1$. Recall from \S \ref{sec:phi} that we need such twisted bundles to generate the oriented and spin cobordism groups, hence they are needed to compute the map $q$. In the following, let $\theta(V)$ be shorthand for $\theta(\pi^1(V), \pi^2(V), \dots)$, so that $\theta$ describes an operation on vector bundles.

The first main change is that when dimensionally reducing the anomaly polynomial to the anomaly theory, the tangent bundle splits up as $TM \rightarrow TM + 1$, so we have
\begin{equation}
    \frac{\Phi_{4k}(i\slashed{D}_{\theta(TM)})}{\gcd(2,k+1)}
    \quad \rightarrow \quad
    \cZ[g, \rho] = \exp\!\left( 2 \pi i\frac{\eta(i\slashed{D}_{\theta(TM + 1)})}{\gcd(2,k+1)} \right)
\end{equation}
Next, upon reducing the anomaly theory further to a $\pin^-$ bordism invariant, the tangent bundle further splits as $TM \rightarrow TM + \xi$ where $\xi$ is the orientation bundle \cite{bahri1987eta}, as motivated in \cref{fig:smith}, so
\begin{equation}
    q\!\left(\frac{\Phi_{4k}(i\slashed{D}_{\theta(TM)})}{\gcd(2,k+1)}\right)
    =
    \exp\!\left( 2 \pi i\frac{\eta(\bar{\gamma}\slashed{D}_{\theta(TM + \xi + 1)})}{\gcd(2,k+1)} \right)
\end{equation}
Note that twisting by $\xi$ negates all bordism invariants \cite[Proposition 3.30]{freed2021}, but it is too early to cancel $\xi + 1$ to zero, because it is still inside a vector bundle operation.

Now we would like to evaluate this on various manifolds. Let us start with $\RP^{4k-2}$. The tangent bundle is $TM = (4k - 1)\xi - 1$, so the total bundle appearing above is
\begin{equation}
    TM + \xi + 1 = 4k \xi
\end{equation}
We need to apply $\theta$ to this. By definition \cite[Appendix B]{freed2021}, $\pi_t(2\xi) = 1 + 2t(\xi - 1)$. Thus $\pi_t(4k\xi) = (1 + 2t(\xi-1))^{2k}$. As far as the eta invariant is concerned, $\xi = -1$, and we are free to make this substitution at this point because we are done with the $\KO$-theory operations. So we can replace $\pi_t(4k\xi) \rightarrow (1 - 4t)^{2k}$. Thus $\pi^r(4k\xi) \rightarrow \binom{2k}{r} (-4)^r$, and
\begin{equation}
    q\!\left(\frac{\Phi_{4k}(i\slashed{D}_{\theta})}{\gcd(2,k+1)}\right)(\RP^{4k-2})
    =
    \exp\!\left( \frac{2 \pi i}{2^{2k+(k \text{ mod } 2)}} \theta(\pi^r = \tbinom{2k}{r} (-4)^r) \right)
\end{equation}
As a consistency check, we should get $1$ when we put $\theta = \pi^r$ with $r > k$, because then $\Phi_{4k}(i\slashed{D}_\theta) = 0$ for degree reasons. This is indeed the case, because the factor of $(-4)^r$ swamps the denominator.

Finally, we assert that when one simultaneously turns on a twisting bundle $\theta$ and evaluates on a product manifold (which we have only done separately so far), the appropriate generalisation is
\begin{multline} \label{eq:omfgbleeurgh}
    q\!\left(\frac{\Phi_{4k}(i\slashed{D}_{\theta})}{\gcd(2,k+1)}\right)(M_J \times \RP^{4l-2}) \\
    =
    \exp\!\left[ \frac{2 \pi i}{2^{2l+(k \text{ mod } 2)}} \int_{M_J} \Phi_{4(k-l)}\bigg(i\slashed{D}_{\theta\left(\pi^r = \sum_{a+b=r} \pi^a(TM_J) \tbinom{2l}{b} (-4)^b\right) }\bigg)\right]
\end{multline}

\subsection{Putting things together} \label{sec:result}

In this section we combine the previous results to give a computation of $H_B^{4k}(\pt)$, in every dimension. The computation is essentially rigorous; in particular, as we shall see there is no issue from some of the pin bordism generators not having explicit representatives. We defer some of the more laborious proofs to \cref{app:proofs}.

The key remaining insight that we need is that only anomaly polynomials with \emph{even coefficients} end up mattering when we compute the bosonisation cohomology. This implies that we do not need to evaluate eta invariants on manifolds in $\pin^-$ bordism that have order 2, at a stroke removing all ambiguities associated with mystery manifolds in the tables in \cref{sec:pin}. This in turn means that the tricks we described in \cref{sec:q}, that let us evaluate eta invariants on product manifolds of the form $\RP^{4l-2}\times M_{\Spin}$, are all we need to compute the kernel of $q$ and thence $H_B^\bullet(\pt)$.

So, we now show why only the ``even'' anomaly polynomials appear in $\ker(q)/\im(p)$.
Fix a $k$, and let $A = \Hom(\Omega_{4k}^\Spin(\pt), \bZ) \cong \bZ^{p(k)}$ be the set of fermionic anomaly polynomials in $4k$ dimensions. We will show that
\begin{equation} \label{eq:kerq_even}
    \ker(q) \subseteq
    \begin{cases}
        2A & k \text{ odd} \\
        2A + \im(p) & k \text{ even}
    \end{cases}
\end{equation}
The argument for this is very different depending on whether $k$ is odd or even, so we split our discussion into these two cases.

\subsubsection*{The case $\bm{k}$ is odd}

In this case, we note from the classification of spin bordism that there exist $p(k)$ spin manifolds of dimension $4k - 2$, denoted $\{ N_J : n(J) \leq k, 1 \notin J \}$, such that
\begin{equation}
    \text{mod-2-index}(i\slashed{D}_{\pi^J}, N_{J'}) = \delta_{J, J'}
\end{equation}
These mod-2 indices generalise the familiar Arf invariant from 2 dimensions. One can straightforwardly prove that
\begin{equation}
    q\Big(\tfrac{1}{2}\Phi_{4k}(i\slashed{D}_{\pi^J})\Big) \, (N_{J'}) = (-1)^{\delta_{J, J'}}
\end{equation}
since the $\pin^-$ bordism invariant on the left hand side becomes the mod-2 index when restricted to a spin manifold. Thus if an integer-linear combination of the $\tfrac{1}{2}\Phi_{4k}(i\slashed{D}_{\pi^J})$ is to have a chance at being in the kernel of $q$, it must be an \emph{even}-integer-linear combination. Since the $\tfrac{1}{2}\Phi_{4k}(i\slashed{D}_{\pi^J})$ form a 2-local basis for $A$, this establishes
\begin{equation}
    \ker(q) \subseteq 2A
\end{equation}

\subsubsection*{The case $\bm{k}$ is even}

Suppose $k$ is even. Then the previous argument is no longer applicable, because a nontrivial mod-2-index no longer exists. Instead, we note that the $p(k) - p(k - 1)$ anomaly polynomials in \eqref{eq:HS-3}
\begin{equation}
    \{ \Phi_{4k}(\slashed{D}_{\pi^J}) : n(J) = k \text{ and } 1 \notin J \}
\end{equation}
are simply given by $p^J$ (the Pontryagin classes), which are trivially in the image of $p$, and therefore guaranteed to map to zero under $q$. Therefore, we only need to consider the remaining $p(k - 1)$ anomaly polynomials of the form
\begin{equation}
    \{ \Phi_{4k}(\slashed{D}_{\pi^J}) : n(J) \leq k - 1 \text{ and } 1 \notin J \}
\end{equation}
We claim that these are perfectly paired with the $p(k - 1)$ $\pin^-$ manifolds
\begin{equation} \label{eq:kevenperfectpairing}
    \{ M_J \times 2^{2l - 1 - (l \text{ mod } 2)} \, \RP^{4l - 2} : 1 \leq l \leq k,\, n(J) = k-l,\, 1 \notin J \}
\end{equation}
which all have order 2. This follows immediately from the Bahri--Gilkey formula \eqref{eq:omfgbleeurgh}, which in the case above reduces to
\begin{equation}
    q\Big(\Phi_{4k}(\slashed{D}_{\pi^{J'}})\Big) \, (M_J \times 2^{2l - 1 - (l \text{ mod } 2)} \, \RP^{4l - 2}) = (-1)^{\int_{M_J} \frac{\scriptstyle\Phi_{4n(J)}(\slashed{D}_{\pi^{J'}})}{\scriptstyle\gcd(2, n(J)+1)}} = (-1)^{\delta_{J,J'}}
\end{equation}
Since the $\Phi_{4k}(\slashed{D}_{\pi^J})$ form a 2-local basis of $A$, this establishes that
\begin{equation}
    \ker(q) \subseteq 2A + \im(p)
\end{equation}
and concludes our proof of~\eqref{eq:kerq_even}.

\subsubsection*{Evaluating the kernel of $\bm{q}$}

The simple result~\eqref{eq:kerq_even} is very powerful: it eliminates any ambiguity that might otherwise arise from the unknown or ``halved'' pin bordism generators in the calculation of $\ker(q)$. This is because our task is reduced to finding all $\Phi \in A$ such that
\begin{equation}
    \Big(q(2\Phi)\Big)(\Omega^{\Pin^-}_{4k-2}(\pt)) = 0
\end{equation}
But this is equivalent to the condition
\begin{equation}
    \Big(q(\Phi)\Big)(2\Omega^{\Pin^-}_{4k-2}(\pt)) = 0
\end{equation}
by linearity. 

Taking the set of $\pin^-$ bordism generators as collated in Tables~(\ref{tab:pin1} -- \ref{tab:pin3}), and multiplying by two, we can infer that a set of generators for $2\Omega^{\Pin^-}_{4k-2}(\pt)$ is given by
\begin{equation}
    \{ M_J \times 2^{(k+l+1) \text{ mod } 2} \, \RP^{4l - 2} : 1 \leq l \leq k,\, n(J) = k-l,\, 1 \notin J \}
\end{equation}
These are all product manifolds, so can be attacked with tools described in the previous subsection, in particular the Bahri--Gilkey factorisation formula. Doing so, from~\eqref{eq:omfgbleeurgh} we have
\begin{multline} \label{eq:qAeval}
    q \Big( \tfrac{1}{\gcd(2,k+1)} \Phi_{4k}(\slashed{D}_{\pi^J}) \Big) (M_{4(k-l)} \times \RP^{4l-2})
    = \\
    \exp\!\left( \frac{2 \pi i}{2^{2l+(k \text{ mod } 2)}} \int_{M_{4(k-l)}} \Phi_{4(k-l)}\big(i\slashed{D}_{\prod_{j \in J} \sum_{a+b=j} \pi^a \tbinom{2l}{b} (-4)^b }\big)\right)
\end{multline}
Recall the pre-factor in the exponential comes from evaluating an eta invariant on the $\RP^{4l-2}$ factor via orbifolding. Now, the subscript of the Dirac operator can be expanded as a polynomial in the $\pi^r$. To avoid getting too many $\pi^a$s, which would make the integral vanish by degree reasons, we have to take a certain number of $(-4)^b$s; specifically, the sum of all $b$ must be $\geq n(J)-(k-l)$. This reduces the denominator outside the integral to
\begin{equation}
    \frac{2^{2(n(J) - (k - l))}}{2^{2l + (k \text{ mod } 2)}} = \frac{1}{2^{2(k - n(J)) + (k \text{ mod } 2)}}
\end{equation}
Because the Dirac indices $\int_{M_{4(k-l)}} \Phi_{4(k-l)}(i\slashed{D}_\theta)$ appearing in \eqref{eq:qAeval} are even integers when $k+l$ is odd, and otherwise integers, this implies
\begin{equation} \label{eq:q_eta}
    q\Big(\tfrac{1}{\gcd(2,k+1)}\Phi_{4k}(\slashed{D}_{\pi^J})\Big)(2\Omega^{\Pin^-}_{4k-2}(\pt)) \subseteq \bZ_{2^{2(k-n(J))-(k+1 \text{ mod } 2)}} \subseteq U(1)
\end{equation}
In other words, the more complicated the anomaly polynomial (that is, the larger $n(J)$ is), the less values its eta invariant can take.

Next, we claim that
\begin{equation} \label{eq:qsurj}
    q : A \rightarrow \Hom(2\Omega^{\Pin^-}_{4k-2}(\pt), U(1))
\end{equation}
is surjective. This is straightforward, because we've already shown it! Or rather, we have in the $k$ even case, when we showed in \eqref{eq:kevenperfectpairing} that the anomaly polynomials with $n(J) \leq k - 1$ and $1 \notin J$ are perfectly paired with the order-2 subgroup of $2\Omega^{\Pin^-}_{4k-2}(\pt)$, which implies the claim. The same result is also true in the $k$ odd case. Now, as an abstract group, the right hand side of \eqref{eq:qsurj} is isomorphic to
\begin{equation}
    \Hom(2\Omega^{\Pin^-}_{4k-2}(\pt), U(1)) \cong \prod_{\substack{n(J) \leq k - 1 \\ 1 \notin J}} \bZ_{2^{2(k-n(J))-(k+1 \text{ mod } 2)}}
\end{equation}
The cyclic factors appearing here are the same as those in \eqref{eq:q_eta}. Together with the surjectivity result~\eqref{eq:qsurj}, this means that the set of eta invariants, namely those for the $\pi^J$ twisted bundles, effect the isomorphism in question. That is,
\begin{equation}
\begin{aligned}
    2\Omega^{\Pin^-}_{4k-2}(\pt) &\xrightarrow{\quad} \prod_{\substack{n(J) \leq k - 1 \\ 1 \notin J}} \bZ_{2^{2(k-n(J))-(k+1 \text{ mod } 2)}} \\
    M &\xmapsto{\quad} \prod_{\substack{n(J) \leq k - 1 \\ 1 \notin J}} q\Big(\tfrac{1}{\gcd(2,k+1)}\Phi_{4k}(i\slashed{D}_{\pi^J})\Big)(M)
\end{aligned}
\end{equation}
is the desired isomorphism.

The above result allows us to read off $\ker(q)$. We find that a 2-local basis for $\ker(q)$ is
\[
    \{ 2^{2(k-n(J))} \Phi_{4k}(i\slashed{D}_{\pi^J}) : n(J) \leq k \text{ and } 1 \notin J \}
\]
Meanwhile, recall from \eqref{eq:HS-3} that a 2-local basis for $\im(p)$ is
\[
    \{ 2^{4(k - n(J)) - \mathrm{bitcount}(k-n(J))} \Phi_{4k}(i\slashed{D}_{\pi^J}) : n(J) \leq k \text{ and } 1 \notin J \}
\]

\subsubsection*{Final result}

It is then straightforward to compute $H_B^{4k}(\pt) = \ker(q) / \im(p)$, where recall that the caveats ``2-locally'' above do not affect the answer.
\begin{shadebox}
\textbf{Theorem:} The bosonisation cohomology groups in dimension $4k$ are given by
\begin{equation}
    H_B^{4k}(\pt) = \prod_{\substack{n(J) \leq k - 1 \\ 1 \notin J}} \bZ_{2^{2(k - n(J)) - \mathrm{bitcount}(k-n(J))}}
\end{equation}
where the $J^\text{th}$ cyclic factor is generated by $2^{2(k-n(J))} \Phi_{4k}(i\slashed{D}_{\pi^J})$.
\end{shadebox}

\section{Examples} \label{sec:examples}

To illustrate the general results of \cref{sec:computing}, here we show how to unpack them into a less abstract form in the physically interesting dimensions $d = 2, 6, 10$. We will give explicit anomaly polynomials for the generators of $H_B^{d+2}(\pt)$, and discuss applications.

\subsection{$d=2$}

In degree 4, as pertains to anomalies of QFTs in 2 dimensions, the standard 2-local bases for the space of fermionic anomaly polynomials and its subspaces $\ker(q)$ and $\im(p)$ are
\begin{flalign} \label{eq:ex:d2}
\rlap{\{\text{2-locally}\footnotemark\}} &&
\begin{alignedat}{2}
    \Hom(\Omega_4^\Spin(\pt), \bZ) &= \text{span} \big\{& \tfrac{1}{2} \Phi_4(i\slashed{D}) \big\} \\
    \ker(q) &= \text{span} \big\{& 2^2 \Phi_4(i\slashed{D}) \big\} \\
    \im(p) &= \text{span} \big\{& 2^3 \Phi_4(i\slashed{D}) \big\}
\end{alignedat} &&
\end{flalign}
where $\frac{1}{2}\Phi_4(i\slashed{D}) = \frac{1}{2}\hat{A}_1(R) = -\frac{p_1}{48} = \frac{\Tr(R^2)}{384\pi^2}$ is the anomaly polynomial of a single Majorana--Weyl fermion. We can therefore read off:
\begin{shadebox}
The bosonisation cohomology group $H_B^4(\pt) = \bZ_2$, generated by $2^2 \Phi_4(i\slashed{D})$, which is the anomaly polynomial of 8 Majorana--Weyl fermions. 
\end{shadebox}
The fact that bosonisation ``backfires'' for such a fermion chain was spelled out in Ref.~\cite{bb}.

\footnotetext{The prefix `2-locally' reminds us that these statements are strictly only true after taking the 2-localisation, as in \cref{sec:p2loc}. We remind the reader that the resulting bosonisation cohomology group is, however, always independent of this 2-local caveat.}

To upgrade the 2-local to an actual basis, we use the Hattori--Stong theorem to get an overcomplete set
\begin{equation}
    \Hom(\Omega_4^\Spin(\pt), \bZ) = \text{span} \big\{ \tfrac{1}{2} \Phi_4(i\slashed{D}), \tfrac{1}{2} \Phi_4(i\slashed{D}_{\pi^1}) \big\}
\end{equation}
We need to refine this to a basis. Since $\Phi_4(i\slashed{D}_{\pi^1}) = \ph_1(\pi^1) = p_1 = -24\Phi_4(i\slashed{D})$, the second generator is redundant. Therefore our 2-local basis was already a basis, so we can actually just delete the caveat `2-locally' from \eqref{eq:ex:d2}.

For illustrative purposes, we also rederive $\ker(q)$ from scratch to demonstrate how various intermediate results from \cref{sec:computing} are used in the general proof. The general fermionic anomaly polynomial is $n \cdot \frac{1}{2} \Phi_4(i\slashed{D})$. We wish to determine when $q$ of this is zero. Applying $q$ and the Smith isomorphism, we get the 2d $\pin^-$ eta invariant $\exp( 2 \pi i n \cdot \frac{1}{2} \eta(\bar{\gamma}\slashed{D}))$. The generator of 2d $\pin^-$ bordism is $\RP^2$. By the orbifolding trick, their pairing is $\exp( 2 \pi i n / 8)$. Thus $\ker(q)$ corresponds to $n = 0$ mod 8, as claimed.

\subsection{$d=6$}

In degree 8, pertaining to anomalies in 6 dimensions, we know that there are $p(2)=2$ independent gravitational anomaly polynomials, corresponding to different linear combinations of the Pontryagin numbers $p_1^2$ and $p_2$.
Applying our general formalism, the standard 2-local bases are
\begin{flalign} \label{eq:ex:d6}
\rlap{\{\text{2-locally}\}} &&
\begin{alignedat}{4}
    \Hom(\Omega_8^\Spin(\pt), \bZ) &= \text{span} \big\{& \Phi_8(i\slashed{D}) ,\, &\Phi_8(i\slashed{D}_{\pi^2}) &\big\} \\
    \ker(q) &= \text{span} \big\{& 2^4 \Phi_8(i\slashed{D}) ,\, &\Phi_8(i\slashed{D}_{\pi^2}) &\big\} \\
    \im(p) &= \text{span} \big\{& 2^7 \Phi_8(i\slashed{D}) ,\, &\Phi_8(i\slashed{D}_{\pi^2}) &\big\}
\end{alignedat} &&
\end{flalign}
Note that this time, there are no $\frac{1}{2}$ factors in the first line, because there are no Majorana--Weyl fermions in 0 mod 8 dimensions. The second basis element cancels between $\ker(q)$ and $\im(p)$ and does not contribute to the bosonisation cohomology. We can therefore read off:
\begin{shadebox}
The bosonisation cohomology group $H_B^8(\pt) = \bZ_8$, generated by $2^4 \Phi_8(i\slashed{D})$, the anomaly polynomial of 16 Weyl fermions.
\end{shadebox}
This is a new result. It says that, while $(-1)^F$ is anomaly-free for 16 Weyl fermions in 6d, we need 128 such Weyls in order for the theory to be properly bosonisable.

Let us give a few more details relevant to this example. Firstly, to upgrade the 2-local basis to an actual basis, by Hattori--Stong we need to find a basis refinement of
\begin{equation}
    \Hom(\Omega_8^\Spin(\pt), \bZ) = \text{span} \big\{ \Phi_8(i\slashed{D}_\theta) \, : \, \theta = 1, \pi^1, (\pi^1)^2, \pi^2 \big\}
\end{equation}
Explicit calculations give $\Phi_8(i\slashed{D})$ and $\Phi_8(i\slashed{D}_{\pi^1})$ as generators. The relation between the 2-local basis and the actual basis is
\begin{equation}
    \begin{pmatrix*}[l]
        \Phi_8(i\slashed{D}) \\
        \Phi_8(i\slashed{D}_{\pi^2})
    \end{pmatrix*}
    =
    \begin{pmatrix}
        1 & 0 \\
        240 & -7
    \end{pmatrix}
    \begin{pmatrix*}[l]
        \Phi_8(i\slashed{D}) \\
        \Phi_8(i\slashed{D}_{\pi^1})
    \end{pmatrix*}
\end{equation}
where the change of basis matrix is invertible 2-locally (has odd determinant), as expected. We now use the following procedure: find a unimodular matrix $U$ such that $\begin{psmallmatrix} 128 & 0 \\ 0 & 1 \end{psmallmatrix} \begin{psmallmatrix} 1 & 0 \\ 240 & -7 \end{psmallmatrix} U^{-1} \begin{psmallmatrix} 128 & 0 \\ 0 & 1 \end{psmallmatrix}^{-1}$ is integral, for which a choice is $U = \begin{psmallmatrix} 1 & 0 \\ -16 & 1 \end{psmallmatrix}$, then transform \eqref{eq:ex:d6} by the substitution
\begin{equation}
    \begin{pmatrix*}[l]
        \Phi_8(i\slashed{D}) \\
        \Phi_8(i\slashed{D}_{\pi^2})
    \end{pmatrix*}
    \rightarrow
    \begin{pmatrix} 1 & 0 \\ -16 & 1 \end{pmatrix}
    \begin{pmatrix*}[l]
        \Phi_8(i\slashed{D}) \\
        \Phi_8(i\slashed{D}_{\pi^1})
    \end{pmatrix*}
\end{equation}
Noting that $\Phi_8(i\slashed{D}_{16 - \pi^1}) = L_2$, the Hirzebruch $L$-genus, this turns \eqref{eq:ex:d6} into the basis
\begin{equation} \label{eq:ex:d6proper}
\begin{alignedat}{2}
    \Hom(\Omega_8^\Spin(\pt), \bZ) &= \text{span} \big\{& \hat{A}_2 ,\, L_2 \big\} \\
    \ker(q) &= \text{span} \big\{& 2^4 \hat{A}_2 ,\, L_2 \big\} \\
    \im(p) &= \text{span} \big\{& 2^7 \hat{A}_2 ,\, L_2 \big\}
\end{alignedat}
\end{equation}
where 
\begin{equation}
    \hat{A}_2(TM) = \frac{7p_1^2 - 4p_2}{5760} 
    \qquad \text{and} \qquad
    L_2 = \frac{7p_2 - p_1^2}{45}
\end{equation}
The latter is manifestly bosonic, being the anomaly polynomial of the signature operator which does not require a spin structure.

We briefly sketch how the computation of $\ker(q)$ goes, following \cref{sec:computing}. The map $p$ already hits $L_2$, so we only need to consider $\hat{A}_2(R)$. The generator of 6d $\pin^-$ bordism is $\RP^6$, and its pairing with $q(\hat{A}_2(R))$ is $\exp(2 \pi i / 16)$ by the orbifolding trick. Thus, $\ker(q)$ is generated by $L_2$ and $16\hat{A}_2(R)$, as claimed.

\subsection{$d=10$ and supergravity} \label{sec:10d}

Moving up a further four dimensions, in degree 12 the standard 2-local bases are
\begin{flalign} \label{eq:ex:d10}
\rlap{\{\text{2-locally}\}} &&
\begin{alignedat}{7}
    \Hom(\Omega_{12}^\Spin(\pt), \bZ) &= \text{span} \big\{& \tfrac{1}{2} \Phi_{12}(i\slashed{D}) &,\,& \tfrac{1}{2} \Phi_{12}(i\slashed{D}_{\pi^2}) &,\,& \tfrac{1}{2} \Phi_{12}(i\slashed{D}_{\pi^3}) &\big\} \\
    \ker(q) &= \text{span} \big\{& 2^6 \Phi_{12}(i\slashed{D}) &,\,& 2^2 \Phi_{12}(i\slashed{D}_{\pi^2}) &,\,& \Phi_{12}(i\slashed{D}_{\pi^3}) &\big\} \\
    \im(p) &= \text{span} \big\{& 2^{10} \Phi_{12}(i\slashed{D}) &,\,& 2^3 \Phi_{12}(i\slashed{D}_{\pi^2}) &,\,& \Phi_{12}(i\slashed{D}_{\pi^3}) &\big\}
\end{alignedat} &&
\end{flalign}
We can therefore read off: 
\begin{shadebox}
The bosonisation cohomology group $H_B^{12}(\pt) = \bZ_{16} \times \bZ_2$, where:
\begin{itemize}
    \item the $\bZ_{16}$ factor is generated by $2^6 \Phi_{12}(i\slashed{D})$, the anomaly polynomial of 128 Majorana--Weyl fermions
    \item the $\bZ_2$ is generated by $2^2 \Phi_{12}(i\slashed{D}_{\pi^2})$, which is related to the anomaly of a gravitino.
\end{itemize}
\end{shadebox}
The precise relation of the second generator to gravitino anomalies shall be elucidated shortly.

Before doing so, let us unpack the customary details regarding basis upgrade. To wit, to upgrade the 2-local basis to an actual basis, by Hattori--Stong we need to find a basis refinement of
\begin{equation}
    \Hom(\Omega_{12}^\Spin(\pt), \bZ) = \text{span} \big\{ \tfrac{1}{2} \Phi_{12}(i\slashed{D}_\theta) \, : \, \theta = 1, \pi^1, (\pi^1)^2, (\pi^1)^3, \pi^2, \pi^2 \pi^1, \pi^3 \big\}
\end{equation}
Explicit calculations give $\theta \in \{1, \pi^1, \pi^3\}$ as generators. The relation between the 2-local basis and the actual basis is
\begin{equation}
    \begin{pmatrix*}[l]
        \Phi_{12}(i\slashed{D}) \\
        \Phi_{12}(i\slashed{D}_{\pi^2}) \\
        \Phi_{12}(i\slashed{D}_{\pi^3})
    \end{pmatrix*}
    =
    \begin{pmatrix}
        1 & 0 & 0 \\
        -504 & -31 & 0 \\
        0 & 0 & 1
    \end{pmatrix}
    \begin{pmatrix*}[l]
        \Phi_{12}(i\slashed{D}) \\
        \Phi_{12}(i\slashed{D}_{\pi^1}) \\
        \Phi_{12}(i\slashed{D}_{\pi^3})
    \end{pmatrix*}
\end{equation}
We now run our procedure from $d=6$, finding $\begin{psmallmatrix*}[l] 1024 & 0 & 0 \\ 0 & 8 & 0 \\ 0 & 0 & 1 \end{psmallmatrix*} \begin{psmallmatrix} 1 & 0 & 0 \\ -504 & -31 & 0 \\ 0 & 0 & 1 \end{psmallmatrix} \begin{psmallmatrix} 1 & 0 & 0 \\ 8 & 1 & 0 \\ 0 & 0 & 1 \end{psmallmatrix}^{-1} \begin{psmallmatrix*}[l] 1024 & 0 & 0 \\ 0 & 8 & 0 \\ 0 & 0 & 1 \end{psmallmatrix*}^{-1}$ is integral, and therefore that we should make the substitution $\Phi_{12}(i\slashed{D}_{\pi^2}) \rightarrow \Phi_{12}(i\slashed{D}_{\pi^1 + 8})$ in \eqref{eq:ex:d10} to turn it into a genuine basis. Hence
\begin{equation} \label{eq:ex:d10proper}
\begin{alignedat}{7}
    \Hom(\Omega_{12}^\Spin(\pt), \bZ) &= \text{span} \big\{& \tfrac{1}{2} \hat{A}_3 &,\;& 2^{-4} L_3 &,\;& \tfrac{1}{2} p_3 &\big\} \\
    \ker(q) &= \text{span} \big\{& 2^6 \hat{A}_3 &,\;& 2^{-1} L_3 &,\;& p_3 &\big\} \\
    \im(p) &= \text{span} \big\{& 2^{10} \hat{A}_3 &,\;& L_3 &,\;& p_3 &\big\}
\end{alignedat}
\end{equation}
using $\Phi_{12}(i\slashed{D}_{\pi^1 + 8}) = 2^{-3} L_3$ and $\Phi_{12}(i\slashed{D}_{\pi^3}) = p_3$.

\subsection{Type IIB anomaly cancellation}

Recall that the fields of Type IIB supergravity which carry perturbative gravitational anomalies are
\begin{itemize}
    \item The dilatino: a positive-chirality complex fermion $\psi$.
    \item The gravitino: a negative-chirality complex fermion $\psi_\mu$.
    \item The self-dual field: a naïvely bosonic 4-form field $C_4$ with self-dual field strength.
\end{itemize}
Consistency of Type IIB string theory requires their anomalies to cancel. Let us review how this works in our notation. The anomaly polynomial of the dilatino is simply $\Phi(\psi) = \Phi_{12}(i\slashed{D})$. The anomaly polynomial of the gravitino is $\Phi(\psi_\mu) = -\Phi_{12}(i\slashed{D}_{TM-3})$, with the sign being due to the negative chirality, and we have subtracted 1 from $TM$ to remove the gauge degrees of freedom, again to pass from the 10d physical theory to the 11d anomaly theory, and again to pass to the 12d anomaly polynomial. Therefore, by anomaly cancellation, the anomaly polynomial of the self-dual field must be
\begin{equation} \label{eq:c4}
    \Phi(C_4) = \Phi_{12}(i\slashed{D}_{TM-4}) = \Phi_{12}(i\slashed{D}_{\pi^1+8}) = 2^{-3} L_3
\end{equation}
using $\pi^1(TM) = TM - 12$. This can also be calculated directly \cite{Alvarez-Gaume:1983ihn, Hsieh:2020jpj}.

Now let us compare this result to \eqref{eq:ex:d10proper}. We see immediately that $C_4$ has a fermion parity anomaly of $(0, 2, 0)$ mod $(128, 8, 2)$. This is strange, as the field $C_4$ is supposedly bosonic, being a 4-form gauge field. Furthermore, even after taking 4 copies of $C_4$ to cancel this anomaly in $(-1)^F$, we still cannot bosonise: there remains a spin structure anomaly of $(0, 1)$ mod $(16, 2) \in H_B^{12}(\pt)$. We would have to take 8 copies of $C_4$ to cancel this anomaly and obtain a truly bosonic object. So, $C_4$ considered on its own is very much a fermionic theory.

Of course, the correct explanation of this puzzle involves a precise formulation of the self-dual field $C_4$, and is known in the literature~\cite{Hsieh:2020jpj} (see also~\cite{Debray:2021vob}). First of all, $C_4$ is not a 4-form gauge field, which would be valued in ordinary differential cohomology $\check{H}^5(M_{10})$; it rather takes values in the differential $K$-theory group $\check{K}^{-1}(M_{10})$. Second, $C_4$ is not a purely 10-dimensional field, but the boundary mode of an 11-dimensional field $\check{A} \in \check{K}^0(M_{11})$. This 11-dimensional bulk is responsible for the gravitational anomaly. It is also fermionic. To see why, we need to know that the action for $\check{A}$ is $e^{2 \pi i \cQ(\check{A})}$ where $\cQ$ is a quadratic refinement of the differential $K$-theory product. To construct $\cQ$ requires a choice of spin structure; this is what introduces hidden fermionic dependence into $C_4$. It is interesting that a hint of these subtleties can already be seen at the level of the perturbative anomaly.

The 2 mod 8 fermion parity anomaly of $C_4$ in many ways mirrors the same anomaly of a 2d Weyl fermion. At some level, this is not surprising: a 2d Weyl fermion is equivalent to a chiral boson,\footnote{This naming, while commonplace, leads to the unfortunate duality chiral fermion $\equiv$ chiral boson. That is, they are not dual under bosonisation/fermionisation, which is outlawed anyway by an anomaly; they are simply the same.} which is a 0-form field $C_0$ with self-dual field strength much like $C_4$. As a further connection, if $M_3 \in \widetilde{\Omega}_3^\Spin(\B\bZ_2)$ is any manifold that detects the 2d anomaly, and $M_8$ is any spin manifold with odd signature, then $M_3 \times M_8$ detects the 10d anomaly. For $M_3$, we could take $\RP^3$, or the mapping torus constructed in \cite[Equation 4.9]{anominter}; for $M_8$ we could take $\HP^2$ with $\sigma(\HP^2) = 1$. This makes it clear that the 10d anomaly is also captured by a mapping torus, \emph{i.e.}\ it is a ``traditional'' global anomaly.

\subsection{A geometric dual}

To conclude this section, we briefly comment on how the groups $H_B^{4k}(\pt)$ can be given a more geometric flavour. It is dual to the cobordism perspective used in the rest of the paper, in the sense that here we consider maps between \emph{manifolds} and bordism classes thereof.

\begin{figure}
    \centering
    \begin{tikzpicture}
        \filldraw[black, fill=cyan!30, fill opacity=1.0]
        (0,0) arc[start angle=45, delta angle=-270, x radius=2, y radius=0.3] coordinate(A)
        arc[start angle=270, delta angle=180, x radius=0.3, y radius=1]
        node[fill opacity=1, anchor=south] {$(M, \rho)$}
        arc[start angle=90, delta angle=90, x radius=1.2, y radius=2]
        arc[start angle=180, delta angle=45, radius=1] coordinate(B)
        coordinate(C) at ($(A)-(B)$)
        .. controls+(0.7,-0.7) and +(-0.7,-0.7) .. (C|-B)
        arc[start angle=-45, delta angle=45, radius=1]
        arc[start angle=0, delta angle=90, x radius=1.2, y radius=2]
        node[fill opacity=1, anchor=south] {$(M, \rho+A)$}
        arc[start angle=90, delta angle=180, x radius=0.3, y radius=1] -- cycle
        ($(A)!0.5!(0,0)-(45:0.6)-(0,0.45)$) coordinate(D) arc[start angle=135, delta angle=-90, x radius=0.6, y radius=0.3] coordinate(E)
        arc[start angle=-45, delta angle=-90, x radius=0.6, y radius=0.3] -- cycle;
        \node[anchor=north] at (0,-0.5) {$N$};
        \draw (D) arc[start angle=225, delta angle=-15, x radius=1, y radius=0.5];
        \draw (E) arc[start angle=-45, delta angle=15, x radius=1, y radius=0.5];
        \fill[red, opacity=0.1]        
        (A) -- (0,0) arc[start angle=270, delta angle=180, x radius=0.3, y radius=1]
        -- +(A)
        arc[start angle=90, delta angle=180, x radius=0.3, y radius=1] -- cycle;
        \draw (0,0) arc[start angle=270, delta angle=180, x radius=0.3, y radius=1];
        \draw (A) arc[start angle=270, delta angle=-180, x radius=0.3, y radius=1];
    \end{tikzpicture}
    \caption{The construction of the manifold $Y$ from $(M, \rho, A)$.}
    \label{fig:ymanifold}
\end{figure}
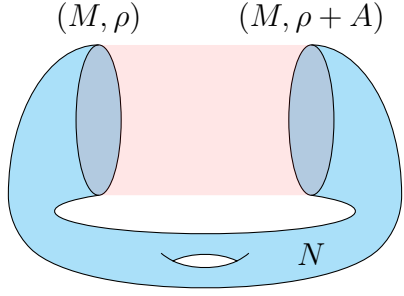

Recall that a theory $\cT_F$ is free of a fermion parity anomaly if and only if its anomaly theory obeys $\frac{\cZ[M, \rho + A]}{\cZ[M, \rho]} = 1$ for every tuple $(M, \rho, A)$. In the special case of $d = 4k-2$ dimensions, this can be translated into a condition purely on the anomaly polynomial $\Phi$. To see how, we invoke the result of Anderson, Brown and Peterson \cite{anderson1966spin} that, as spin manifolds, there is a null-bordism
\begin{equation}
    (M, \rho + A) - (M, \rho) = \partial N
\end{equation}
when $M$ has dimension $4k-1$.\footnote{ABP showed that a complete set of $n$-dimensional spin bordism invariants are the $\KO$-characteristic numbers (valued in $\KO^{-n}(\pt)$) and the Stiefel--Whitney numbers. Since $\KO^{-(4k-1)}(\pt) = 0$, and the Stiefel--Whitney numbers are independent of spin structure, this proves the claim.} So far, all the manifolds and bordisms involved have respected the spin structure. Now glue the two boundaries of $N$ together to form a closed manifold $Y$, as in \cref{fig:ymanifold}. This process does not respect the spin structure, so $Y$ is only oriented, not spin. (In some sense, $Y$ is a generalised mapping torus for a shift in the spin structure.) It follows that
\begin{equation}
    \frac{\cZ[M, \rho + A]}{\cZ[M, \rho]} = \exp\!\big( 2 \pi i \int_Y \Phi \big)
\end{equation}
So $\cT_F$ is free of a fermion parity anomaly if and only if $\int_Y \Phi \in \bZ$ for all manifolds $Y$ one can construct in this way. By contrast, $\cT_F$ is free of a spin structure anomaly if and only if $\int_Y \Phi \in \bZ$ on \emph{all} oriented manifolds. Thus the bosonisation cohomology group $H_B^{4k}(\pt)$ measures the failure of $Y$ to generate all oriented manifolds.

We can make this precise by formalising the construction of $Y$ as a map
\begin{align*}
    \widetilde{\Omega}_{4k-1}^\Spin(\B\bZ_2) \;&\overset{f}{\longrightarrow}\; \frac{\Omega_{4k}^\SO(\pt)}{\Omega_{4k}^\Spin(\pt)} \\
    (M, \rho, A) \;&\longrightarrow\; Y
\end{align*}
Here the right hand side is all oriented bordism classes modulo those with a spin representative. Then $H_B^{4k}(\pt) = \coker(f)$, a measure of the failure of surjectivity of $f$. For example, in low dimensions, the explicit form of this map is as follows
\begin{equation}
\begin{alignedat}{7}
d=2: \qquad& \bZ_8& &\{\RP^3\} \quad&\overset{\times 2}{\xrightarrow{\hspace{5ex}}}&\quad \bZ_{16}& &\{\CP^2\} \\
    d=6: \qquad& \bZ_{16}& &\{\RP^7\} \quad&\overset{\times 8 \cdot 11}{\xrightarrow{\hspace{5ex}}}&\quad \bZ_{128}& &\{\CP^4\}
\end{alignedat}
\end{equation}
using $\frac{1}{2}\hat{A}_1(\CP^2) = \frac{1}{16}$ and $\hat{A}_2(\CP^4) = \frac{3}{128}$.

\section{Discussion and Outlook} \label{sec:discussion}

To finish, we list some open questions from our work, including comments on the SymTFT interpretation.

First of all, we strongly suspect that it is possible to give a shorter, more abstract calculation of $H_B^n(\pt)$ by computing the action of $p$ and $q$ on the cohomology of the spectra and deducing the action on cobordism. However we lack the topological chops to do so. Although one can use the Adams spectral sequence to deduce the action on bordism, one cannot use the split exact sequence \eqref{eq:UCT} to turn this into an action on cobordism, because the splitting is not natural. In any case, we would still prefer our eta invariant based calculation for its clear connection to the physics of gapped fermion SPT phases.

\subsection*{Symmetry TFT}

Second, all the features of $H_B^{d+2}(\pt)$ we have calculated should be encoded in a $(d+1)$-dimensional TQFT called the \emph{SymTFT}. Conversely, $H_B^{d+2}(\pt)$ can be used to glean a small amount of information about the SymTFT.

For $d$-dimensional fermionic theories without a gravitational anomaly, the SymTFT is known to take the form \cite{Gaiotto:2015zta, Cappelli:2025ecm}
\begin{equation} \label{eq:tft}
    \cZ_\text{SymTFT}[M_{d+1}] =
    \sum_{\substack{
        a_1 \in C^1(M_{d+1}; \bZ_2) \\
        a_{d-1} \in C^{d-1}(M_{d+1}; \bZ_2)
    }}
    (-1)^{
    \int_{M_{d+1}} (\delta a_1 + w_2) \smile a_{d-1}
    }
\end{equation}
The most important feature for us is the shift of $\delta a_1$ by (a cocycle representative of) the second Stiefel--Whitney class $w_2$. If not for this, \eqref{eq:tft} would be an ordinary $\bZ_2$ gauge theory written in BF form.

Now let us ask: what boundary conditions does \eqref{eq:tft} have? The easiest way to answer this question is to integrate out fields. Integrating out $a_1$ renders $\delta a_{d-1} = 0$, leaving us with a twisted $\bZ_2^{[d-2]}$ gauge theory. Alternatively, integrating out $a_{d-1}$ sets $\delta a_1 = w_2$ and so $a_1$ is a spin structure, which we duly rename $\rho$, and we find an untwisted `spin structure gauge theory'. Thus we have the equivalences
\begin{equation} \label{eq:tft2}
    \eqref{eq:tft}
    \qquad \leftrightarrow \qquad
    \sum_{a_{d-1} \in H^{d-1}(M_{d+1}; \bZ_2)} (-1)^{\int_{M_{d+1}} w_2 \smile a_{d-1}}
    \qquad \leftrightarrow \qquad
    \sum_\rho 1
\end{equation}
up to normalisation factors. It is then clear that we have a Dirichlet boundary condition $\ket{A_{d-1}}$ for $a_{d-1}$, and a Dirichlet boundary condition $\ket{\rho}$ for $\rho$.

By colliding these states with a dynamical boundary condition $\ket{\cZ}$, we obtain the two sets of partition functions
\begin{equation}
    \cZ_F[\rho] = \braket{\rho|\cZ}
    \qquad \xrightleftharpoons[\;\text{fer}\;]{\text{bos}} \qquad
    \cZ_B[A_{d-1}] = \braket{A_{d-1}|\cZ}
\end{equation}
which can be converted into each other via the change of basis matrix between the states $\ket{\rho}$ and $\ket{A_{d-1}}$. This is the statement of $d$-dimensional bosonisation and fermionisation. (Note the bosonic theory has a $\bZ_2^{[d-2]}$ symmetry with anomaly $(-1)^{\int w_2 \smile A_{d-1}}$.)

\subsection*{Adding anomalies} 
It is simple to generalise \eqref{eq:tft2} to anomalous theories. Suppose we consider fermionic theories with an anomaly specified by $\cZ_\text{AnomSPT} \in \mho_\Spin^{d+2}(\pt)$. Then simply replace the trivial action for $\rho$ in \eqref{eq:tft2} with $\cZ_\text{AnomSPT}$. This yields a bosonic TQFT
\begin{equation} \label{eq:tft3}
    \cZ_\text{SymTFT}[M_{d+1}] = \sum_\rho \cZ_\text{AnomSPT}[M_{d+1}, \rho]
\end{equation}
(We have suppressed an additional metric dependence for simplicity.) Such gauged SPTs are often called \emph{Dijkgraaf--Witten} theories, whether or not they actually follow the original construction in \cite{Dijkgraaf:1989pz}.

Of primary interest to us is the fate of the boundary conditions $\ket{\rho}$ and $\ket{A_{d-2}}$. Clearly, $\ket{\rho}$ survives for any value of the anomaly. But what about $\ket{A_{d-2}}$? In general, we expect this boundary condition to persist if and only if $\cZ_\text{AnomSPT} \in \im(p)$, or when $\cT_F$ is bosonisable. It would be very interesting to derive this directly from the TQFT by constructing all possible boundary conditions.

At specific, other values of the anomaly, namely the nontrivial classes in $H_B^{d+2}(\pt)$, we expect to see more exotic behaviour. For example when $d=2$, at generic values of the anomaly, there is only one boundary condition $\ket{\rho}$. But at the nontrivial class in $H_B^4(\pt) \cong \bZ_2$, one finds the emergence of two new boundary conditions $\ket{\rho}_S$ and $\ket{\rho}_C$.

One can ask similar questions at the level of bulk topological operators. For every $d$-dimensional topological term $\cZ_\text{stack} \in \Hom(\Omega^\Spin_d(\pt), U(1))$ there is a stacking operation $\cT_F \rightarrow \cZ_\text{stack} \times \cT_F$. This corresponds to an invertible topological operator $U$ of the SymTFT with the action on states
\begin{equation}
    U \ket{\rho} = \cZ_\text{stack}[\rho] \ket{\rho}
\end{equation}
although nontrivial operators only exist in $d = 1, 2$ mod 8, being counted by \eqref{eq:countingfn3}. These are the operators present for any value of the anomaly.

However, just as for boundary conditions, the spectrum of operators is typically enriched at the various classes in $H_B^{d+2}(\pt)$. For example, in $d=2$, at the class $0 \in \bZ_2$ there is a $D(\bZ_2)$ non-invertible symmetry, while at the class $1 \in \bZ_2$ there is a $\text{Vec}(\mathbb{S}_3)$ grouplike symmetry. Such symmetries indicate the presence of further boundary conditions and dualities. Again, it would be interesting to explore these operators in the general TQFT \eqref{eq:tft3}.

\subsection*{Other symmetry types}

The curious reader might notice that, while we formulated bosonisation cohomology $H_B^\bullet(X)$ for theories with maps to some internal space $X$ in \cref{sec:smashX}, we only ever computed it for the case $X=\pt$.
We found that $H_B^n(\pt)$ is always a finite abelian group, and moreover is 2-torsion. 
As a purely mathematical question, one can ask if the same is true of $H_B^n(X)$. Is it always 2-torsion, or even finite?

A well-motivated stack of examples to consider is those where $X=\B G$ is the classifying space of some symmetry group. As a particular example, we expect that the observations made in \cref{sec:10d} regarding anomalies in supergravity, and the connection to quadratic refinements, should have an analogous story for 6d theories with discrete internal symmetries~\cite{Dierigl:2025rfn}. It would thus be interesting to compute $H_B^8(B\bZ_k)$ to see how the bosonisation cohomology class is trivialised in these theories.

Finally, we comment that in addition to imposing internal symmetries as in \cref{sec:smashX}, we can also impose discrete spacetime symmetries such as time-reversal. This leads to a variation of the sequence \eqref{eq:int:seq}
\begin{center}
\begin{tikzcd}
    \mho^n_\Oo(\pt) \arrow[r] &
    \mho^n_{\Pin^\pm}(\pt) \arrow[r] &
    \widetilde{\mho}^n_{\Pin^\pm}(\B\bZ_2)
\end{tikzcd}
\end{center}
which measures whether spin structure summation differs from gauging $(-1)^F$ in the presence of a time-reversal symmetry. Unlike \eqref{eq:int:seq}, this sequence can have interesting cohomology in degrees other than $d = 2$ mod 4. We leave a more complete investigation to future work.

\section*{Acknowledgements}

The authors thank Markus Dierigl, Pavel Putrov, and Yunqin Zheng for many useful discussions, and are grateful to Yuji Tachikawa for helpful comments on the manuscript. 
PBS is supported by the ERC-COG grant NP-QFT No.~864583, by the MUR-FARE2020 grant No.~R20E8NR3HX, and for part of this work also by WPI Initiative, MEXT, Japan at Kavli IPMU, the University of Tokyo.

\appendix

\section{Generating Functions for Bordism Groups} \label{app:gen}

We list various generating functions which dictate the structure of oriented, spin, and $\pin^-$ cobordism, hence the structure of the cobordism sequence \eqref{eq:int:seq}. These generating functions are implemented in the \texttt{Mathematica} notebook \texttt{counting-fns.nb}, uploaded as ancillary material with the \texttt{arXiv} version of this manuscript.

The mod-2 cohomology of a spectrum forms a module over the Steenrod algebra $\cA$. For various spectra of interest, the $\cA$-module structure and Hilbert--Poincaré series are
\begin{center}
\begin{tabular}{l|l|l}
Spectrum & $\cA$-module & Hilbert--Poincaré series \\
\hline
$M\SO$ & & $\prod_{n \geq 2} (1-t^n)^{-1}$ \\
$M\Spin$ & & $\prod_{n \geq 4 ,\, n \neq 2^r+1} (1-t^n)^{-1}$ \\
$H\bZ_2$ & $\cA$ & $\prod_{n = 2^r-1 ,\, r \geq 1} (1-t^n)^{-1}$ \\
$H\bZ$ & $\cA/\cA\Sq^1$ & $\prod_{n = 2^r-1 ,\, r \geq 2} (1-t^n)^{-1} (1-t^2)^{-1}$ \\
$\ko$ & $\cA/\cA(\Sq^1,\Sq^2)$ & $\prod_{n = 2^r-1 ,\, r \geq 3} (1-t^n)^{-1} (1-t^4)^{-1} (1-t^6)^{-1}$ \\
$\ko\langle2\rangle$ & $\cA/\cA\Sq^3$ & $\prod_{n = 2^r-1 ,\, r \geq 3} (1-t^n)^{-1} (1-t^4)^{-1} (1-t^6)^{-1} (1+t+t^2+t^3+t^4)$
\end{tabular}
\medskip
\end{center}
It is known that $M\SO$ has the 2-local splitting
\begin{equation}
    M\SO = A(\Sigma^4) H\bZ + B(\Sigma) H\bZ_2
\end{equation}
where $A(t)$ and $B(t)$ are polynomials, and recall that $\Sigma$ is the suspension operator. In order to reproduce the structure of perturbative gravitational anomalies, we must have
\begin{equation}
    A(t^4) = \prod_{n \geq 1} (1-t^{4n})^{-1}
\end{equation}
Then by comparing Hilbert--Poincaré series, the polynomial $B(t)$ takes the value
\begin{equation}
    B(t) = \frac{P_{M\SO}(t) - A(t^4) P_{H\bZ}(t)}{P_{H\bZ_2}(t)} = t^5 + 2t^9 + t^{10} + \dots
\end{equation}
Similarly, $M\Spin$ has the 2-local splitting
\begin{equation}
    M\Spin = C(\Sigma^8) \ko + D(\Sigma^8) \ko\langle2\rangle + E(\Sigma) H\bZ_2
\end{equation}
for polynomials $C(t)$, $D(t)$, and $E(t)$.\footnote{The physical significance of this splitting is that the first part (involving $\ko$ and $\ko\langle2\rangle$) encodes free fermion anomalies, while the last part (involving $H\bZ_2$) encodes beyond-free-fermion anomalies.} In order to match perturbative gravitational anomalies, we must have
\begin{equation}
    C(t^8) (1-t^4)^{-1} + D(t^8) t^4 (1-t^4)^{-1} = \prod_{n \geq 1} (1-t^{4n})^{-1}
\end{equation}
which uniquely determines $C(t)$ and $D(t)$. Then by matching Hilbert--Poincaré series, the polynomial $E(t)$ takes the value
\begin{equation}
    E(t) = \frac{P_{M\Spin}(t) - C(t^8) P_{\ko}(t) - D(t^8) t^2 P_{\ko\langle2\rangle}(t)}{P_{H\bZ_2}(t)} = t^{20} + t^{22} + 2t^{28} + \dots
\end{equation}
Since the bordism groups $\Omega_n^\SO(\pt)$ and $\Omega_n^\Spin(\pt)$ are known to have no odd torsion, the above results allow us to read off both of them: they are
\begin{align}
    \sum_{n \geq 0} t^n \Omega_n^\SO(\pt) &= A(t^4) \bZ + B(t) \bZ_2 \label{eq:app_Om_SO} \\
    \sum_{n \geq 0} t^n \Omega_n^\Spin(\pt) &= A(t^4) \bZ + F(t) \bZ_2 \label{eq:app_Om_Spin}
\end{align}
where
\begin{equation}
    F(t) = C(t^8) (t + t^2) (1-t^8)^{-1} + D(t^8) (t^2 + t^9) (1-t^8)^{-1} + E(t)
\end{equation}
and we have used the stable homotopy groups $\pi_{n \geq 0}(\ko) = \pi_{n \geq 2}(\ko\langle2\rangle) = \KO^{-n}(\pt) = \{\bZ, \bZ_2, \bZ_2, 0, \bZ, 0, 0, 0\}$ for $n$ mod 8 taking the values $\{0, \dots, 7\}$. Explicitly, in the lowest degrees this reads
\begin{equation}
    \Omega_{n\geq0}^{\SO}(\pt)=\{\bZ,0,0,0,\bZ,\bZ_2,0,0,\bZ^2,\bZ_2^2,\bZ_2,\bZ_2,\bZ^3,\bZ_2^4,\bZ_2^2,\dots\}
\end{equation}
for oriented bordism,
and
\begin{equation}
    \Omega_{n\geq0}^\Spin(\pt)=\{\bZ,\bZ_2,\bZ_2,0,\bZ,0,0,0,\bZ^2,\bZ_2^2,\bZ_2^3,0,\bZ^3,0,0,\dots \}
\end{equation}
for spin bordism~\cite[Sec.\ 10]{bunke2009secondary}.

For the bordism groups $\Omega_n^{\Pin^-}(\pt)$, the computation of \cite{kirbytaylor} gives
\begin{equation}
    \sum_{n \geq 0} t^n \Omega_n^{\Pin^-}(\pt) = C(t^8) \sum_{l \geq 1} t^{4l-2} \bZ_{2^{2l + (l \text{ mod } 2)}} + D(t^8) t^4 \sum_{l \geq 1} t^{4l-2} \bZ_{2^{2l + (l+1 \text{ mod } 2)}} + G(t) \bZ_2
\end{equation}
where
\begin{equation}
    G(t) = C(t^8) (1 + t) (1-t^8)^{-1} + D(t^8) \big[ t^2 + (t^3 + t^4 + t^7 + 2t^8 + t^9) (1-t^8)^{-1} \big] + (1-t)^{-1} E(t)
\end{equation}
This reproduces the following bordism groups
\begin{equation}
    \Omega_{n\geq0}^{\Pin^-}(\pt)=\{\bZ_2,\bZ_2,\bZ_8,0,0,0,\bZ_{16},0,\bZ_2^2,\bZ_2^2, \bZ_2 \times \bZ_8 \times \bZ_{128}, \bZ_2\dots \}
\end{equation}
in low degrees~\cite{Kapustin:2014dxa}.

There is one more piece of information we need: from \cref{sec:4k}, the kernel of the forgetful map $\Omega_n^\Spin(\pt) \xlongrightarrow{p_n} \Omega_n^\SO(\pt)$ is
\begin{equation}
    \sum_{n \geq 0} t^n \ker(p_n) = H(t) \bZ_2
\end{equation}
where
\begin{equation}
    H(t) = [C(t^8) + D(t^8) t^8] (t + t^2) (1 - t^8)^{-1}
\end{equation}
We can now give the proper definitions of the series mentioned in \cref{sec:int}. Using the Smith isomorphism $\widetilde{\Omega}_n^\Spin(\B\bZ_2) = \Omega_{n-1}^{\Pin^-}(\pt)$, and the fact cobordism groups $\mho^n_H(X)$ are obtained from bordism groups $\Omega_n^H(X)$ by shifting the torsion up one degree, we find
\begin{itemize}

\item Isolated bosonic anomalies:
\begin{equation} \label{eq:countingfn1}
\begin{split}
    [B(t) - F(t) + H(t)]t &= [B(t) - E(t)]t - D(t^8) t^3 \\
    &= t^6 + 2t^{10} + t^{12} + \dots
\end{split}
\end{equation}

\item Fermionic anomalies that are images of bosonic anomalies:
\begin{equation} \label{eq:countingfn2}
\begin{split}
    [F(t) - H(t)]t &= D(t^8) t^3 + E(t) t \\
    &= t^{11} + 2t^{19} + t^{21} + \dots
\end{split}
\end{equation}

\item Fermionic anomalies that give rise to an anomaly for $(-1)^F$:
\begin{equation} \label{eq:countingfn3}
\begin{split}
    H(t)t &= [C(t^8) + D(t^8) t^8] (t^2 + t^3) (1 - t^8)^{-1} \\
    &= t^2 + t^3 + 2t^{10} + \dots
\end{split}
\end{equation}

\item Isolated fermionic $\bZ_2$ anomalies:
\begin{equation} \label{eq:countingfn4}
\begin{split}
    G(t)t^2 - H(t)t - D(t^8)t^4 &= D(t^8) (t^5+t^6+t^7+t^{10}) (1-t^8)^{-1} + E(t) t^2 (1-t)^{-1} \\
    &= t^{13} + t^{14} + \dots
\end{split}
\end{equation}

\item Copies of the first infinite structure:
\begin{equation} \label{eq:countingfn5}
    C(t^8) = 1 + t^8 + 2t^{16} + \dots
\end{equation}

\item Copies of the second infinite structure:
\begin{equation} \label{eq:countingfn6}
    D(t^8)t^4 = t^{12} + 2t^{20} + \dots
\end{equation}

\end{itemize}

\section{Stiefel--Whitney SPT Phases} \label{app:stiefelwhitneys}

For completeness, we list the technology we used to determine the SPT phases involving Stiefel--Whitney numbers stated in \cref{sec:int}. The \texttt{Sage} notebook \texttt{stiefel-whitneys.ipynb} was used to generate these results, and is shared as ancillary material with the \texttt{arXiv} submission of this paper.

Given a smooth unoriented $n$-manifold $M$, the Stiefel--Whitney numbers of $M$ are subject to certain relations called the \emph{Wu relations}. To derive them, start with the defining relation of the Wu classes
\begin{equation}
    \nu_i \smile w^J = \Sq^i(w^J)
\end{equation}
where $w^J$ represents a monomial in the Stiefel--Whitney classes, and $i + n(J) = n$ so that both sides have degree $n$. The right hand side can be reduced to a polynomial in the Stiefel--Whitney classes by recursively applying the Cartan formula
\begin{equation}
    \Sq^i(ab) = \sum_{j+k=i} \Sq^j(a) \Sq^k(b)
\end{equation}
and the Wu formula
\begin{equation}
    \Sq^i(w_j) =
    \begin{cases}
        0 & i > j \\
        w_j^2 & i = j \\
        \sum_{t=0}^i \binom{j+t-i-1}{t} w_{i-t} w_{j+t} & i < j
    \end{cases}
\end{equation}
The left hand side can be similarly reduced using additionally the recursive definition of the Wu classes
\begin{equation}
    \nu_i = w_i + \sum_{j = 1}^i \Sq^j(\nu_{i - j})
\end{equation}
Thus, for each $J$ with $n(J) \leq n$, we obtain a (generally nontrivial) relation among Stiefel--Whitney numbers. Dold \cite{Dold1956} proved that these give \emph{all} relations among Stiefel--Whitney numbers of an unoriented manifold.

For other tangential structures $H$, there are additional relations among Stiefel--Whitney numbers. They are
\begin{center}
\begin{tabular}{l|cccc}
    $H$ & $\SO$ & $\Spin$ & $\Pin^+$ & $\Pin^-$ \\
    \hline
    additional relations & $w_1 = 0$ & $w_1 = w_2 = 0$ & $w_2 = 0$ & $w_2 + w_1^2 = 0$
\end{tabular}
\end{center}
By supplementing these with the Wu relations, one can determine all Stiefel--Whitney numbers of oriented, spin, and $\pin^\pm$ manifolds. It is then straightforward to prove:
\begin{itemize}

\item The only nonzero Stiefel--Whitney number for $\Omega_5^\SO(\pt)$ is $w_2 w_3$.

\item The only nonzero Stiefel--Whitney number for $\Omega_{10}^\SO(\pt)$ is $w_4 w_6$.

\item The nonzero Stiefel--Whitney numbers for $\Omega_{11}^{\Pin^-}(\pt)$ are $w_1 w_4 w_6 = w_5 w_6$.

Under the Smith isomorphism to $\widetilde{\Omega}_{12}^\Spin(\B\bZ_2)$, the first of these maps to $A^2 w_4 w_6$.

\item The only nonzero Stiefel--Whitney number for $\Omega_{20}^\Spin(\pt)$ is $w_8 w_{12}$.

This is the first non-free-fermion SPT phase, namely the $t^{20}$ in $E(t)$.

\end{itemize}

\section{Clearing the powers of two in Anomaly Polynomials} \label{app:proofs}

In order to complete our computation of the map $p$, which goes from bosonic anomaly polynomials into fermionic ones, we need to prove some technical properties related to the normalisation of the fermionic anomaly polynomials; in particular, we need to know how to clear the 2-torsion in a general polynomial. We fill in these somewhat laborious details in this appendix.

To accompany and verify the proofs, we upload the \texttt{Sage} notebook \texttt{anom-polys.ipynb} with the \texttt{arXiv} version of this paper, which uses the formalism in this appendix to generate anomaly polynomials up to high degrees.

Recall that a fermionic polynomial for a general twisted Dirac operator, as discussed at length in the main text, takes the form
\begin{equation*}
    \Phi(i\slashed{D}_\theta) = \hat{A}(TM) \, \theta(\ph(\pi^r(TM)))
\end{equation*}
with $\hat{A}(TM)$ and $\ph(\pi^r(TM))$ given by Eqs.~\eqref{eq:Ahat} and~\eqref{eq:ph-1} respectively. We need to analyse the powers of 2 appearing in the denominators of both parts of this general anomaly polynomial.

\noindent
\textbf{Claim 1: } The $\hat{A}$-genus has the following property:
\begin{equation}
    \hat{A}_n(TM) = \frac{1}{24^n n!} \Big( p_1^n + 2\text{(terms with odd denominator)} \Big)
\end{equation}
This equation says that if we look at the $\hat{A}$-genus in dimension $4n$ as a polynomial in the Pontryagin classes, then the term with the most powers of two in the denominator is $p_1^n$, and that power of two is read off from the equation.

Note that the factor of $n!$ appearing in the denominator of this $p_1^n$ term is the origin of the bitcount function appearing in many of our formulae, such as our theorem for the 2-local version of Hattori--Stong~\eqref{eq:HS-4}, since $\nu_2(n!) = n - \text{bitcount}(n)$. Here $\nu_2(n!)$ denotes the 2-adic valuation, namely the exponent of the highest power of two that divides $n!$ (for example, $\nu_2(12)=2$ because $12$ is divisible by 4 but not 8).

\noindent
\textit{Proof: } This follows directly from the explicit formula \cite[(5.2)]{expliciths} for the coefficient of $p^J$ in the $\hat{A}$-genus, which reads
\begin{equation}
    \hat{A}(TM)\big|_{p^J} =
    \frac{(-1)^{|J|}}{\prod_j m_j(J)!}
    \sum_{S \in \Pi(|J|)}
    (-1)^{|S|}
    \prod_{s\in S} (|s|-1)! \, h_{n(J_s)}
\end{equation}
Here $|J|$ is the number of elements of $J$, $m_{j \geq 1}(J)$ is the multiplicity of $j$ in $J$, $\Pi(n)$ denotes all partitions of $\{1,\dots,n\}$ into disjoint subsets, and $J_{s \subseteq \{1,\dots,|J|\}} = \{J_i : i \in s\}$. For example if $J=\{6,2,2,1\}$, then $|J| = 4$, $m_2(J) = 2$, a typical partition $S \in \Pi(4)$ is $S=\{\{1,3\},\{2,4\}\}$, for which $J_{\{1,3\}} = \{6,2\}$ and $J_{\{2,4\}} = \{2,1\}$. The coefficients $h_n$ appearing above are defined by
\begin{equation}
    h_{n \geq 1}
    = (-1)^{n-1} \left(\frac{d}{dx}\log\frac{\sqrt{x}/2}{\sinh(\sqrt{x}/2)}\right)_{x^{n-1}}
    = \frac{(-1)^n B_{2n}}{2(2n)!}
    = \{-\tfrac{1}{24},-\tfrac{1}{1440},\dots\}
\end{equation}
where $B_n$ are the Bernoulli numbers. Using the fact $B_{2n} = \frac{1}{2\cdot\text{odd}}$, we have $\nu_2(h_n) = \text{bitcount}(n) - 2(n+1) \geq -3n$ with equality iff $n = 1$. Therefore, when we compute the coefficient of $p_1^n$, the $S=\{\{1\},\dots,\{1\}\}$ term is the unique term in the sum with the most powers of 2 in the denominator, and hence
\begin{equation}
    \hat{A}(TM)\big|_{p_1^n} = \frac{1}{24^n n!} + \text{(terms with fewer powers of 2 in denominator)}
\end{equation}
Using the exact same logic, the coefficient of any other Pontryagin monomial $p^J$ has strictly fewer powers of two in the denominator than this.

Having dealt with the $\hat{A}(TM)$ part, we now turn to the Pontryagin character part.

\noindent
\textbf{Claim 2: } The $\ph(\pi^r)$ have the following property:
\begin{equation}
    \ph_n(\pi^r) =
    \begin{cases}
        0 & n < r \\
        p_r & n = r \\
        \frac{1}{2^{2(n-r)}} \text{(terms with odd denominator)} & n > r
    \end{cases}
\end{equation}

\noindent
\textit{Proof: } 
The first two of these properties were proven in the main text. 
Recapping, we start 
from the generating function for $\ph(\pi^r)$ given in Eq.~\eqref{eq:ph-gen}.
The $t^r$ coefficient is
    $\ph(\pi^r(TM)) = 2^r e_r(\{\cosh x_i-1\})$. 
Taylor expanding gives 
$\ph(\pi^r(TM)) = p_r + \mathcal{O}(x^{2r+2})$
using $p_r=e_r(\{x_i^2\})$. 

To prove the third property, start by considering the first subleading term \emph{i.e.}\ the $n=r+1$ term, which is
\begin{equation}
    2^r \frac{1}{2^{r-1}}\frac{1}{4!}\sum_j x_j^4 \cdot \left[\text{coeff.\ of $t^{r-1}$ in~} \prod_{i\neq j} (1+x_i^2 t) \right]
\end{equation}
The prefactor is
    $2^r \frac{1}{2^{r-1}}\frac{1}{4!} = \frac{1}{2^2 \times \text{odd}}$.
We can write
\begin{equation}
    \prod_{i\neq j} (1+x_i^2 t) = \frac{1}{1+x_j^2 t} \prod_{i} (1+x_i^2 t) = \sum_{k\geq 0}(-1)^k x_j^{2k} t^k \times \sum_{m\geq 0} e_m(\{x_i^2\}) t^m
\end{equation}
Extracting the $t^{r-1}$ term, the part in large square brackets above is the finite sum $\sum_{k=0}^{r-1}(-1)^k e_{r-1-k}(\{x_i^2\}) x_j^{2k}$. Thus, the $\mathcal{O}(x^{2r+2})$ term we want is
\begin{equation}
    \ph(\pi^r(TM)) \supset \frac{1}{2^2 \times \text{odd}} \sum_{k=0}^{r-1}(-1)^k p_{r-1-k} \sum_j x_j^{2(k+2)}
\end{equation}
Finally, one can relate the power sum $\sum_j x_j^{2k+4}$ to elementary symmetric polynomials via the following partition formula, in which $J$ runs over partitions of $k+2$:
\begin{align}
    \sum_j (x_j^2)^{k+2} =
    \sum_{J\, \vdash\, k+2} (-1)^{k-|J|} (k+2) \frac{(|J|-1)!}{\prod_j m_j!} e_{\lambda}(\{x_i^2\})
    = p_1^{k+2} + \text{integral}
\end{align}
where $m_j$ is, as above, the multiplicity of $j$ occurring in the partition $J$.
The first term that we write explicitly on the RHS sets the normalisation of this power sum. The important thing is that it is integral, so
\begin{equation}
    \ph(\pi^r(TM)) \supset \frac{1}{2^2 \times \text{odd}} \left( \text{integral stuff}\right)
\end{equation}
This matches the property we sought to prove, for the case $n-r=1$.

This generalises going to higher order. Being a little schematic, we again pick out integral combinations of symmetric polynomials, with the same factor of $2^r$ in the numerator but larger factorials in the denominator. In particular, the piece of order $\mathcal{O}(x^{2r+2l})$ with the fewest number of powers of 2 in the denominator is that with $l$ factors of $x_i^4$ pulled out and $r-l$ factors of $x_i^2$, which has coefficient
\begin{equation}
    \frac{2^r}{2^{r-l}4!^{l} \times \text{odd}} \times \text{integral} \sim \frac{1}{4^l \times \text{odd}} \times \text{integral}
\end{equation}
Noting $l=n-r$, this proves the claim.

\noindent
\textbf{Claim 3: }
We now present the key claim that we needed in the main text, which is that
\begin{equation}
    \hat{A}(TM) \, \ph(\pi^J) |_n = \frac{1}{2^{4(n-n(J))-g(n-n(J))}} p_1^{n-n(J)} p^J + \text{(less powers of two in denominator)}
\end{equation}

\noindent
\textit{Proof:} Put together the previous two results. The basic point is that you can take higher-degree terms from the expansion of $\ph(\pi^r)$, instead of $p_r$, and you get a factor of $\frac{1}{4}$ for each degree you go up; but then you get a factor of $8$ from going down one degree in $\hat{A}(TM)$, so the overall result is more powers of 2 in the numerator.

This result proves the structure of the map $p$ that we state in the main text, which is a key ingredient in computing the bosonisation cohomology $H_B^{4k}=\ker(q_{4k})/\im(p_{4k})$.

\bibliographystyle{jhep}
\bibliography{refs}

@article{Coleman:1974bu,
    author = "Coleman, Sidney R.",
    editor = "Stone, M.",
    title = "{The Quantum Sine-Gordon Equation as the Massive Thirring Model}",
    reportNumber = "Print-74-1722 (HARVARD)",
    doi = "10.1103/PhysRevD.11.2088",
    journal = "Phys. Rev. D",
    volume = "11",
    pages = "2088",
    year = "1975"
}

@article{Mandelstam:1975hb,
    author = "Mandelstam, S.",
    editor = "Stone, M.",
    title = "{Soliton Operators for the Quantized Sine-Gordon Equation}",
    reportNumber = "Print-75-0198 (UC,BERKELEY)",
    doi = "10.1103/PhysRevD.11.3026",
    journal = "Phys. Rev. D",
    volume = "11",
    pages = "3026",
    year = "1975"
}

@article{Karch:2016sxi,
    author = "Karch, Andreas and Tong, David",
    title = "{Particle-Vortex Duality from 3d Bosonization}",
    eprint = "1606.01893",
    archivePrefix = "arXiv",
    primaryClass = "hep-th",
    doi = "10.1103/PhysRevX.6.031043",
    journal = "Phys. Rev. X",
    volume = "6",
    number = "3",
    pages = "031043",
    year = "2016"
}

@article{Chen:2017fvr,
    author = "Chen, Yu-An and Kapustin, Anton and Radi{\v{c}}evi{\'c}, {\DJ}or{\dj}e",
    title = "{Exact bosonization in two spatial dimensions and a new class of lattice gauge theories}",
    eprint = "1711.00515",
    archivePrefix = "arXiv",
    primaryClass = "cond-mat.str-el",
    doi = "10.1016/j.aop.2018.03.024",
    journal = "Annals Phys.",
    volume = "393",
    pages = "234--253",
    year = "2018"
}

@article{Chen:2019wlx,
    author = "Chen, Yu-An",
    title = "{Exact bosonization in arbitrary dimensions}",
    eprint = "1911.00017",
    archivePrefix = "arXiv",
    primaryClass = "cond-mat.str-el",
    doi = "10.1103/PhysRevResearch.2.033527",
    journal = "Phys. Rev. Res.",
    volume = "2",
    number = "3",
    pages = "033527",
    year = "2020"
}

@article{Tantivasadakarn:2020lhq,
    author = "Tantivasadakarn, Nathanan",
    title = "{Jordan-Wigner Dualities for Translation-Invariant Hamiltonians in Any Dimension: Emergent Fermions in Fracton Topological Order}",
    eprint = "2002.11345",
    archivePrefix = "arXiv",
    primaryClass = "cond-mat.str-el",
    doi = "10.1103/PhysRevResearch.2.023353",
    journal = "Phys. Rev. Res.",
    volume = "2",
    number = "2",
    pages = "023353",
    year = "2020"
}

@article{Chen:2021ppt,
    author = "Chen, Yu-An and Tata, Sri",
    title = "{Higher cup products on hypercubic lattices: Application to lattice models of topological phases}",
    eprint = "2106.05274",
    archivePrefix = "arXiv",
    primaryClass = "cond-mat.str-el",
    doi = "10.1063/5.0095189",
    journal = "J. Math. Phys.",
    volume = "64",
    number = "9",
    pages = "091902",
    year = "2023"
}

@article{Su:2025wbe,
    author = "Su, Lei and Martin, Ivar",
    title = "{Bosonization and Kramers-Wannier dualities in general dimensions}",
    eprint = "2508.20167",
    archivePrefix = "arXiv",
    primaryClass = "cond-mat.str-el",
    month = "8",
    year = "2025"
}

@article{bb,
    author = "Boyle Smith, Philip and Zheng, Yunqin",
    title = "{Backfiring Bosonisation}",
    eprint = "2403.03953",
    archivePrefix = "arXiv",
    primaryClass = "hep-th",
    month = "3",
    year = "2024"
}

@article{Seiberg_2024,
    title={Majorana chain and Ising model - (non-invertible) translations, anomalies, and emanant symmetries},
    volume={16},
    ISSN={2542-4653},
    url={http://dx.doi.org/10.21468/SciPostPhys.16.3.064},
    DOI={10.21468/scipostphys.16.3.064},
    number={3},
    journal={SciPost Physics},
    publisher={Stichting SciPost},
    author={Seiberg, Nathan and Shao, Shu-Heng},
    year={2024},
    month=mar
}

@article{bunke2009secondary,
    title={Secondary Invariants for String Bordism and tmf}, 
    author={Ulrich Bunke and Niko Naumann},
    year={2009},
    eprint={0912.4875},
    archivePrefix={arXiv},
    primaryClass={math.KT},
    url={https://arxiv.org/abs/0912.4875}, 
}

@article{KOchar,
    ISSN = {0003486X, 19398980},
    URL = {http://www.jstor.org/stable/1970470},
    author = {D. W. Anderson and E. H. Brown and F. P. Peterson},
    journal = {Annals of Mathematics},
    number = {1},
    pages = {54--67},
    publisher = {[Annals of Mathematics, Trustees of Princeton University on Behalf of the Annals of Mathematics, Mathematics Department, Princeton University]},
    title = {{SU}-corbodism, {KO}-characteristic Numbers, and the {K}ervaire Invariant},
    urldate = {2025-08-24},
    volume = {83},
    year = {1966}
}

@article{anominter,
    author = "Davighi, Joe and Lohitsiri, Nakarin",
    title = "{The algebra of anomaly interplay}",
    eprint = "2011.10102",
    archivePrefix = "arXiv",
    primaryClass = "hep-th",
    doi = "10.21468/SciPostPhys.10.3.074",
    journal = "SciPost Phys.",
    volume = "10",
    number = "3",
    pages = "074",
    year = "2021"
}

@article{lorentzfrac,
    author = "Hsin, Po-Shen and Shao, Shu-Heng",
    title = "{Lorentz Symmetry Fractionalization and Dualities in (2+1)d}",
    eprint = "1909.07383",
    archivePrefix = "arXiv",
    primaryClass = "cond-mat.str-el",
    reportNumber = "CALT-TH-2019-035",
    doi = "10.21468/SciPostPhys.8.2.018",
    journal = "SciPost Phys.",
    volume = "8",
    pages = "018",
    year = "2020"
}

@article{Callan:1984sa,
    author = "Callan, Jr., Curtis G. and Harvey, Jeffrey A.",
    title = "{Anomalies and Fermion Zero Modes on Strings and Domain Walls}",
    reportNumber = "Print-84-0860 (PRINCETON)",
    doi = "10.1016/0550-3213(85)90489-4",
    journal = "Nucl. Phys. B",
    volume = "250",
    pages = "427--436",
    year = "1985"
}

@article{Adler:1969gk,
    author = "Adler, Stephen L.",
    title = "{Axial vector vertex in spinor electrodynamics}",
    doi = "10.1103/PhysRev.177.2426",
    journal = "Phys. Rev.",
    volume = "177",
    pages = "2426--2438",
    year = "1969"
}

@article{Bell:1969ts,
    author = "Bell, J. S. and Jackiw, R.",
    title = "{A PCAC puzzle: $\pi^0 \to \gamma \gamma$ in the $\sigma$ model}",
    doi = "10.1007/BF02823296",
    journal = "Nuovo Cim. A",
    volume = "60",
    pages = "47--61",
    year = "1969"
}

@article{Witten:1982fp,
    author = "Witten, Edward",
    editor = "Shifman, Mikhail A.",
    title = "{An SU(2) Anomaly}",
    doi = "10.1016/0370-2693(82)90728-6",
    journal = "Phys. Lett. B",
    volume = "117",
    pages = "324--328",
    year = "1982"
}

@article{Wang:2018qoy,
    author = "Wang, Juven and Wen, Xiao-Gang and Witten, Edward",
    title = "{A New SU(2) Anomaly}",
    eprint = "1810.00844",
    archivePrefix = "arXiv",
    primaryClass = "hep-th",
    doi = "10.1063/1.5082852",
    journal = "J. Math. Phys.",
    volume = "60",
    number = "5",
    pages = "052301",
    year = "2019"
}

@inproceedings{Witten:2019bou,
    author = "Witten, Edward and Yonekura, Kazuya",
    title = "{Anomaly Inflow and the $\eta$-Invariant}",
    booktitle = "{The Shoucheng Zhang Memorial Workshop}",
    eprint = "1909.08775",
    archivePrefix = "arXiv",
    primaryClass = "hep-th",
    month = "9",
    year = "2019"
}

@article{Atiyah:1975jf,
    author = "Atiyah, M. F. and Patodi, V. K. and Singer, I. M.",
    title = "{Spectral asymmetry and Riemannian Geometry 1}",
    doi = "10.1017/S0305004100049410",
    journal = "Math. Proc. Cambridge Phil. Soc.",
    volume = "77",
    pages = "43",
    year = "1975"
}

@article{Atiyah:1976qjr,
    author = "Atiyah, M. F. and Patodi, V. K. and Singer, I. M.",
    title = "{Spectral asymmetry and Riemannian geometry. III}",
    doi = "10.1017/S0305004100052105",
    journal = "Math. Proc. Cambridge Phil. Soc.",
    volume = "79",
    pages = "71--99",
    year = "1976"
}

@article{Freed:2004yc,
    author = "Freed, Daniel S. and Moore, Gregory W.",
    title = "{Setting the quantum integrand of M-theory}",
    eprint = "hep-th/0409135",
    archivePrefix = "arXiv",
    doi = "10.1007/s00220-005-1482-7",
    journal = "Commun. Math. Phys.",
    volume = "263",
    pages = "89--132",
    year = "2006"
}

@article{Freed:2012bs,
    author = "Freed, Daniel S. and Teleman, Constantin",
    title = "{Relative quantum field theory}",
    eprint = "1212.1692",
    archivePrefix = "arXiv",
    primaryClass = "hep-th",
    doi = "10.1007/s00220-013-1880-1",
    journal = "Commun. Math. Phys.",
    volume = "326",
    pages = "459--476",
    year = "2014"
}

@article{Freed:2014iua,
    author = "Freed, Daniel S.",
    editor = "Donagi, Ron and Douglas, Michael R. and Kamenova, Ljudmila and Rocek, Martin",
    title = "{Anomalies and Invertible Field Theories}",
    eprint = "1404.7224",
    archivePrefix = "arXiv",
    primaryClass = "hep-th",
    doi = "10.1090/pspum/088/01462",
    journal = "Proc. Symp. Pure Math.",
    volume = "88",
    pages = "25--46",
    year = "2014"
}

@article{Dierigl:2025rfn,
    author = "Dierigl, Markus and Tartaglia, Michelangelo",
    title = "{(Quadratically) Refined discrete anomaly cancellation}",
    eprint = "2504.02934",
    archivePrefix = "arXiv",
    primaryClass = "hep-th",
    reportNumber = "CERN-TH-2025-068, IFT-25-35",
    doi = "10.1007/JHEP08(2025)145",
    journal = "JHEP",
    volume = "08",
    pages = "145",
    year = "2025"
}

@book{gilkey1989geometry,
  title={The geometry of spherical space form groups},
  author={Gilkey, Peter B},
  year={1989},
  publisher={World Scientific}
}

@article{Debray:2021vob,
    author = "Debray, Arun and Dierigl, Markus and Heckman, Jonathan J. and Montero, Miguel",
    title = "{The anomaly that was not meant IIB}",
    eprint = "2107.14227",
    archivePrefix = "arXiv",
    primaryClass = "hep-th",
    reportNumber = "LMU-ASC 24/21",
    doi = "10.1002/prop.202100168",
    journal = "Fortsch. Phys.",
    volume = "70",
    number = "1",
    pages = "2100168",
    year = "2022"
}

@book{freed2019lectures,
  title={Lectures on field theory and topology},
  author={Freed, Daniel S},
  volume={133},
  year={2019},
  publisher={American Mathematical Soc.}
}

@book{anderson1970universal,
  title={Universal coefficient theorems for K-theory},
  author={Anderson, Donald W},
  year={1970},
  publisher={MIT Department of Mathematics}
}

@article{brown1965abstract,
  title={Abstract homotopy theory},
  author={Brown, Edgar H},
  journal={Transactions of the American Mathematical Society},
  volume={119},
  number={1},
  pages={79--85},
  year={1965},
  publisher={JSTOR}
}

@article{Grady:2023sav,
    author = "Grady, Daniel",
    title = "{Deformation classes of invertible field theories and the Freed--Hopkins conjecture}",
    eprint = "2310.15866",
    archivePrefix = "arXiv",
    primaryClass = "math.AT",
    month = "10",
    year = "2023"
}

@article{Fidkowski:2009dba,
    author = "Fidkowski, Lukasz and Kitaev, Alexei",
    title = "{The effects of interactions on the topological classification of free fermion systems}",
    eprint = "0904.2197",
    archivePrefix = "arXiv",
    primaryClass = "cond-mat.str-el",
    doi = "10.1103/PhysRevB.81.134509",
    journal = "Phys. Rev. B",
    volume = "81",
    pages = "134509",
    year = "2010"
}

@article{Ryu:2012he,
    author = "Ryu, Shinsei and Zhang, Shou-Cheng",
    title = "{Interacting topological phases and modular invariance}",
    eprint = "1202.4484",
    archivePrefix = "arXiv",
    primaryClass = "cond-mat.str-el",
    doi = "10.1103/PhysRevB.85.245132",
    journal = "Phys. Rev. B",
    volume = "85",
    pages = "245132",
    year = "2012"
}

@article{Atiyah:1976jg,
    author = "Atiyah, M. F. and Patodi, V. K. and Singer, I. M.",
    title = "{Spectral asymmetry and Riemannian geometry 2}",
    doi = "10.1017/S0305004100051872",
    journal = "Math. Proc. Cambridge Phil. Soc.",
    volume = "78",
    pages = "405",
    year = "1976"
}

@article{Dai:1994kq,
    author = "Dai, Xian-zhe and Freed, Daniel S.",
    title = "{eta invariants and determinant lines}",
    eprint = "hep-th/9405012",
    archivePrefix = "arXiv",
    doi = "10.1063/1.530747",
    journal = "J. Math. Phys.",
    volume = "35",
    pages = "5155--5194",
    year = "1994",
    note = "[Erratum: J.Math.Phys. 42, 2343--2344 (2001)]"
}

@article{Witten:1985xe,
    author = "Witten, Edward",
    editor = "Salam, A. and Sezgin, E.",
    title = "{Global gravitational anomalies}",
    reportNumber = "PRINT-85-0246 (PRINCETON)",
    doi = "10.1007/BF01212448",
    journal = "Commun. Math. Phys.",
    volume = "100",
    pages = "197",
    year = "1985"
}

@article{Gaiotto:2015zta,
    author = "Gaiotto, Davide and Kapustin, Anton",
    editor = "Dokshitzer, Yuri L. and Levai, Peter and Nyiri, Julia",
    title = "{Spin TQFTs and fermionic phases of matter}",
    eprint = "1505.05856",
    archivePrefix = "arXiv",
    primaryClass = "cond-mat.str-el",
    doi = "10.1142/S0217751X16450445",
    journal = "Int. J. Mod. Phys. A",
    volume = "31",
    number = "28n29",
    pages = "1645044",
    year = "2016"
}

@article{Kapustin:2014dxa,
    author = "Kapustin, Anton and Thorngren, Ryan and Turzillo, Alex and Wang, Zitao",
    title = "{Fermionic Symmetry Protected Topological Phases and Cobordisms}",
    eprint = "1406.7329",
    archivePrefix = "arXiv",
    primaryClass = "cond-mat.str-el",
    doi = "10.1007/JHEP12(2015)052",
    journal = "JHEP",
    volume = "12",
    pages = "052",
    year = "2015"
}

@article{Cappelli:2025ecm,
    author = "Cappelli, Andrea and Villa, Riccardo",
    title = "{Bosonizations and dualities in 2+1 dimensions}",
    eprint = "2503.02801",
    archivePrefix = "arXiv",
    primaryClass = "hep-th",
    month = "3",
    year = "2025"
}

@article{bahri1987eta,
    title={The eta invariant, $\text{Pin}^c$ bordism, and equivariant $\text{Spin}^c$ bordism for cyclic 2-groups},
    author={Bahri, Anthony and Gilkey, Peter},
    journal={Pacific journal of mathematics},
    volume={128},
    number={1},
    pages={1--24},
    year={1987},
    publisher={Mathematical Sciences Publishers}
}

@article{Hsieh:2020jpj,
    author = "Hsieh, Chang-Tse and Tachikawa, Yuji and Yonekura, Kazuya",
    title = "{Anomaly Inflow and p-Form Gauge Theories}",
    eprint = "2003.11550",
    archivePrefix = "arXiv",
    primaryClass = "hep-th",
    reportNumber = "IPMU-20-0028, TU-1098",
    doi = "10.1007/s00220-022-04333-w",
    journal = "Commun. Math. Phys.",
    volume = "391",
    number = "2",
    pages = "495--608",
    year = "2022"
}

@article{spin,
    ISSN = {0003486X, 19398980},
    URL = {http://www.jstor.org/stable/1970690},
    author = {D. W. Anderson and E. H. Brown and F. P. Peterson},
    journal = {Annals of Mathematics},
    number = {2},
    pages = {271--298},
    publisher = {[Annals of Mathematics, Trustees of Princeton University on Behalf of the Annals of Mathematics, Mathematics Department, Princeton University]},
    title = {The Structure of the Spin Cobordism Ring},
    volume = {86},
    year = {1967}
}

@article{anderson1966spin,
    title={Spin cobordism},
    author={Anderson, DW and Brown Jr, EH and Peterson, FP},
    journal={Bulletin of the American Mathematical Society},
    volume={72},
    number={2},
    pages={256--261},
    year={1966},
    publisher={American Mathematical Society (AMS)}
}

@Article{anderson1969pin,
    author={Anderson, D. W.
    and Brown, E. H.
    and Peterson, F. P.},
    title={Pin cobordism and related topics},
    journal={Commentarii Mathematici Helvetici},
    year={1969},
    month={Dec},
    day={01},
    volume={44},
    number={1},
    pages={462-468},
    issn={1420-8946},
    doi={10.1007/BF02564545},
    url={https://doi.org/10.1007/BF02564545}
}

@Article{kirbytaylor,
    author={Kirby, R. C. and Taylor, L. R.},
    title={A calculation of $\text{Pin}^+$ bordism groups},
    journal={Commentarii Mathematici Helvetici},
    year={1990},
    month={Dec},
    day={01},
    volume={65},
    number={1},
    pages={434-447},
    issn={1420-8946},
    doi={10.1007/BF02566617},
    url={https://doi.org/10.1007/BF02566617}
}

@article{hs1,
    title = {Relations among characteristic numbers—{I}},
    journal = {Topology},
    volume = {4},
    number = {3},
    pages = {267-281},
    year = {1965},
    issn = {0040-9383},
    doi = {https://doi.org/10.1016/0040-9383(65)90011-X},
    url = {https://www.sciencedirect.com/science/article/pii/004093836590011X},
    author = {R.E. Stong}
}

@article{hs2,
    title = {Relations among characteristic numbers—{II}},
    journal = {Topology},
    volume = {5},
    number = {2},
    pages = {133-148},
    year = {1966},
    issn = {0040-9383},
    doi = {https://doi.org/10.1016/0040-9383(66)90014-0},
    url = {https://www.sciencedirect.com/science/article/pii/0040938366900140},
    author = {R.E. Stong}
}

@article{expliciths,
    title={Explicit formulas for the {H}attori-{S}tong theorem and applications},
    author={Ping Li and Wangyang Lin},
    year={2024},
    eprint={2409.05107},
    archivePrefix={arXiv},
    primaryClass={math.DG},
    url={https://arxiv.org/abs/2409.05107},
}

@article{Tachikawa_2019,
   title={Why are fractional charges of orientifolds compatible with {D}irac quantization?},
   volume={7},
   ISSN={2542-4653},
   url={http://dx.doi.org/10.21468/SciPostPhys.7.5.058},
   DOI={10.21468/scipostphys.7.5.058},
   number={5},
   journal={SciPost Physics},
   publisher={Stichting SciPost},
   author={Tachikawa, Yuji and Yonekura, Kazuya},
   year={2019},
   month=nov
}

@article{Witten_2016,
    title={Fermion path integrals and topological phases},
    volume={88},
    ISSN={1539-0756},
    url={http://dx.doi.org/10.1103/RevModPhys.88.035001},
    DOI={10.1103/revmodphys.88.035001},
    number={3},
    journal={Reviews of Modern Physics},
    publisher={American Physical Society (APS)},
    author={Witten, Edward},
    year={2016},
    month=jul
}

@article{freed2021,
    title={Consistency of {M}-{T}heory on nonorientable manifolds},
    author={Daniel S. Freed and Michael J. Hopkins},
    year={2021},
    eprint={1908.09916},
    archivePrefix={arXiv},
    primaryClass={hep-th},
    url={https://arxiv.org/abs/1908.09916},
}

@article{Freed:2016rqq,
    author = "Freed, Daniel S. and Hopkins, Michael J.",
    title = "{Reflection positivity and invertible topological phases}",
    eprint = "1604.06527",
    archivePrefix = "arXiv",
    primaryClass = "hep-th",
    doi = "10.2140/gt.2021.25.1165",
    journal = "Geom. Topol.",
    volume = "25",
    pages = "1165--1330",
    year = "2021"
}

@article{Hosseini:2025oka,
    author = "Hosseini, Saghar S. and Tachikawa, Yuji and Zhang, Hao Y.",
    title = "{Type I anomaly cancellation revisited}",
    eprint = "2505.07933",
    archivePrefix = "arXiv",
    primaryClass = "hep-th",
    month = "6",
    year = "2025"
}

@article{Dold1956,
    author={Dold, Albrecht},
    title={Vollst{\"a}ndigkeit der Wuschen Relationen zwischen den Stiefel-Whitneyschen Zahlen differenzierbarer Mannigfaltigkeiten},
    journal={Mathematische Zeitschrift},
    year={1956},
    month={Dec},
    day={01},
    volume={65},
    number={1},
    pages={200-206},
    issn={1432-1823},
    doi={10.1007/BF01473879},
    url={https://doi.org/10.1007/BF01473879}
}

@article{Alvarez-Gaume:1983ihn,
    author = "Alvarez-Gaume, Luis and Witten, Edward",
    editor = "Salam, A. and Sezgin, E.",
    title = "{Gravitational Anomalies}",
    reportNumber = "HUTP-83/A039",
    doi = "10.1016/0550-3213(84)90066-X",
    journal = "Nucl. Phys. B",
    volume = "234",
    pages = "269",
    year = "1984"
}

@article{Dijkgraaf:1989pz,
    author = "Dijkgraaf, Robbert and Witten, Edward",
    title = "{Topological Gauge Theories and Group Cohomology}",
    reportNumber = "THU-89-9, IASSNS-HEP-89-33",
    doi = "10.1007/BF02096988",
    journal = "Commun. Math. Phys.",
    volume = "129",
    pages = "393",
    year = "1990"
}

\end{document}